\documentclass[preprint,3p]{elsarticle}
\makeatletter
\def\ps@pprintTitle{%
  \let\@oddhead\@empty
  \let\@evenhead\@empty
  \let\@oddfoot\@empty
  \let\@evenfoot\@oddfoot
}
\makeatother
\usepackage{color}
\definecolor{light-gray}{gray}{0.80}
\usepackage{listings}
\usepackage{xcolor}
\usepackage{soul}
\definecolor{codegreen}{rgb}{0.95,0.95,0.92}
\usepackage{tabularx}    
\usepackage{siunitx}  
\usepackage{amsfonts}
\usepackage{amsmath}
\usepackage{bm}
\usepackage{amssymb}
\usepackage{amsthm}
\usepackage{pgfgantt}
\usepackage{tikz}
\usepackage{float}
\usepackage[colorlinks,citecolor=red,urlcolor=blue,bookmarks=false,hypertexnames=true]{hyperref} 
\usepackage{multirow}
\usepackage{graphicx}
\usepackage{lineno}
\usepackage{enumerate}
\usepackage{multirow}
\usepackage{array}
\usepackage{booktabs} 

\usepackage{multicol}
\usepackage{mathtools}

\usepackage{caption}
\usepackage{subcaption}
\usepackage{algorithm} 
\usepackage{algpseudocode}

\newcommand{\degree}{^{\circ}}
\newcommand{\lb}{l_{\text{b}}}
\newcommand{\Tb}{T_{\text{b}}}
\newcommand{\ub}{\mathbf{u}}
\newcommand{\Ub}{\mathbf{U}}
\newcommand{\fb}{\mathbf{f}}
\newcommand{\xm}{\mathbf{x}}
\newcommand{\qb}{\mathbf{q}}
\newcommand{\wb}{\mathbf{w}}
\newcommand{\pb}{\mathbf{p}}
\newcommand{\thetab}{\boldsymbol{\theta}}
\newcommand{\It}{\mathcal{I}_{\mathrm{therm}}(\phi)}
\newcommand{\Ie}{\mathcal{I}_{\mathrm{elas}}(\phi)}
\newcommand{\sigmB}{\boldsymbol{\sigma}^\mathrm{B}}
\newcommand{\sigmC}{\boldsymbol{\sigma}^\mathrm{C}}
\newcommand{\epsElasSym}{\boldsymbol{\epsilon}_\mathrm{elas}^{\mathrm{sym}}}
\newcommand{\epsElasSkew}{\boldsymbol{\epsilon}_\mathrm{elas}^{\mathrm{skew}}}
\newcommand{\epsElas}{\boldsymbol{\epsilon}_{\mathrm{elas}}}
\newcommand{\epstherm}{\boldsymbol{\epsilon}_{\mathrm{th}}}
\newcommand{\mb}{\mathbf{m}}
\newcommand{\tm}{\mathbf{t}_{\mathrm{m}}}
\newcommand{\boldb}{\mathbf{b}}
\newcommand{\boldI}{\mathbf{I}}
\newcommand{\tsigma}{\mathbf{t}_{\sigma}}
\newcommand{\boldn}{\mathbf{n}}

\newcommand{\Nthetax}{\mathcal{N}_{\theta}(\mathbf{x})}
\newcommand{\Nux}{\mathcal{N}_{u}(\mathbf{x})}
\newcommand{\Ntx}{\mathcal{N}_{T}(\mathbf{x})}
\newcommand{\Ntheta}{\mathcal{N}_{\theta}}
\newcommand{\Nu}{\mathcal{N}_{u}}
\newcommand{\Nt}{\mathcal{N}_{T}}

\newcommand{\Bskewx}{\mathcal{B}_{u}^{\mathrm{skew}}(\mathbf{x})}
\newcommand{\Bsymx}{\mathcal{B}_{u}^{\mathrm{sym}}(\mathbf{x})}
\newcommand{\Bthetax}{\mathcal{B}_\theta(\mathbf{x})}
\newcommand{\Btx}{\mathcal{B}_T(\mathbf{x})}
\newcommand{\Btheta}{\mathcal{B}_\theta}
\newcommand{\Bt}{\mathcal{B}_T}

\usetikzlibrary{calc,arrows.meta}
\definecolor{darkteal}{rgb}{0.0, 0.4, 0.4}
\definecolor{slateblue}{rgb}{0.42, 0.35, 0.80}
\definecolor{deeporange}{rgb}{0.85, 0.37, 0.01}
\usepackage[normalem]{ulem} 
\usepackage{xcolor}          

\usetikzlibrary{shapes.arrows,shadows,decorations.pathmorphing}
\tikzstyle{startstop} = [rectangle, rounded corners, minimum width=3cm, minimum height=1cm, align=center, text width=4cm, draw=black, fill=red!30 ]
\tikzstyle{io} = [trapezium, trapezium left angle=70, trapezium right angle=110, minimum width=3cm, minimum height=1cm, align=center, text width=4cm, draw=black, fill=blue!30]
\tikzstyle{process} = [rectangle, minimum width=3cm, minimum height=1cm, align=center, text width=4cm, draw=black, fill=orange!30]
\tikzstyle{process1} = [rectangle, minimum width=5cm, minimum height=1cm, align=center, text width=5cm, draw=black, fill=orange!30]
\tikzstyle{decision} = [diamond, minimum width=1cm, minimum height=1cm, align=center, text width=4cm, draw=black, fill=green!30]
\tikzstyle{arrow} = [thick,->,>=stealth]
\hypersetup{pdfauthor={Name}}
\usepackage{setspace}

\newif\ifdarkmode
\darkmodefalse
\ifdarkmode
	\definecolor{commentcolor}{rgb}{1,0,0}
	\definecolor{sourcecolor}{rgb}{1,0,0}
	\pagecolor{black}\color{white} 
	\definecolor{newblue}{rgb}{0.7,0.7,1}
\else
	\definecolor{commentcolor}{rgb}{1,0,0}
	\definecolor{sourcecolor}{rgb}{0,.5,0}
	\definecolor{newblue}{rgb}{0,0,1}
\fi
\begin{document}
\onehalfspacing
\begin{frontmatter}


\title{Level set-based topology optimization of micropolar solids under thermo-mechanical loading}



\author[1]{Mayank Shekhar}
\author[1]{Ayyappan Unnikrishna Pillai}
\author[2]{Subhayan De}
\author[1]{Mohammad Masiur Rahaman\corref{cor1}}
\ead{masiurr@iitbbs.ac.in}
\address[1]{School of Infrastructure, Indian Institute of Technology Bhubaneswar, Odisha - 752050, India}
\cortext[cor1]{Corresponding author}
\address[2]{Department of Mechanical Engineering, Northern Arizona University, Flagstaff, AZ 86011, USA
}

\begin{abstract}
We propose a novel level set-based topology optimization for micropolar solids subjected to thermo-mechanical loading. To capture the size effects, we have incorporated the microstructural length-scale information into the level set-based topology optimization method by adopting a micropolar theory. The proposed non-local topology optimization method can provide accurate topology optimization for size-dependent solids under thermo-mechanical loading. We have demonstrated the effectiveness of the proposed method through a few representative two-dimensional benchmark problems. The numerical results reveal the substantial influence of underlying micro-structures, incorporated in the model through micropolar parameters, and temperature on topology optimization, highlighting the necessity of the proposed thermo-mechanical micropolar formulation for materials with pronounced non-local effects. For the numerical implementation of the proposed model, we have used open-source finite element libraries, \texttt{Gridap.jl}, and \texttt{GridapTopOpt.jl}, available in Julia, to ensure transparency and reproducibility of the reported computational results. 
\end{abstract}

\end{frontmatter}

\section{Introduction}
\label{sec:Introduction} 
Thermo-mechanical loading is a critical factor influencing the performance, safety, and longevity of engineering structures. In many practical scenarios, materials undergo thermal expansion or contraction in response to temperature variations, which can significantly impact their strength, durability, and stability. Typical examples include civil engineering structural components subjected to fire \cite{madsen2016topology}, thermo-chemical heat storage \cite{chen2020topology}, automobile exhaust thermo-electric generator system \cite{he2024optimized}, and so on. To ensure the safety and reliability of such systems under combined thermal and mechanical effects, meticulous design strategies are essential 
\cite{fan2025thermo,ooms2023compliance}. A central challenge in this context is achieving lightweight structures that can safely withstand severe thermal and mechanical loads. Reducing material usage not only enhances efficiency but also lowers production costs and environmental impact \cite{hostos2021computational, fang2022topology, zheng2023stress}. However, lightweight designs must balance material reduction with sufficient stiffness and thermal conductivity, especially in high-risk applications \cite{ooms2024thermoelastic, xue2024thermoelastic,onodera2025topology}. This challenge can be effectively addressed through topology optimization, a powerful computational approach that has emerged. By systematically distributing material within a given design domain, it produces highly efficient and innovative structural geometries that are tailored to specific loading conditions, constraints, and material properties \cite{bendsoe1989optimal}. The optimized geometries from this approach offer superior performance compared to conventional designs based solely on engineering intuition \cite{wu2021topology}. Over the years, classical mechanics-based topology optimization methods have been developed to effectively address macroscale thermo-mechanical problems in fields such as aeronautics, civil, and automotive engineering \cite{urso2024thermomechanical,wang2025multi}. However, the rapid advancements in additive manufacturing, which enable the fabrication of intricate geometries at the microscale, have shifted research interests toward microscale devices \cite{shin2025topology,wu2024machine,xu2024topology}. Prominent examples include thermal actuators \cite{sigmund2001design}, Bio-MEMS (micro-electro-mechanical systems) heaters \cite{pandey2023recent,heo2008minimum}, micro-resonators \cite{fu2023efficient}, and micro-heat-exchangers \cite{fawaz2022topology}. At small scales, conventional topology optimization often fails to capture size-dependent and micro-rotational effects, necessitating non-local continuum-based methods for micro- and nano-scale systems under coupled thermo-mechanical loading.

Modern topology optimization builds on Michell’s optimal truss theory \cite{michell1904lviii}, with key advancements from homogenization-based approaches by \citet{bendsoe1988generating} and the simplified solid isotropic material with penalization (SIMP) method \cite{bendsoe1989optimal}. While density-based methods remain prevalent, boundary-based techniques offer sharper and more well-defined structural interfaces. Among these, the \textit{level set method} (LSM) \cite{allaire2002level,allaire2004structural} has emerged as a leading approach. The LSM represents material boundaries implicitly as iso-contours of a level set function, transforming topology optimization into a shape evolution problem. This framework ensures smooth boundary representation and avoids numerical instabilities like checkerboarding. Originally developed for image segmentation and multiphase flow modeling \cite{sussman1994level}, the method was later refined for structural optimization by \citet{allaire2002level,allaire2004structural,wang2003level}. Today, LSM is widely adopted for generating crisp, manufacturable designs with clear structural boundaries. 

Following these developments, the LSM has been extended to address coupled multiphysics problems, such as thermo-elastic topology optimization. In this context, \citet{rodrigues1995material} pioneered the formulation for thermo-elastic structures using the homogenization approach with the objective of minimizing structural compliance under the combined influence of mechanical and thermal loads, providing benchmark examples for future work. Later, \citet{xia2008topology} applied the LSM to minimize compliance while considering both volume and thermal constraints. \citet{yamada2011level} focused on steady-state heat transfer problems using the level set approach. Sigmund \cite{sigmund2001design,sigmund2001design2} studied thermal actuators driven by thermal expansion. \citet{cheng2025thermomechanical} proposed a compatibility-driven topology optimization framework to overcome the limitations of transformation-based methods in designing multifunctional thermo-mechanical cloaks. \citet{deng2017topology} investigated thermo-elastic buckling using the level set method, relevant to supersonic aircraft.  Recently, robust topology optimization has been used to design meta-materials with negative thermal expansion under design uncertainties \cite{li2020robust,muayad2025thermo,akamatsu2024optimal}. Reliability-based topology optimization of thermo-elastic structures using the bi-directional evolutionary structural optimization method was studied by \citet{habashneh2023reliability}. Wang et al. \cite{wang2025level} improved the level set approach for geometrically nonlinear structures with thermo-mechanical coupling and derived a new formula for element coupling stresses caused by combined thermo-mechanical loads.  \citet{torisaki2024micro} proposed a homogenization-based micro-scale shape optimization method that uses adjoint sensitivities to reduce thermal stresses in periodic porous microstructures under heat conduction. Recently, \citet{wang2025topology} presents an efficient, compactly supported radial basis function-based parameterized level set method for smooth and stable thermo-mechanical topology optimization. \citet{wang2025multiscale} couples transient heat conduction with thermo-mechanical loading using SIMP and homogenization to capture time-dependent effects on macro- and micro-structures. For a detailed review of thermo-elastic topology optimization, readers can refer to \citet{deaton2014survey,guibert2025introducing,van2013level,gibou2018review}. 

While conventional topology optimization has been successful at larger scales, its applicability at smaller scales is limited by the assumptions of classical continuum mechanics. Classical continuum theories assume stress transfer only between adjacent points, with rotations entirely governed by the displacement field, neglecting independent micro-rotations and associated micro-structural effects. Classical theories also assume continuous distribution of matters, neglecting complex microstructures such as granular materials \cite{merkel2011experimental,chang1991micromechanical}, cellular solids \cite{zhang2006scale}, lattices \cite{wu2019topology,yan2025multiscale}, porous media \cite{diebels2014micromechanical,rezaei2024equivalent}, and fiber-reinforced composites \cite{huang2025multiscale}.  Several studies \cite{zhang2005quadrilateral,burgueno2005hierarchical} have shown that these theories inaccurately predict the flexural stiffness of cellular beams when the beam height is comparable to the cell size. To address these issues, \citet{voigt1887theoretische} introduced the concept of \textit{couple stresses}, which represents moment-based interactions in addition to conventional force stresses, which gives rise to the asymmetric couple stress theory. \citet{cosserat1909theorie} further generalized this by introducing independent micro-rotations, allowing each material point six degrees of freedom: three displacements and three rotations. Although their formulation was general, it was mathematically complex, and later works \cite{gunther1958statik, toupin1964theories, eringen1964nonlinear, mindlin1964micro} provided refinements in more practical settings. A linear version was established in studies such as \cite{pal1964fundamental, eringen1966linear, mindlin1965stress}. Eringen \cite{eringen1964nonlinear,eringen1966theory} then expanded the theory to include microinertia effects and renamed it the micropolar theory of elasticity, recognizing micro-rotation as independent of translation. Cowin \cite{cowin1969singular,cowin1970stress} later introduced the dimensionless \textit{micropolar coupling number} $N$ ($0 \leq N \leq 1$), which bridges these formulations: $N=1$ recovers couple-stress theory, $N=0$ reduces to classical elasticity, and intermediate values represent the micropolar framework.

Based on these generalized continuum theories, such as couple stress, micropolar, and strain-gradient formulations, researchers have incorporated them into topology optimization frameworks to account for size-dependent effects. \citet{liu2010topology} and \citet{cong2024topology} applied couple-stress-based approaches to topology optimization, with the latter employing a modified couple-stress model to design periodic composite plates exhibiting vibration band gaps. Later, \citet{zhao2024concurrent} applied the modified couple stress theory for multiscale design involving multi-phase materials. \citet{li2024topology} presents a topology optimization approach for microscale structures using integral nonlocal theory to account for size effects. Similarly, \citet{singh2025nonlocal} developed a nonlocal integral model for thickness-dependent stiffness in additive manufacturing and integrated it into topology optimization to prevent unrealistically thin features. \citet{li2015topology} integrated Mindlin’s microstructure theory \cite{mindlin1964micro} with topology optimization to study size effects on structural designs. Strain-gradient theory uses higher-order terms to capture microstructural behavior and reduce stress concentrations \cite{lin2023cross,li2017topology,li2022isogeometric}. However, these formulations require higher-order partial differential equations and shape functions, reducing computational efficiency. In comparison, micropolar models avoid these requirements and offer a physically intuitive representation through rotational degrees of freedom. The first micropolar topology optimization model was introduced by \citet{rovati2007optimal}, who studied optimal topologies using isotropic micropolar continua and reported results that differed from those of conventional models. Since then, micropolar solids have been used in topology optimization to investigate stiffness, vibration modes, and auxetic behaviors \cite{veber2012topology,bruggi2012maximization}. Furthermore, \citet{chen2021parameterized} used a parameterized level set-based topology optimization method for Cosserat solids under elastic loading, with the goal of minimizing compliance. Recently, \citet{zhou2025machine,zhou2026topology} employed a machine learning-driven approach to perform topology optimization of micropolar solids under elastic loading conditions for two- and three-dimensional problems. While previous studies explored non-local and size-dependent effects in topology optimization, their influence under coupled thermo-mechanical loading, which is critical for the performance of micro- and nanoscale devices, remains largely unaddressed.

In this study, we propose a nonlocal topology optimization method that integrates micropolar theory within the level set-based topology optimization framework to capture size-dependent material behaviors under coupled thermo-mechanical loading effectively. This approach enables the design of structures that are not only optimized for stiffness and thermal performance but also incorporate the size effects in the continuum model. To the best of the authors’ knowledge, this work represents the first effort to develop a non-local topology optimization-based design methodology specifically tailored for solids where thermo-mechanical effects and size-dependent phenomena are simultaneously significant. We have used open-source finite element libraries called \texttt{Gridap.jl} \cite{badia2020gridap, verdugo2022software} and \texttt{GridapTopOpt.jl} \cite{wegert2025gridaptopopt} in Julia to implement the proposed model. The remainder of this article is organized as follows: Section~\ref{sec:Background} describes the background of micropolar thermo-elasticity and the level set method for topology optimization. Section~\ref{sec:proposedmodel} details the proposed micropolar level set-based topology optimization for thermo-elastic solids. Section~\ref{sec:numericalresults} provides representative numerical examples to demonstrate the effectiveness of the proposed approach. Finally, Section~\ref{sec:conclusions} summarizes the main findings, discusses the advantages and limitations of the methodology, and outlines potential avenues for future research.

\section{Background}
\label{sec:Background}
In this section, we provide the essential background on micropolar thermo-elasticity and the level set method. Micropolar thermo-elasticity extends classical thermo-elasticity by incorporating microrotations and couple stresses, making it important for materials where size effects are prominent. The level set method represents shapes implicitly using a scalar function $\phi$, where the interface of the material domain is characterized by the level set $\phi = c$, where   $c$ is a constant in $\mathbb{R}$.
\subsection{Micropolar thermo-elasticity}
\label{sec:micro-polarthermoelasticity}
In the framework of micropolar continuum mechanics, the strain tensor $\boldsymbol{\epsilon}$ at a material point $\xm \in \mathcal{D}$ and time $t \in \mathbb{R}_{\geq 0}$ is expressed as a function of the displacement field $\ub$ and microrotation field $\thetab$ as
\begin{equation}
\boldsymbol{\epsilon}(\ub, \thetab) = \nabla \ub^\top - \overset{3}{\mathbb{E}}\,\thetab,
\end{equation}
where $\nabla$ denotes the spatial gradient operator, $(\cdot)^\top$ indicates the transpose, and $\overset{3}{\mathbb{E}}$ represents the third-order Levi-Civita permutation tensor, which facilitates the appropriate coupling between microrotations and the strain field. Moreover, under thermo-mechanical loading, the total strain is additively decomposed into an elastic part $\epsElas$ and a thermal part $\epstherm$, \textit{i.e.}, 
\begin{equation} \boldsymbol{\epsilon} = \epsElas + \epstherm, \label{eq:strainDecomposition} \end{equation}
where the thermal strain $\epstherm$ is given by
\begin{equation} \epstherm = \alpha_t\,(T - T_0)\,\boldI, \label{eq:ThermalStrain} \end{equation}
with $\alpha_t$ denoting the coefficient of thermal expansion, $T$ is the temperature field, $T_0$ is the reference temperature, and $\boldI$ is the second-order identity tensor. The elastic part of the micropolar strain tensor can be decomposed into its symmetric and skew-symmetric parts, capturing classical and non-classical deformation modes, respectively, as
\begin{equation}
\epsElas(\ub,\boldsymbol{\theta}) = \underbrace{\frac{1}{2}(\nabla\ub^{\top} + \nabla\ub) -\epstherm}_{= \epsElasSym} + \underbrace{\frac{1}{2}(\nabla\ub^{\top} - \nabla\ub) - \overset{3}{\mathbb{E}}\,\thetab}_{= \epsElasSkew},
    \label{eq:sym_skewsym_strain_decomp}
\end{equation}
where $\epsElasSym$ corresponds to the symmetric component of the elastic strain associated with the classical continuum contribution, while $\epsElasSkew$ corresponds to the skew-symmetric part driven by microrotational effects. Note that the thermal strain $\epstherm$ is a second-order symmetric tensor. Furthermore, the curvature tensor $\boldsymbol{\varkappa}$
is defined as the gradient of the microrotation field variable, \textit{i.e.}, $\boldsymbol{\varkappa}(\thetab) = \nabla \thetab$. 

\begin{figure}[H]
    \centering
\includegraphics[width=0.6\linewidth]{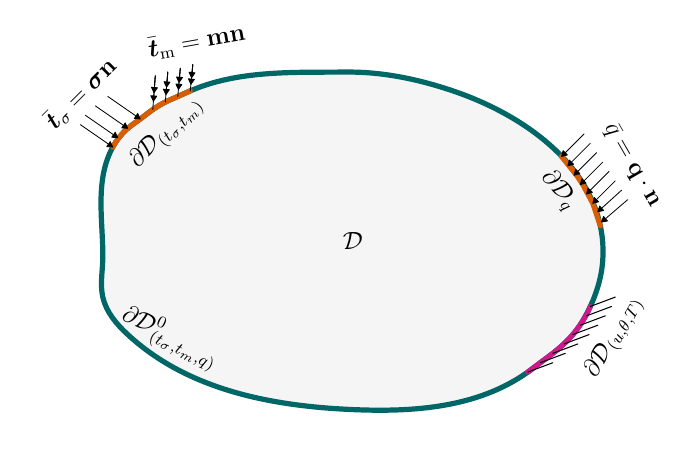}
\caption{Schematic representation of a material domain, $\mathcal{D}$, with a smooth boundary,  $\partial  \mathcal{D}$, subjected to specified mechanical and thermal boundary conditions. Dirichlet conditions are prescribed on $\partial  \mathcal{D}_{(u,\theta, T)}$ (indicated in magenta), whereas the Neumann-type conditions—namely, the applied traction $\bar{\mathbf{t}}_\sigma = \boldsymbol{\sigma}\,\boldn$, $\bar{\mathbf{t}}_\sigma  = \mb\,\boldn$ and heat flux $\bar{q} = \qb \cdot \boldn$—are enforced on $\partial  \mathcal{D}_{(t_{\sigma},t_m)}$ and $\partial  \mathcal{D}_{q}$ (shown in orange). The remainder of the boundary, $\partial  \mathcal{D}^{0}_{(t_{\sigma},t_{m},q)}$ (highlighted in green), is subjected to homogeneous Neumann conditions, implying vanishing tractions and heat flux. These boundary subsets satisfy the partitioning relations $\partial  \mathcal{D}_{(t_{\sigma},t_\mathrm{m},q)} = \partial  \mathcal{D}_{(t_{\sigma},t_\mathrm{m})} \cup \partial  \mathcal{D}_q \cup \partial  \mathcal{D}^{0}_{(t_{\sigma},t_{m},q)}$ and $\partial  \mathcal{D} = \partial  \mathcal{D}_{(u,\theta,T)} \cup \partial  \mathcal{D}_{(t_{\sigma},t_\mathrm{m},q)}$.}
    \label{fig:DesignDomainIllustrate}
\end{figure}

In the micropolar continuum framework, the mechanical response of the system incorporates both classical and non-classical stress measures. Specifically, the internal forces comprise the couple-stress tensor $\mb$ and the microrotation-induced traction $\tm$, in addition to the symmetric force-stress tensor $\sigmB$ and the coupling stress tensor $\sigmC$. The symmetric force-stress tensor $\sigmB$ corresponds to the symmetric elastic strain $\epsElasSym$, while the coupling stress tensor $\sigmC$ is linked to its skew-symmetric counterpart $\epsElasSkew$. The system is further subjected to body forces $\boldb$ and boundary tractions $\tsigma$. The schematic representation of a miocropolar continuum body is illustrated in Fig.~\ref{fig:DesignDomainIllustrate}.

The mechanical strain energy density, $\Psi_{\mathrm{elas}}$, is given by  
\begin{equation} \Psi_{\mathrm{elas}}\left(\epsElasSym\right) = \frac{1}{2} \biggl[\lambda\, (\mathrm{tr}(\epsElasSym))^2 + (2\mu + \kappa)\,(\epsElasSym : \epsElasSym)\biggr].
\label{eq:PosPartElastEnerg_Vol-Dev} \end{equation}
The coupling strain energy density, $\Psi_{\mathrm{coup}}$ and the rotational strain energy density, $\Psi_{\mathrm{rot}}$ are expressed as 
\begin{equation}
    \Psi_{\mathrm{coup}}(\epsElasSkew) = \frac{1}{2}\biggl(\kappa\, \epsElasSkew: \epsElasSkew\biggr),
    \label{eq:Coupling-Energy}
\end{equation}
and
\begin{equation}
\Psi_{\mathrm{rot}}(\boldsymbol{\varkappa}) = \frac{1}{2}\biggl( \alpha\,(\boldsymbol{\varkappa}:\boldI)^2 + \beta\,(\boldsymbol{\varkappa}:\boldsymbol{\varkappa}^\top)+\gamma\,(\boldsymbol{\varkappa}:\boldsymbol{\varkappa})\biggr),
    \label{eq:Rotational-Energy}
\end{equation}
respectively. The parameters $\lambda$ and $\mu$ are the Lame's parameters while $\alpha,~\beta,~\gamma$, and $\kappa$ are the micropolar constants. The constitutive relations for the stress and couple-stress measures in terms of the strain and curvature tensors are defined below.
\begin{equation}
\sigmB =
\lambda\,\text{tr}\left(\epsElasSym \right) \boldI
+ (2\mu + \kappa) \,\epsElasSym,
\label{eq:Stress-Expression_Vol-Dev}
\end{equation}
\begin{equation}
\sigmC =
\kappa\,\epsElasSkew,
\label{eq:ElasticCouplingStress-Expression}
\end{equation}
\begin{equation}
\mb =
\alpha (\boldsymbol{\varkappa} : \boldI)\,\boldI + \beta\,\boldsymbol{\varkappa}^\top + \gamma\,\boldsymbol{\varkappa},
\label{eq:CoupledStress}
\end{equation}
where $\sigmB$ denotes the symmetric stress part associated with classical strain energy, while $\sigmC$ in Eq.~\eqref{eq:ElasticCouplingStress-Expression} corresponds to the antisymmetric (or coupling) stress arising from microrotation interactions. The couple-stress tensor $\mb$ in Eq.~\eqref{eq:CoupledStress} accounts for curvature-dependent contributions to the internal moment.  One can refer to \citet{eringen1984theory} for a detailed derivation of the expression. To facilitate interpretation and material modeling, the following relationships express standard mechanical constants and characteristic micropolar parameters in terms of the primary material coefficients \cite{yavari2002fractal, lakes2015bending}: 
\begin{equation}
\begin{multlined}
E = \frac{(2\mu + \kappa)(3\lambda + 2\mu + \kappa)}{2\lambda +2\mu + \kappa}, \qquad
G = \frac{2\mu + \kappa}{2},\qquad  \nu = \frac{\lambda}{2\lambda + 2\mu + \kappa},\\
l_{\text{t}} = \sqrt{\frac{\beta + \gamma}{2\mu + \kappa}},\qquad \lb = \sqrt{\frac{\gamma}{2(2\mu + \kappa)}},\qquad N = \sqrt{\frac{\kappa}{2(\mu+\kappa)}}, \qquad \chi = \frac{\beta + \gamma}{\alpha + \beta + \gamma},
\end{multlined}
\label{eq:MaterialConstants}
\end{equation}
where $E$, $G$, $\nu$, $l_{\text{t}}$, $\lb$, $\chi$, and $N$ are Young’s modulus, shear modulus, Poisson’s ratio, torsional characteristic length, bending characteristic length, polar ratio, and the micropolar coupling number, respectively. In a general three-dimensional (3D) setting, the microrotation field is described by three independent rotational degrees of freedom. However, in two-dimensional (2D) problems, this vector field reduces to a scalar variable, with only the microrotation component $\theta_3$ being necessary to characterize the rotational behavior of the material. Consequently, for 2D formulations, the higher-order material constants $\alpha$ and $\beta$—associated with specific curvature modes—can be neglected without loss of generality, significantly simplifying the constitutive modeling. Under this simplification, the relevant set of material properties reduces to the $E$, $G$, $\nu$, $\lb$, and $N$. Among these, $\lb$ serves as a micro-structural length scale parameter that governs the nonlocal bending behavior of the continuum. It plays a critical role in distinguishing micropolar formulations from classical elasticity models, where such scale effects are absent \cite{bavzant1999size}. 
The micropolar coupling number $N$ quantifies the extent of micropolarity present in the material. The value of $N$ ranges from $0$ to $1$, where $N = 0$ corresponds to the classical Cauchy continuum (\textit{i.e.}, zero couple stress), whereas $N = 1$ represents a fully coupled stress theory. This parametric control allows the proposed framework to seamlessly interpolate between classical and micropolar models, enhancing its versatility in capturing both local and size-dependent effects. For this reason, in the numerical examples presented in section~\ref{sec:numericalresults}, both the micropolar coupling number $N$ and the characteristic bending length scale $\lb$ will be systematically varied to demonstrate their influence on the optimized structural responses.

\subsubsection{Governing PDEs and the strong form}
\label{sec:BVP}
In this section, we present the governing partial differential equations (PDEs) of the micropolar thermo-elasticity problem. Assuming the absence of body forces, \textit{i.e.}, $\boldb = \boldsymbol{0}$, the linear momentum balance equation can be expressed in terms of the kinematic variables as
\begin{equation} 
\nabla \cdot \left( \sigmB +  \kappa\, \epsElasSkew \right) = \boldsymbol{0}, \label{eq:Disp-StrongForm} 
\end{equation}
with macroscopic traction vector acting on a boundary surface with outward normal $\boldn$ is given by $\tsigma ( \boldn ) = \left( \sigmB + \sigmC \right) \boldn$, where the stress tensor $\sigmB$ is expressed as given in Eq.~\eqref{eq:Stress-Expression_Vol-Dev}. By utilizing the expressions for $\sigmC$ and $\mb$ as given in Eq.~\eqref{eq:ElasticCouplingStress-Expression} and Eq.~\eqref{eq:CoupledStress}, the angular momentum balance equation can also be defined in terms of the kinematic variables as: 
\begin{equation}
\nabla \cdot \left( \alpha (\boldsymbol{\varkappa} : \boldI)\, \boldI + \beta\, \boldsymbol{\varkappa}^{\top} + \gamma \,\boldsymbol{\varkappa} \right) + \overset{3}{\mathbb{E}} \,\kappa\, \epsElasSkew = 0, \label{eq:ElectricPotential-StrongForm} 
\end{equation}
with the microrotation-induced traction vector acting on the boundary is given by $\tm ( \boldn ) = \mb \,\boldn$.

The temperature evolution equation can be written for a steady state problem \cite{rahaman2017dynamic} as 
\begin{equation}
  -\nabla\cdot \qb + Q_{\text{h}} = 0,
    \label{eq:TempStrongForm}
\end{equation}
where $\qb = -k\,\nabla T$ is the heat flux with $k$ representing the thermal conductivity of the material. The PDEs given by Eq.~\eqref{eq:Disp-StrongForm}, Eq.~\eqref{eq:ElectricPotential-StrongForm}, and Eq.\,\eqref{eq:TempStrongForm} are coupled equations with specified $\ub$, applied traction, $T$, and $\thetab$ on $\partial \mathcal{D}_u$, $\partial \mathcal{D}_{t}$, $\partial \mathcal{D}_{T}$, and $\partial \mathcal{D}_{\theta}$, respectively. These PDEs, along with the boundary conditions, are typically referred to as the strong form of the problem, as given below:
\begin{equation}
 \begin{aligned}
\nabla\cdot\left(\sigmB + \kappa\,\epsElasSkew \right) = {\bf{0}}  \,\,\,\,\text{in}\,\,\,\,\mathcal{D},\\
\nabla\cdot \left(\alpha\, (\boldsymbol{\varkappa}:\boldI)\,\boldI + \beta\, \boldsymbol{\varkappa}^{\top} + \gamma\,\boldsymbol{\varkappa}\right) + \overset{3}{\mathbb{E}}\,\kappa\,\epsElasSkew = {\bf{0}}\,\,\,\,\text{in}\,\,\,\,\mathcal{D},\\
 - \nabla \cdot (k\,\nabla T) = Q_{\text{h}}\,\,\,\,\text{in}\,\,\,\,\mathcal{D},\\
\ub = \bar{\ub} \,\,\,\,\text{on}\,\,\,\,\partial \mathcal{D}_u,\\
\boldsymbol\theta = \boldsymbol{\bar\theta} \,\,\,\,\text{on}\,\,\,\,\partial \mathcal{D}_{\theta},\\
 {T} = \bar{{T}} \,\,\,\,\text{on}\,\,\,\,\partial \mathcal{D}_{T},\\
(\sigmB+\sigmC)\,\boldn = {\bf{0}}\,\,\,\,\text{on}\,\,\,\,\partial \mathcal{D}\backslash\partial \mathcal{D}_u,\\
 \mb\,\boldn = {\bf{0}}\,\,\,\,\text{on}\,\,\,\,\partial \mathcal{D}\backslash\partial \mathcal{D}_{\theta}, \\
  \qb\cdot\boldn =0\,\,\,\,\text{on}\,\,\,\,\partial \mathcal{D}\backslash\partial \mathcal{D}_T.
\end{aligned}
\label{eq:StrongFormDisplacement}
\end{equation}
The weak form of the strong formulation can be derived to solve for displacement, micro-rotation, and temperature within a multi-field finite element framework. 

\subsection{Topology optimization based on level set method}
\label{sec:levelsetTopOPt} 
In this section, we provide a description of the level set method for topology optimization \cite{allaire2002level,sethian2001evolution,wei201888}. In this approach, the material domain is represented by a scalar function whose evolution governs the shape and topology of the structure. This method provides an implicit description of structural boundaries, which makes it well-suited for capturing complex geometries and topological changes during the optimization process.

\subsubsection{Level set method}
\begin{figure}[H]
\centering
\begin{subfigure}[t]{0.48\linewidth}
\centering
\caption{Initial level set function $\phi_0$ for a two-dimensional problem domain.}
\includegraphics[width=\linewidth]{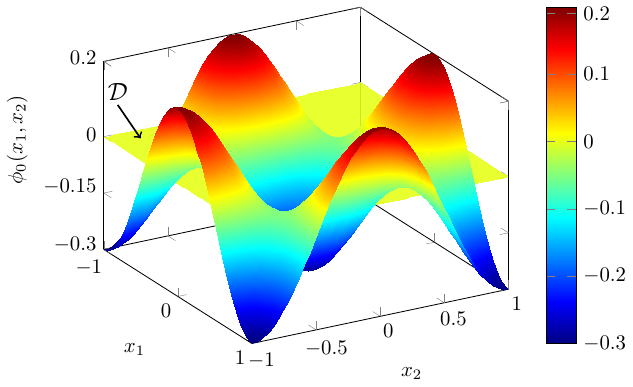}
\label{fig:levelset}
\end{subfigure}
\hfill
\begin{subfigure}[t]{0.5\linewidth}
\centering
\caption{Initialization of the problem domain through intersection of $\phi_0$ with the $\phi_0 = 0$ plane.}
\includegraphics[width=\linewidth]{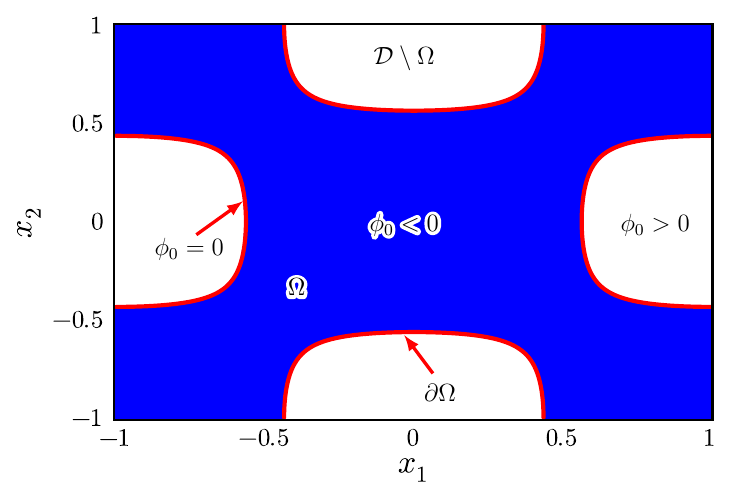}
\label{fig:levelsetcontour}
\end{subfigure}    \caption{Initial level set function $\phi_0(x_1,x_2)$ in a two-dimensional domain for $\zeta = 1$ and $b = 0.2$. (a) Initial level set function $\phi_0(x_1,x_2)$ with the reference plane at $\phi_0(x_1,x_2) = 0$, representing the design domain $\mathcal{D}$. (b) Intersection of $\phi_0$ with the reference plane at $\phi_0(x_1,x_2) = 0$. The blue region ($\phi_0(x_1,x_2) < 0$) denotes the material domain $\Omega$, the red line ($\phi_0(x_1,x_2) = 0$) indicates the evolving material boundary $\partial\Omega$, and the white region ($\phi_0(x_1,x_2) > 0$) corresponds to the void domain, $\mathcal{D}\setminus\Omega$.}
\label{fig:LevelsetFunction}
\end{figure}

The level set method is a powerful tool for implicitly representing the boundary of a sub-domain $\Omega$ within a bounding domain $\mathcal{D} \subset \mathbb{R}^d$. Instead of explicitly tracking the boundary, the method represents it through a scalar function, known as the level set function, $\phi: \mathcal{D} \rightarrow \mathbb{R}$. The interface $\partial \Omega$ is characterized by the level set $\phi = c$, with $c=0$ adopted here, while the sign of $\phi$ distinguishes whether a point lies inside or outside the material domain \cite{osher2001level}. Thus, function $\phi(\xm)$ is defined as 
\begin{equation}
   \begin{cases} 
  \phi(\xm) < 0 & \text{if } \xm \in \Omega, \\
  \phi(\xm) = 0 & \text{if } \xm \in \partial \Omega, \\
  \phi(\xm) > 0 & \text{if } \xm \in \mathcal{D} \setminus {\Omega}.
\end{cases} 
\end{equation}
In this formulation, the level set function distinguishes the material and void regions: $\phi(\xm) < 0$ corresponds to the material domain, $\phi(\xm) > 0$ represents the void, and $\phi(\xm) = 0$ defines the boundary separating the two. Several strategies exist for initializing the level set function to start the optimization. In this work, the initial level set function $\phi_0$ is defined using a trigonometric expression, following \citet{wegert2025gridaptopopt},
\begin{equation}
\label{eq:LSF}
    \phi_0(\xm) = -\frac{1}{4}\prod_{i = 1}^{d} \big(\cos(\zeta \pi x_i)\big) - \frac{b}{4},
\end{equation} 
where $\{x_i\}_{i=1}^d$ are the components of the spatial coordinate vector $\xm$, and $d$ denotes the spatial dimension ($d=2$ for two-dimensional problems, $d=3$ for three-dimensional problems).  The parameters $\zeta$ and $b$ govern the initial periodicity of the level set function and the size of the initial holes, respectively. For instance, an increase in $b$ shifts the level set function downward, thereby reducing the radius of the holes. These values are considered to create an equally spaced sufficient number of holes in the design domain $\mathcal{D}$. The initial level set function $\phi_0$ and its intersection with the reference plane at $\phi_0 = 0$ are shown in Fig.~\ref{fig:LevelsetFunction}. The figure also indicates the corresponding solid material domain, material boundary, and void region.
Having defined the initial level set function, the next step is to formulate a strategy for integrating over the evolving material domain $\Omega$ and its boundary $\partial \Omega$ while working within a fixed computational domain $\mathcal{D}$. This can be achieved by expressing any function $f:\mathbb{R}^d \to \mathbb{R}$ in terms of the level set function as
\begin{equation}
    \int_{\Omega} f \, \mathrm{d}V 
    = \int_{\mathcal{D}} f \,\big(1 - H_{\chi}(\phi)\big) \, \mathrm{d}V,
    \label{fig:OmegatoD}
\end{equation}  
where $H_{\chi}(\phi)$ is the Heaviside function associated with the level set field, defined as  
\begin{equation}
    H_{\chi}(\phi) =
    \begin{cases}
        0 & \text{if } \phi \leq 0, \\[6pt]
        1 & \text{if } \phi > 0.
    \end{cases}
    \label{eq:Hfunc}
\end{equation}  
To ensure smooth integration, $H_{\chi}(\phi)$ is replaced by a regularized Heaviside function $H_\eta(\phi)$, as given by:  
\begin{equation}
    H_{\eta}(\phi) =  \begin{cases}
0 & \text{if}~\phi < -\eta, \\
\frac{1}{2} + \frac{\phi}{2\eta} + \frac{1}{2\pi}\sin\left(\frac{\pi \phi}{\eta}\right) & \text{if} ~ |\phi| \leq \eta, \\
1 & \text{if}~\phi > \eta, 
    \end{cases}
\end{equation}
where $\eta$ is the smoothing radius, equal to half the transition width between solid ($H_{\eta}(\phi)=0$) and void ($H_{\eta}(\phi)=1$). In the level set framework, the weak form is relaxed over the full domain $\mathcal{D}$ (Eq.~\eqref{fig:OmegatoD}), introducing void regions that make the stiffness matrix singular \cite{allaire2004structural}. To avoid this, the \textit{ersatz material approximation} replaces voids with a fictitious soft material, enabling PDE solution without body-fitted meshes. This is achieved by a smooth characteristic function $\mathcal{I}:\mathbb{R}\to[\varepsilon,1]$.
\begin{equation}
    \mathcal{I}(\phi) = (1 - H_{\eta}(\phi)) + \varepsilon H_{\eta}(\phi),
\label{eq:CharacteristicFunc}
\end{equation}
where $\varepsilon \ll 1$ is a numerical parameter related to void properties. An additional function called \emph{density function} is introduced, defined as  
\begin{equation}
    \rho(\phi) = 1 - H_\eta(\phi),
\end{equation}
which smoothly distinguishes between material and void. Unlike $\mathcal{I}(\phi)$, this function is used to restrict source terms like body force, heat generation, and heat flux so that they act only within the evolving material domain $\Omega$. Furthermore, it provides a convenient way to evaluate the current material volume via $\int_{\mathcal{D}} \rho(\phi)~\mathrm{d}V$, during optimization.

\subsubsection{Level set function updation}
The evolution of the boundary $\partial\Omega$ for small pseudo time $t \in (0,\mathcal{T})$ under a normal velocity field $v_{\mathrm{n}}$ is described by Hamilton-Jacobi evolution equation \cite{osher2001level,peng1999pde} as
\begin{equation}
    \begin{cases}
        \dfrac{\partial \phi(t,\xm)}{\partial t} + v_{\mathrm{n}}(\xm) \left| \nabla \phi(t,\xm) \right| = 0, \\
        \phi(0,\xm) = \phi_0(\xm), \\
        \xm \in \mathcal{D},\quad t \in (0, \mathcal{T}),
    \end{cases}
    \label{eq:HJ_eq}
\end{equation}
where $t$ denotes the pseudo-time and $\mathcal{T}>0$ the total evolution duration. The level set, initialized by $\phi_0$, evolves with the normal velocity $v_{\mathrm{n}}$, obtained from shape sensitivities using a Hilbertian extension-regularization framework as discussed in \ref{sec:AugmentLag} to ensure smooth descent. Here $v_{\mathrm{n}}$ is the normal component of the velocity vector $\mathbf{v}$, defined at each point in the domain. In this work, the descent direction is driven by the augmented Lagrangian functional, which enforces equality constraints through multiplier updates and penalty terms, iteratively updating both the domain and multipliers toward the optimum. Among the several available approaches for constrained level set optimization--such as the Lagrangian method \citep{allaire2004structural}, sequential linear programming, and Hilbertian projection methods \citep{wegert2023hilbertian}--the present work adopts the augmented Lagrangian framework as discussed in \ref{sec:AugmentLag}.

\subsubsection{Level set function reinitialization}
To maintain a well-behaved level set function $\phi$, particularly around interfaces $\partial\Omega$, it is desirable that the gradient magnitude \textit{i.e.}, $\lvert \nabla \phi \rvert$ remains close to unity. This is achieved by reinitializing the level set function $\phi$ as a signed distance function $d_{\Omega}$ \cite{wegert2025gridaptopopt}. 
For the sub-domain $\Omega$, the signed distance function is defined as:
\begin{equation}
d_{\Omega} = 
    \begin{cases}
        -d(\xm,\Omega) & \text{if}~~\xm\in \Omega, \\
        0 &\text{if}~~\xm\in \partial\Omega, \\
        d(\xm,\Omega) & \text{if}~~\xm\in \mathcal{D} \setminus {\Omega},
    \end{cases}
\end{equation}
where, $d(\xm,\partial\Omega):= \text{min}_{\pb \in \partial\Omega} |\xm - \pb| $ is the minimum Euclidean distance from $\xm$ to the boundary $\partial\Omega$. For reinitialization from a pre-existing level set function $\phi_0(\xm)$ toward a signed distance function, is performed using following reinitialization equation \citep{wegert2025gridaptopopt}:
\begin{equation}
    \begin{cases}
        \dfrac{\partial \phi(t,\xm)}{\partial t} + \text{sign}(\phi_0(\xm)) \left( \lvert \nabla \phi(t,\xm) \rvert - 1 \right) = 0, \\
        \phi(t=0, \xm) = \phi_0(\xm), \\
        \xm \in \mathcal{D},\quad t > 0,
    \end{cases}
    \label{eq:HJ_Signed}
\end{equation}
where $\mathrm{sign}(\cdot)$ is a regularized sign function given by $\mathrm{sign}(\phi) = {\phi}/{\sqrt{\phi^2 + \mathrm{h}^2 \lvert \nabla \phi \rvert^2}}$, which ensures stability during evolution with $\mathrm{h}$ denoting the finite element mesh size parameter of the discretized domain. The Hamilton–Jacobi evolution equation (Eq.~\eqref{eq:HJ_eq}) and the signed-distance reinitialization equation (Eq.~\eqref{eq:HJ_Signed}) are solved using standard first-order Godunov upwind finite-difference schemes \cite{peng1999pde,allaire2004structural}. The function $\text{sign}(\phi)$ is updated at every time step, and Eq.~\eqref{eq:HJ_Signed} is iteratively solved until a near-steady state solution is achieved.

\section{Proposed level set-based topology optimization for thermo-elastic micropolar solids}
\label{sec:proposedmodel} 
In this section, we present the proposed level set-based topology optimization framework for micropolar solids subjected to thermo-elastic loading. The formulation non-trivially extends conventional thermo-elastic topology optimization \cite{xia2008topology,yamada2011level,deng2017topology} by incorporating micropolar kinematics and constitutive relations to capture size-dependent effects. In particular, the governing PDEs, the optimization problem formulation, and the sensitivity analysis using an adjoint formulation are developed for the coupled thermo-elastic micropolar solids. The proposed formulation ensures that rotational degrees of freedom and microstructural length-scale effects are consistently integrated into the topology optimization process.

\subsection{Governing PDEs for topology optimization of micropolar thermo-elastic solids}
\label{sec:Interpolation}
In order to incorporate the evolving material distribution within the fixed computational domain, the classical strong form of the governing equations must be modified. This is achieved by introducing additional smooth characteristic functions, which interpolate the material properties between solid and void regions without requiring a body-fitted mesh, as described in Eq.~\eqref{eq:CharacteristicFunc}. 
For thermo-elastic micropolar problems, the material properties are interpolated using two characteristic functions: 
\begin{align}
    \Ie &= (1 - H_{\eta}(\phi)) + {\varepsilon}_{\mathrm{elas}} H_{\eta}(\phi), \\
     \It &= (1 - H_{\eta}(\phi)) + {\varepsilon}_{\mathrm{therm}} H_{\eta}(\phi),
\end{align}
where $\Ie$ is employed to interpolate the elastic and micro-rotation properties between solid and void, with $\varepsilon_{\mathrm{elas}}$ denoting the void stiffness parameter. Similarly, $\It$ interpolates the thermal properties, where $\varepsilon_{\mathrm{therm}}$ represents the void thermal conductivity parameter. By incorporating these interpolations into the material definitions, the original strong form given in Eq.~\eqref{eq:StrongFormDisplacement} is modified accordingly. The resulting governing PDEs in their modified strong form can be expressed as,
\begin{equation}
\begin{aligned}
\nabla \cdot \left( \Ie\left( \sigmB + \kappa \,\epsElasSkew\right) \right) 
  &= \mathbf{0} \quad\quad\quad\,\, \text{in } \mathcal{D}, \\
\nabla \cdot \Big( \Ie( \alpha (\boldsymbol{\varkappa}:\boldI)\,\boldI 
  + \beta \boldsymbol{\varkappa}^\top + \gamma \boldsymbol{\varkappa} ) \Big) 
  + \Ie(\overset{3}{\mathbb{E}}\,\kappa \,\epsElasSkew) 
  &= \mathbf{0} \quad\quad\quad\,\, \text{in } \mathcal{D}, \\
- \nabla \cdot \Big( \It\, k \nabla T \Big) 
  &= \rho(\phi) \,Q_{\text{h}} \quad \text{in } \mathcal{D}, \\
\ub &= \bar{\ub} \quad\quad\quad\,\, \text{on } \partial \mathcal{D}_u, \\
\thetab &= \bar{\thetab} \quad\quad\quad\,\, \text{on } \partial \mathcal{D}_{\theta}, \\
T &= \bar{T} \quad\quad\quad\,\, \text{on } \partial \mathcal{D}_{T}, \\
\left( \Ie(\sigmB+\boldsymbol{\sigma}^c) \right)\,\boldn 
  &= \mathbf{0} \quad\quad\quad\,\, \text{on } \partial \mathcal{D} \setminus \partial \mathcal{D}_u, \\
(\Ie\,\mb)\,\boldn 
  &= \mathbf{0} \quad\quad\quad\,\, \text{on } \partial \mathcal{D} \setminus \partial \mathcal{D}_{\theta}, \\
(\It\,\qb)\cdot\boldn 
  &= 0 \quad\quad\quad\,\, \text{on } \partial \mathcal{D} \setminus \partial \mathcal{D}_T.
\end{aligned}
\label{eq:StrongFormDisplacementModified}
\end{equation}
The weak form corresponding to the above strong form in Eq.~\eqref{eq:StrongFormDisplacementModified} is derived next for the finite element formulation of the proposed model. 

\subsection{Weak form and finite element formulation}
\label{sec:Weakform}
The weak form is derived herein using a standard Galerkin finite element procedure and serves as the basis for the numerical implementation of the proposed micropolar topology optimization framework. Let $\wb$, $\boldsymbol{\vartheta}$ and $\tau$ denote suitable test functions corresponding to the displacement field $\ub$, the microrotation field $\thetab$, and temperature field $T$, respectively. The trial spaces for the primary variables are defined as: 
\begin{equation}
\begin{aligned}
    \mathbb{H}_\text{u} &= \{\ub \in \mathbb{H}^1(\mathcal{D}) ; \; \ub = \bar{\ub} \; \text{on} \; \partial \mathcal{D}_{u} \}, \\
    \mathbb{H}_{\theta} &= \{\boldsymbol\theta \in \mathbb{H}^1( \mathcal{D}) ; \; \boldsymbol \theta = \boldsymbol{\bar{\theta}} \; \text{on} \; \partial \mathcal{D}_{\theta} \}, \\
    \mathbb{H}_\text{T} &= \{T \in \mathbb{H}^1( \mathcal{D}) ; \; T= \bar{T} \; \text{on} \; \partial \mathcal{D}_{T}\},
\end{aligned}
\label{eq:TrialSpaceuthetaT} 
\end{equation} 
where $\mathbb{H}^1(\mathcal{D})$ is the Hilbert space, $\bar{\ub}$, $\bar{\thetab}$ and $\bar{T}$ denote the prescribed displacement, rotation and temperature fields on the Dirichlet boundary $\partial \mathcal{D}_{u}$, $\partial \mathcal{D}_{\theta}$ and $\partial \mathcal{D}_{T}$, respectively. Correspondingly, the test function spaces are defined as
\begin{equation}
\begin{aligned}
    \mathbb{V}_\text{u} &= \{\wb \in \mathbb{H}^1( \mathcal{D}); \; \wb = \boldsymbol{0} \; \text{on} \; \partial \mathcal{D}_{u} \}, \\
   \mathbb{V}_{\theta} &= \{\boldsymbol{\vartheta} \in \mathbb{H}^1( \mathcal{D}); \; \boldsymbol{\vartheta} = \boldsymbol{0} \; \text{on} \; \partial \mathcal{D}_{\theta} \}, \\
   \mathbb{V}_\text{T} &= \{\tau \in \mathbb{H}^1( \mathcal{D}); \; \tau = \boldsymbol{0} \; \text{on} \; \partial \mathcal{D}_{T} \}.
\end{aligned}
\label{eq:TestSpaceuthetaTemp}
\end{equation}
Adopting a multi-field formulation, the weak form corresponding to the balance of linear and angular momentum equations under external loading $\bar{\fb}$ is given by
\begin{equation}
\begin{aligned}
a_{\text{MP}}((\ub,\thetab,T),(\wb,\boldsymbol{\vartheta},\tau),\phi) &= 
\int_{\mathcal{D}} \Ie\biggl[\epsilon(\wb):\sigmB 
+ \epsElasSkew(\wb,\boldsymbol{\vartheta}) :\kappa\epsElasSkew \\
&\qquad + \nabla\boldsymbol{\vartheta}:\bigl(\alpha (\boldsymbol{\varkappa}:\boldI)\,\boldI 
+ \beta \,\boldsymbol{\varkappa}^{\top} + \gamma\,\boldsymbol{\varkappa}\bigr)
- \boldsymbol{\vartheta}\cdot \kappa\,\overset{3}{\mathbb{E}}\,\epsElasSkew\biggr] \,\mathrm{d}V \\
&\qquad + \int_{\mathcal{D}} \It \left[\nabla \tau \cdot (k \,\nabla T)\right]\mathrm{d}V, \\[6pt]
b_{\text{MP}}((\wb, \boldsymbol{\vartheta},\tau),\phi) &= 
\int_\Gamma (\wb \cdot \bar{\fb})\,\mathrm{d}S 
+ \int_{\mathcal{D}}\Ie\left( \boldsymbol{\epsilon}(\wb):\boldsymbol{\sigma}_F \right)\, \mathrm{d}V \\
&\qquad + \int_{\mathcal{D}}\rho(\phi)\left(\tau\,Q_{\text{h}}\right)\, \mathrm{d}V,
\end{aligned}
\label{eq:cc}
\end{equation}

where $\bar{\fb}$ is the traction vector on the traction boundary $\Gamma$. Here, $\boldsymbol{\sigma}_F$ is defined as
\begin{equation}
  \boldsymbol{\sigma}_F = -\lambda\text{tr}\left(\alpha_t T_0 \boldI\right) \boldI
- (2\mu + \kappa) \alpha_t T_0 \boldI,
 \label{eq:StressTotAndThermal}
\end{equation}
where $T_0$ is the initial temperature of the domain.
In the numerical implementation, we employ a multi-field Galerkin finite element discretization using the \texttt{Gridap.jl} package \cite{badia2020gridap, verdugo2022software}, which enables the solution of coupled displacement and rotation fields directly through user-defined weak forms. The weak formulation presented in Eq.~\eqref{eq:cc} serves as a direct input to Gridap’s variational cell assembly framework, facilitating efficient and modular simulation of micropolar mechanics for topology optimization.
\subsection{Finite element implementation}
\label{sec:FEImplementation}
This section presents the finite element procedure for solving the weak form presented in Eq.~\eqref{eq:cc}. The discretization described below is formulated in a general setting so that it is independent of the choice of element type, interpolation order, and problem dimension. For the $e^{\mathrm{th}}$ element, the primary
variables $\ub_h^e(\xm)$, $\thetab_h^e(\xm)$, and $T_h^e(\xm)$ are written using finite element interpolation as
\begin{equation}
\ub^e_h(\xm) = \Nux \,\ub^e,\quad \thetab^e_h(\xm) = \Nthetax\,\thetab^e,\,\,\mathrm{and}\quad T^e_h(\xm) = \Ntx\,{T}^e,
\end{equation}
where $\Nu,\Ntheta,\text{and}~ \Nt$ are the interpolation functions corresponding to the degree of freedom of displacement, rotation, and temperature, respectively, and $\ub^e$, $\thetab^e$, and $T^e$ are the corresponding vectors of nodal degrees of freedom for the $e^{\mathrm{th}}$ element. The corresponding derivatives can be expressed as
\begin{gather}
\epsilon_{\mathrm{elas},h}^\mathrm{sym} = \Bsymx\,\mathbf{u}^e 
- \alpha_t\!\left(\Ntx\,T^e - T_0\right)\mathbf{I}, \quad
\epsilon_{\mathrm{elas},h}^\mathrm{skew} = \Bskewx\,\ub^e 
- \overset{3}{\mathbb{E}}\,\Nthetax\,\thetab^e,\\
\boldsymbol{\varkappa}_h = \Bthetax\,\thetab^e, \,\mathrm{and} \quad \nabla T_h = \Btx\,{T}^e,
\end{gather}
where, $\mathcal{B}_u^\mathrm{sym}$, and $\mathcal{B}_u^\mathrm{skew}$ are the symmetric and skew-symmetric part of $\mathcal{B}_u = \nabla \mathcal{N}_u$, $\Btheta = \nabla \Ntheta$, and $\Bt = \nabla \Nt$. Similarly, the test function and its derivatives can be expressed as
\begin{gather}
\tilde{\ub}^e_h(\mathbf{x}) =  \Nux \,\tilde{\ub}^e,\quad \tilde{\thetab}^e_h(\xm) =  \Nthetax\,\tilde{\thetab}^e,\,\mathrm{and}\,\,\tilde{T}^e_h(\xm) = \Ntx\,\tilde{T}^e, \\
\tilde{\epsilon}_{\mathrm{elas},h}^{\mathrm{sym}} = \Bsymx\,\tilde{\ub}^e 
- \alpha_t\!\left(\Ntx\,{\tilde{T}}^e - T_0\right)\mathbf{I}, \quad
\tilde{\epsilon}_{\mathrm{elas},h}^\mathrm{skew} = \Bskewx\,\tilde{\ub}^e 
- \overset{3}{\mathbb{E}}\Nthetax\,\tilde{\thetab}^e,\\
\boldsymbol{\tilde{\varkappa}}_h = \Bthetax\,\tilde{\thetab}^e, \,\mathrm{and} \quad \nabla \tilde{T}_h = \Btx\,\tilde{T}^e,
\end{gather}
where $\tilde{\ub}^e$, $\tilde{\thetab}^e$, and $\tilde{T}^e$ are the test functions of the $e^\mathrm{th}$ element. The domain for each element is defined by an open set $\mathcal{D}_e \subset \mathcal{D}$ such that $\cup_{\text{e}=1}^{N_e}\mathcal{D}_e = \mathcal{D}$, where $N_e$ is the total number of elements. Thus, for the sub-domain $\mathcal{D}_e$, the weak form (see Eq.~\eqref{eq:cc}) in the discretized form can be expressed as 

\begin{equation}
\begin{aligned}
\int_{\mathcal{D}_e} &\Ie\biggl[\Bsymx:\mathbb{D}\,\Bsymx\,\ub^e -\Bsymx:\mathbb{D}\,\alpha_t\,\Ntx\, T_e\,\mathbf{I} 
+  \Bskewx:\kappa\,\Bskewx\,\ub^e \\
&-  \Bskewx:\kappa\, \overset{3}{\mathbb{E}}\,\Nthetax\,\thetab^e
- \overset{3}{\mathbb{E}}\,\Nthetax:\kappa\,\Bskewx\,\ub^e 
+ \overset{3}{\mathbb{E}}\,\Nthetax:\kappa\,
\overset{3}{\mathbb{E}}\,\Nthetax\,\thetab^e 
\\
&+\Bthetax:\mathbb{R}\,\Bthetax\,\thetab^e
- \Nthetax\cdot \kappa\,\overset{3}{\mathbb{E}}\,\Bskewx\,\ub^e  
+\Nthetax\cdot \kappa\,\overset{3}{\mathbb{E}}\,\overset{3}{\mathbb{E}}\,\Nthetax\,\thetab^e\biggr] \,\mathrm{d}V_e \\
& + \int_{\mathcal{D}_e} \It \biggl[\Btx\, \cdot k \,\Btx\,{T}^e\biggr]\mathrm{d}V_e 
=
\int_\Gamma (\Nux \cdot \bar{\fb})\,\mathrm{d}S_e \\
&- \int_{\mathcal{D}_e}\Ie\left( \Bsymx:\mathbb{D} \,\alpha_t \, T_0 \, \mathbf{I}  \right)\, \mathrm{d}V_e  + \int_{\mathcal{D}_e}\rho(\phi)\left(\Ntx\,Q_{\text{h}}\right)\, \mathrm{d}V_e,
\end{aligned}
\label{eq:cc3}
\end{equation}
where $\mathbb{D} = \lambda (\boldI\otimes \boldI) + (2\mu + \kappa)\mathbb{I}$ and $\mathbb{R} = \alpha\,(\boldI\otimes \boldI) + \beta\,\mathbb{I} + \gamma\,\mathbb{I}$. Here, $\mathbb{I}$ is a fourth-order identity tensor. For $e^{\mathrm{th}}$ element, the element stiffness matrix $\mathsf{K}^e$ and element load vector $\mathsf{F}^e$ can be reduce from the discretized weak form given in Eq.~\eqref{eq:cc3} as 
\begin{equation}
    \mathsf{K}^e\mathbf{U}^e = \mathsf{F}^e,
\end{equation}
where $\mathsf{K}^e$ is the coupled element stiffness matrix, $\mathbf{U}^e$ is the generalized displacement vector containing elemental displacement, micro-rotations, and temperature, and $\mathsf{F}^e$ is the element forcing matrix from force, moment traction,
and thermal heat flux at all the degree of freedom. It can be written in matrix form as 
\begin{equation}
    \mathsf{K}^e = \begin{bmatrix}
        \mathsf{K}_{uu}^e & \mathsf{K}_{u\theta}^e & \mathsf{K}_{uT}^e \\ 
        \mathsf{K}_{\theta u}^e & \mathsf{K}_{\theta \theta}^e & \mathsf{K}_{\theta T}^e \\
        \mathsf{K}_{Tu}^e & \mathsf{K}_{T\theta}^e & \mathsf{K}_{TT}^e
    \end{bmatrix}, \quad \mathbf{U}^e = \begin{bmatrix}
        \ub^e \\ \thetab^e \\ T^e
    \end{bmatrix}, \quad \quad \mathsf{F}^e = \begin{bmatrix}
        \mathsf{f}_\mathrm{u}^e \\ \mathsf{f}_{\theta}^e  \\ \mathsf{f}_{T}^e
    \end{bmatrix}, 
\end{equation}
where, 
\begin{equation}
    \begin{aligned}
    &\mathsf{K}_{uu}^e = \int_{\mathcal{D}_e} \Ie\biggl[\Bsymx:\mathbb{D}\,\Bsymx +  \Bskewx:\kappa\,\Bskewx \biggr]\,\mathrm{d}V_e, \\
    &\mathsf{K}_{u\theta}^e = \int_{\mathcal{D}_e} \Ie\biggl[-  \Bskewx:\kappa\, \overset{3}{\mathbb{E}}\,\Nthetax\biggr] \,\mathrm{d}V_e, \\
    &\mathsf{K}_{uT}^e = \int_{\mathcal{D}_e} \Ie\biggl[-\Bsymx:\mathbb{D}\,\alpha_t\,\Ntx\, \mathbf{I} \biggr] \,\mathrm{d}V_e, \\
    &\mathsf{K}_{\theta u}^e = \int_{\mathcal{D}_e} \Ie\biggl[- \overset{3}{\mathbb{E}}\,\Nthetax:\kappa\,\Bskewx - \Nthetax\cdot \kappa\,\overset{3}{\mathbb{E}}\,\Bskewx \biggr] \,\mathrm{d}V_e, \\
    &\mathsf{K}_{\theta \theta}^e = \int_{\mathcal{D}_e} \Ie\biggl[\overset{3}{\mathbb{E}}\,\Nthetax:\kappa\, \overset{3}{\mathbb{E}}\,\Nthetax +\Bthetax:\mathbb{R}\,\Bthetax + \Nthetax\cdot \kappa\,\overset{3}{\mathbb{E}}\,\overset{3}{\mathbb{E}}\,\Nthetax \biggr] \,\mathrm{d}V_e, \\
    &\mathsf{K}_{\theta T}^e = \mathbf{0}, \qquad
    \mathsf{K}_{Tu}^e = \mathbf{0}, \qquad
    \mathsf{K}_{T\theta}^e = \mathbf{0}, \\
    &\mathsf{K}_{TT}^e = \int_{\mathcal{D}_e} \It \biggl[\Btx\, \cdot k \,\Btx\,T^e\biggr]\mathrm{d}V_e,
    \end{aligned}
\end{equation}
and
\begin{equation}
    \begin{aligned}
    &\mathsf{f}_\mathrm{u}^e = \int_\Gamma (\Nux \cdot \bar{\fb})\,\mathrm{d}S_e
- \int_{\mathcal{D}_e}\Ie\left( \Bsymx:\mathbb{D} \,\alpha_t \, T_0 \, \mathbf{I}  \right)\, \mathrm{d}V_e, \\
&\mathsf{f}_{\theta}^e = \mathbf{0}, \qquad
\mathsf{f}_{T}^e = \int_{\mathcal{D}_e}\rho(\phi)\left(\Ntx\,Q_{\text{h}}\right)\, \mathrm{d}V_e. 
   \end{aligned}
\end{equation}
By mapping each element’s local degrees of freedom for displacement, micro-rotation, and temperature to the corresponding global degrees of freedom, the element matrices and vectors are assembled into the global system. Introducing global stiffness matrix $\mathsf{K}$, and the global force vector $\mathsf{F}$, the assembly operation can be written as 
\begin{equation}
    \mathsf{K}(I,J) = \sum_{e=1}^{N_e} \mathsf{K}^e(i,j)\,\text{and}\,\,\,\mathsf{F}(I)= \sum_{e=1}^{N_e} \mathsf{F}^e(i),
    \label{eq:GlobalMatrix}
\end{equation}
where $i \,\,\mathrm{and}\,\,j$ refer to the local degree of freedom of each element, corresponding to displacement, micro-rotation, temperature, while $I\,\,\mathrm{and}\,\,J$ denote associated global degrees of freedom obtained through the finite element assembly procedure. The assembled global system of equations can then be expressed as
\begin{equation}
\mathsf{K}\,\mathbf{U} = \mathsf{F}.
\end{equation}
After global assembly and enforcement of boundary conditions, the discrete system above is obtained, from which the global vector of unknowns $\mathbf{U}$ can be solved.

\subsection{Generalized topology optimization problem for micropolar solids}
\label{sec:GeneralTopOpt}
We propose the generalized topology optimization problem for micropolar solids under thermo-mechanical loading as  
\begin{equation}
\begin{aligned}
\min_{\phi} \quad 
  & \mathcal{J}(\phi) \\[6pt]
\text{subject to} \quad 
  & \mathcal{C}(\phi) \leq 0, \\[4pt]
& a_{\text{MP}}\big((\ub,\thetab,T),(\wb,\boldsymbol{\vartheta},\tau),\phi\big) = b_{\text{MP}}\big((\wb,\boldsymbol{\vartheta},\tau),\phi\big),
\end{aligned}
\label{eq:TopOptProblemThermoElas}
\end{equation} 
where $\mathcal{J}(\phi)$ denotes any general objective functional to be minimized, while $\mathcal{C}(\phi)$ represents the in-equality constraint. The forms $a_{\text{MP}}$ and $b_{\text{MP}}$ correspond to the bilinear and linear weak formulations of the governing PDEs, as shown in Eq.~\eqref{eq:cc}. The aim of the PDE-constrained optimization problem in Eq.~\eqref{eq:TopOptProblemThermoElas} is to find an optimal subdomain $\Omega \subset \mathcal{D}$ that minimizes the objective functional while satisfying the governing PDEs and associated constraints. Here, the integration over the Neumann boundary $\Gamma$ is retained exactly, since this part of the boundary is not altered during optimization. For the remainder of this article, $\mathcal{C}(\phi)$ is used to define the volume constraint in terms of the density function $\rho(\phi)$, leading to 
\begin{align}
\mathcal{C}(\phi) &= \frac{1}{\mathrm{Vol}(\mathcal{D})} \int_{\mathcal{D}} \left( \rho(\phi) - V_f \right) \, \mathrm{d}V,
\label{eq:VolConst}
\end{align}
where $V_f$ denotes the desired volume fraction of the solid material in the domain, and $\mathrm{Vol}(\mathcal{D}) = \int_{\mathcal{D}}\mathrm{d}V$ is the volume of the design domain $\mathcal{D}$.

\subsection{Sensitivity analysis using adjoint formulation}
In topology optimization, updating the design variable, \textit{i.e.}, the level set function $\phi$ within the augmented Lagrangian framework (see \ref{sec:AugmentLag}) requires the computation of shape derivatives \cite{allaire2002level} or sensitivities of the objective functional $\mathcal{J}$ and the constraint functional $\mathcal{C}$ with respect to $\phi$. Since the material domain $\Omega$ evolves with $\phi$, performing differentiation directly on $\Omega$ would complicate the analysis. Instead, the formulation is carried out on a fixed design domain $\mathcal{D}$, where the evolving geometry is implicitly represented by $\phi$. Within the optimization scheme, this sensitivity analysis is performed at each iteration after solving the governing PDEs, and the resulting sensitivities provide the descent direction for updating the $\phi$ through the Hamilton-Jacobi evolution equation \eqref{eq:HJ_eq}, as outlined in the optimization algorithm in \ref{sec:TopOptAlgo}. These sensitivities can be determined as described below.

For a fixed computational domain  $\mathcal{D}$, differentiation of any functional $\mathcal{F}$ with respect to the level set function $\phi$, based on an adjoint formulation, is closely related to Céa’s method \cite{cea1986conception}. Functional $\mathcal{F}$ can be an objective function $\mathcal{J}$ or a constraint $\mathcal{C}$.  
Consider any generalized functional  $\mathcal{F}(\phi, \Ub),$ where $\Ub$ is a state vector that represents all the unknown fields of the coupled thermo-elastic micropolar problem, \textit{i.e.}, $[\Ub] = [\ub,  \thetab, T]^{\top}$, 
where $\ub, \thetab,\,\mathrm{and}\,\, T$ satisfy the PDEs governed by the 
weak-form residual
\begin{equation}
\begin{aligned}
\mathcal{R}(\phi, \Ub) = a_{\text{MP}} - b_{\text{MP}} = 0.
\end{aligned}
\label{eq:WeakFormResiduals}
\end{equation}
To determine the gradient of the functional $\mathcal{F}$, one can apply the chain rule of differentiation on $\mathcal{F}$ as:
\begin{equation}
\frac{\mathrm{d}\mathcal{F}}{\mathrm{d}\phi}
= \frac{\partial \mathcal{F}}{\partial \phi}
+ \frac{\partial \mathcal{F}}{\partial \Ub}
\frac{\partial \Ub}{\partial \phi}.
\label{eq:dFdphi}
\end{equation} 
Moreover, imposing the condition of $\frac{\mathrm{d}\mathcal{R}}{\mathrm{d}\phi}=0$ on the residual \(\mathcal{R} = 0\) yields:
\begin{equation}
\frac{\partial \mathcal{R}}{\partial \Ub}
\frac{\partial \Ub}{\partial \phi}
+ \frac{\partial \mathcal{R}}{\partial \phi} = 0.
\label{eq:drdphi}
\end{equation}
Rearranging the terms in Eq.~\eqref{eq:drdphi}, one can obtain \begin{equation}
\frac{\partial \Ub}{\partial \phi}
= -\left( \frac{\partial \mathcal{R}}{\partial \Ub} \right)^{-1}
\frac{\partial \mathcal{R}}{\partial \phi}.
\label{eq:doUdophi}
\end{equation}
Substituting Eq.~\eqref{eq:doUdophi} into Eq.~\eqref{eq:dFdphi} yields:
\begin{equation}
\frac{\mathrm{d}\mathcal{F}}{\mathrm{d}\phi}
= \frac{\partial \mathcal{F}}{\partial \phi}
- \frac{\partial \mathcal{F}}{\partial \Ub}
\left( \frac{\partial \mathcal{R}}{\partial \Ub} \right)^{-1}
\frac{\partial \mathcal{R}}{\partial \phi}.
\label{eq:dFdphiNew}
\end{equation}
To avoid the explicit inversion of 
$\tfrac{\partial \mathcal{R}}{\partial \Ub}$, 
we introduce the adjoint vector $\boldsymbol{\lambda}$, defined as $[\boldsymbol{\lambda}] = [\boldsymbol{\lambda}_{\mathrm{u}},\boldsymbol{\lambda}_{\theta},\lambda_{\mathrm{T}}]^{\top}$ 
where $\boldsymbol{\lambda}_{\mathrm{u}}$, $\boldsymbol{\lambda}_{\theta}$, 
and $\lambda_{\mathrm{T}}$ represent the adjoint variables associated with 
displacement, rotation, and temperature, respectively. The adjoint system is then formulated as
\begin{equation}
\left( \frac{\partial \mathcal{R}}{\partial \Ub} \right)^{\top}
\boldsymbol{\lambda}
= \left( \frac{\partial \mathcal{F}}{\partial \Ub} \right)^{\top}.
\label{eq:Adjoint}
\end{equation}
The matrix conjugate transpose, also called the matrix adjoint, and for this reason, the vector $\boldsymbol{\lambda}$ is called the vector of adjoint variables and the linear Eq.~\eqref{eq:Adjoint} is called the adjoint equation. 
Once \(\boldsymbol{\lambda}\) is obtained, the sensitivity can be determined as
\begin{equation}
\frac{\mathrm{d}\mathcal{F}}{\mathrm{d}\phi}
= \frac{\partial \mathcal{F}}{\partial \phi}
- \boldsymbol{\lambda}^{\top} \frac{\partial \mathcal{R}}{\partial \phi}.
\label{eq:dFdphiFinal}
\end{equation}


In summary, this section provides the full formulation and implementation of the proposed topology optimization framework. The strong and weak forms of the governing PDEs and the finite element implementation are introduced, followed by the general topology optimization problem under a volume constraint and the adjoint-based sensitivity analysis for elastic and coupled thermo-mechanical loading. These components are integrated within the \texttt{Gridap.jl}~\cite{badia2020gridap, verdugo2022software} and \texttt{GridapTopOpt.jl} ~\cite{wegert2025gridaptopopt} packages, which use automatic differentiation through \texttt{ForwardDiff.jl}~\cite{RevelsLubinPapamarkou2016} to link discretization with efficient sensitivity evaluation. For completeness, the full algorithmic workflow is summarized in \ref{sec:TopOptAlgo}. Although implemented in \texttt{Julia}, the formulation is general and can be adapted to any finite element or numerical framework, highlighting the flexibility of the proposed methodology.

\section{Numerical results}
\label{sec:numericalresults}
The efficacy of the proposed topology optimization framework is demonstrated through three numerical examples, delineated as elastic and thermo-elastic cases. Note that the proposed method is validated using a cantilever beam subjected to elastic loading under isothermal conditions, and compared against the result from  \citet{rovati2007optimal}, as detailed in \ref{sec:Validationmicro-polar}.
 In the elastic case, Example I considers a half-MBB beam to investigate the influence of the micropolar coupling number $N$ and the bending length scale $\lb$ on the optimized topology, as well as to examine size-dependent micropolar effects by varying the specimen dimensions. The thermo-elastic case is addressed through Example II, involving a fixed beam, and Example III, which considers a half-MBB beam configuration. Collectively, these examples illustrate the elastic and coupled thermo-mechanical responses and demonstrate the role of micropolarity in shaping optimized structures under both mechanical and thermal loading.

\subsection{Example I: Half-MBB beam under elastic loading} \label{sec: example1}
\begin{figure}[H]
    \centering
    \begin{minipage}{0.49\textwidth}
    \begin{subfigure}[t]{\linewidth}
        \centering
          \caption{}
      \includegraphics[width=\linewidth]{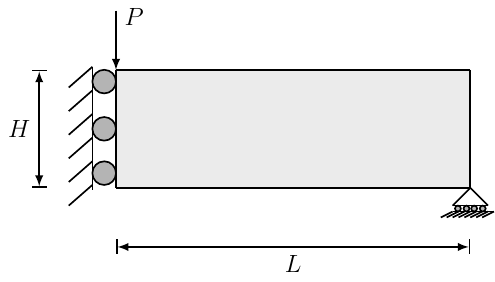}
    \end{subfigure}
    \end{minipage}
    \begin{minipage}{0.49\textwidth}
    \centering
    \begin{subfigure}[t]{\linewidth}
        \centering
          \caption{}
         \vspace{12mm}
     \includegraphics[width=0.75\linewidth]{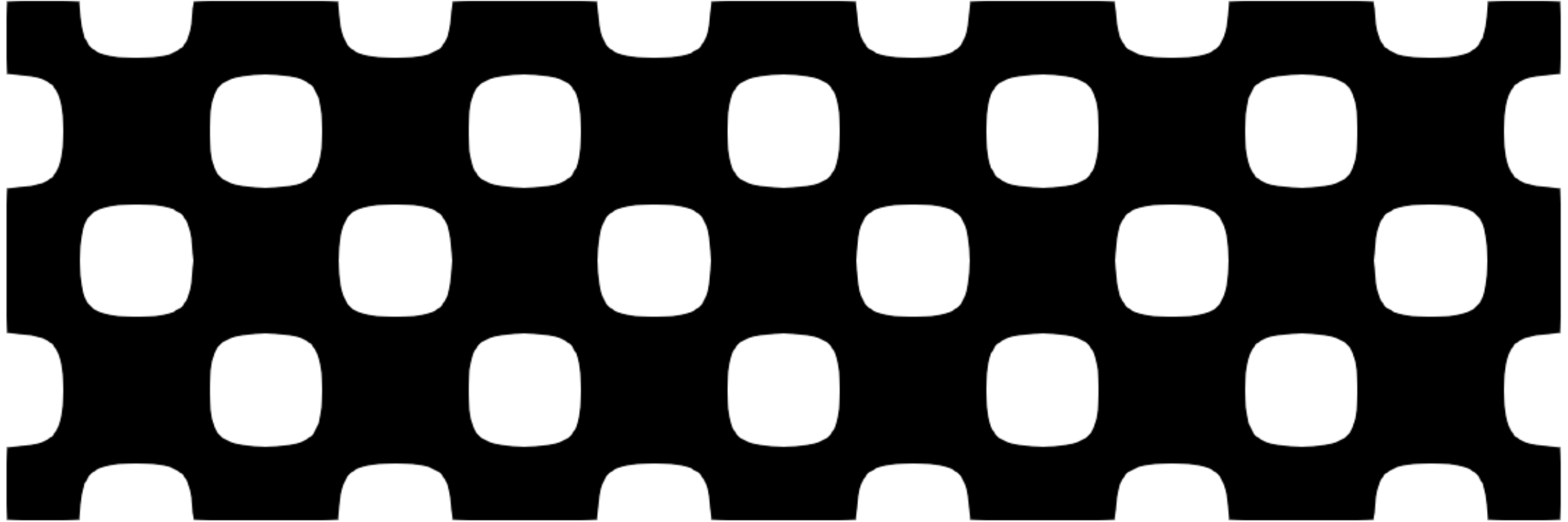}
        \vspace{14mm}
    \end{subfigure}
    \end{minipage}
    \caption{In (a), the design domain and boundary conditions are shown for a half-MBB (Messerschmitt–Bölkow–Blohm) beam in Example I under elastic loading. A symmetry condition is applied along the vertical axis to model half of the simply supported beam subjected to three-point bending. $L$ and $H$ denote the length and height of the half-MBB beam, respectively. The beam is subjected to a mechanical load $P$ at the top-left corner, and (b) shows the initial design.}
\label{fig:HalfMBBDesignDomain}
\end{figure}

In this section, the advantages of the micropolar topology optimization method are demonstrated in comparison with the conventional topology optimization methods for elastic solids. A two-dimensional simply supported half-MBB (Messerschmitt–Bölkow–Blohm) beam is considered. The analysis is performed under isothermal conditions, neglecting thermal effects and considering purely mechanical loading. Unless otherwise stated, plane stress conditions are assumed throughout. The topology optimization problem for the considered test case is formulated as  
\begin{equation}
\begin{aligned}
\min_{\phi} \quad & \mathcal{J}(\phi) = \int_{\Gamma} \ub\cdot\bar{\fb}\, \mathrm{d}S \\[6pt]
\text{subject to} \quad & \mathcal{C}(\phi) \leq 0, \\[4pt]
& a_{\text{MP}}\big((\ub,\thetab),(\wb,\boldsymbol{\vartheta}),\phi\big) 
   = b_{\text{MP}}\big((\wb,\boldsymbol{\vartheta}),\phi\big), 
\end{aligned}
\label{eq:TopOptElasticProblem}
\end{equation}
where $\mathcal{J}(\phi)$ is the work performed by the external traction $\bar{\fb}$, and $\mathcal{C}(\phi)$ is computed using Eq.~\eqref{eq:VolConst}. Note that $a_{\text{MP}}$ and $b_{\text{MP}}$ are originally defined for a micropolar thermo-elastic medium; under isothermal conditions, it reduces to a micropolar elastic medium as a special case. Therefore, the problem considered herein is a compliance minimization subject to a volume constraint with the prescribed volume fraction $V_f$ at 30\% of the total design domain. The design domain is discretized using a uniform mesh of $300 \times 100$ bilinear quadrilateral finite elements.  A concentrated load of magnitude $P = 1~\mathrm{N}$ is applied at the location indicated in Fig.~\ref{fig:HalfMBBDesignDomain}. For all configurations, the aspect ratio of length $(L)$ to height $(H)$ is maintained at $L/H = 3$. The beam is modeled with a constant thickness of $t = 1~\text{mm}$ along its out-of-plane direction. The initial holes are generated using the parameters $\zeta = 12/L$ and $b = 0.2$, as defined in Eq.~\eqref{eq:LSF}. The relevant material properties used herein are summarized in Table~\ref{tab:ElasticMatProp}.
\begin{table}[H]
\centering
\caption{Material properties for the half-MBB beam in Example I.}
\label{tab:ElasticMatProp}
\renewcommand{\arraystretch}{1.2} 
\begin{tabular}{@{\extracolsep{20pt}} l c c @{}}
\toprule
\textbf{Property} & \textbf{Value} & \textbf{Units} \\
\midrule
Young's modulus (solid), $E$ & $1.0$ & $\mathrm{N/mm^{2}}$ \\
Void stiffness parameter, $\varepsilon_{\mathrm{elas}}$ & $1.0 \times 10^{-9}$ & -- \\
Poisson's ratio, $\nu$ & $0.30$ & -- \\
\bottomrule
\end{tabular}
\end{table}

\subsubsection{Influence of $\lb$ and $N$}
\begin{figure}[H]
\centering
\renewcommand{\arraystretch}{2.5} 
\setlength{\tabcolsep}{6pt}     

\begin{tabular}{>{\centering\arraybackslash}m{1.75cm} | c | c | c}
\hline
 & $\lb=8$\,mm & $\lb=3$\,mm & $\lb=2$\,mm \\
 \hline
 \begin{minipage}[c][\height][c]{1.75cm}
  \centering
  \vspace{-1.0cm}
  $N=0.0$
\end{minipage} &
\begin{subfigure}[b]{0.25\linewidth}
\vspace{2mm} 
    \includegraphics[width=\linewidth]{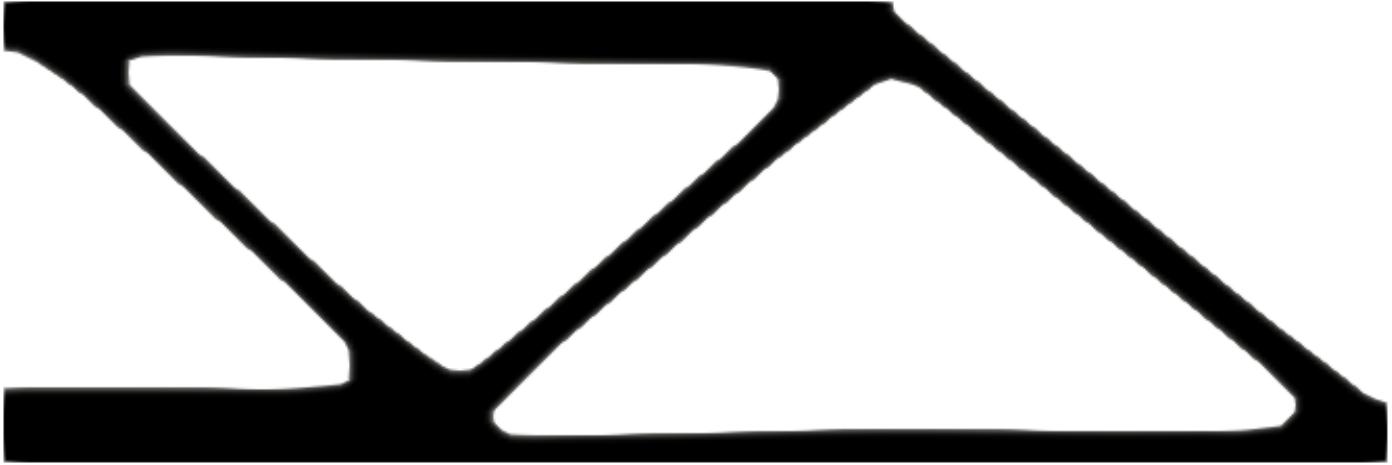}
\end{subfigure} &
\begin{subfigure}[b]{0.25\linewidth}
    \includegraphics[width=\linewidth]{Figures/HalfMBB10times30N0.5Lb0.1.pdf}
\end{subfigure} &
\begin{subfigure}[b]{0.25\linewidth}
    \includegraphics[width=\linewidth]{Figures/HalfMBB10times30N0.5Lb0.1.pdf}
\end{subfigure} \\

\hline 

\begin{minipage}[c][\height][c]{1.75cm}
  \centering
  \vspace{-1.0cm}
  $N=0.5$
\end{minipage} &
\begin{subfigure}[b]{0.25\linewidth}
\vspace{2mm} 
    \includegraphics[width=\linewidth]{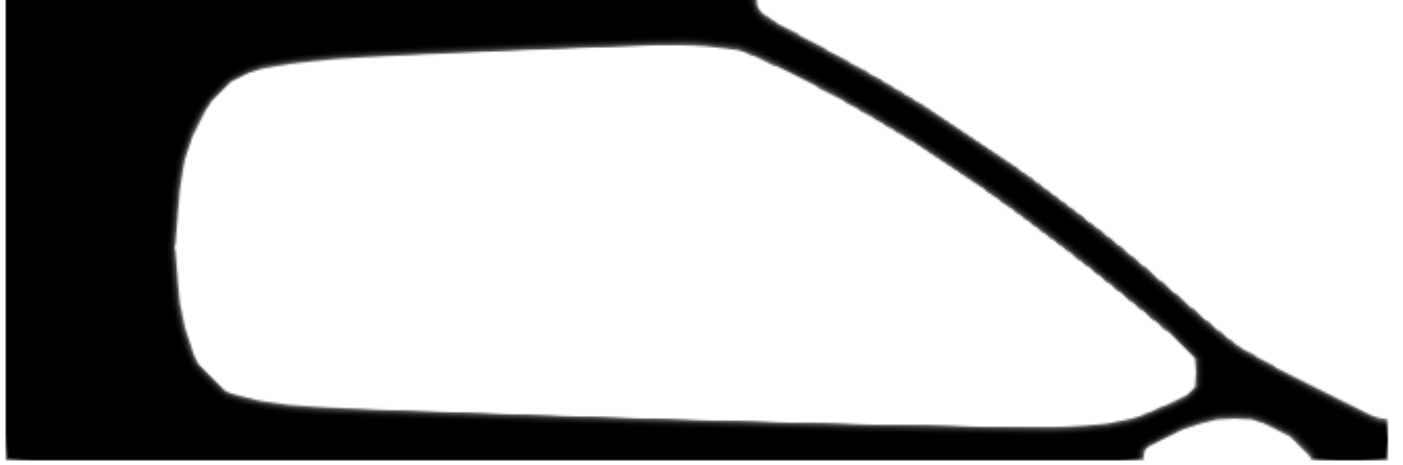}
\end{subfigure} &
\begin{subfigure}[b]{0.25\linewidth}
    \includegraphics[width=\linewidth]{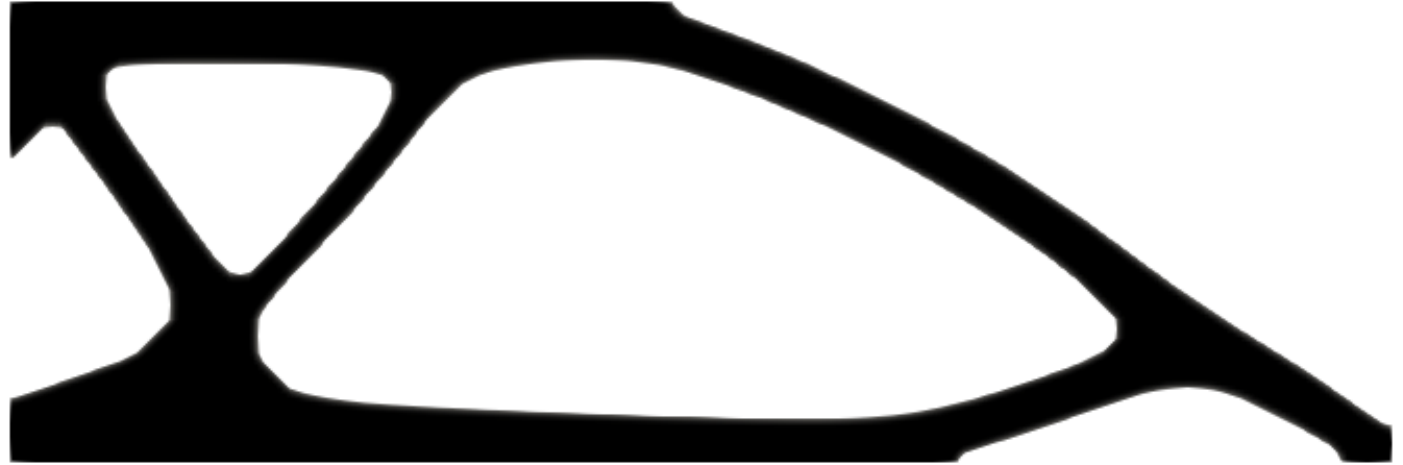}
\end{subfigure} &
\begin{subfigure}[b]{0.25\linewidth}
    \includegraphics[width=\linewidth]{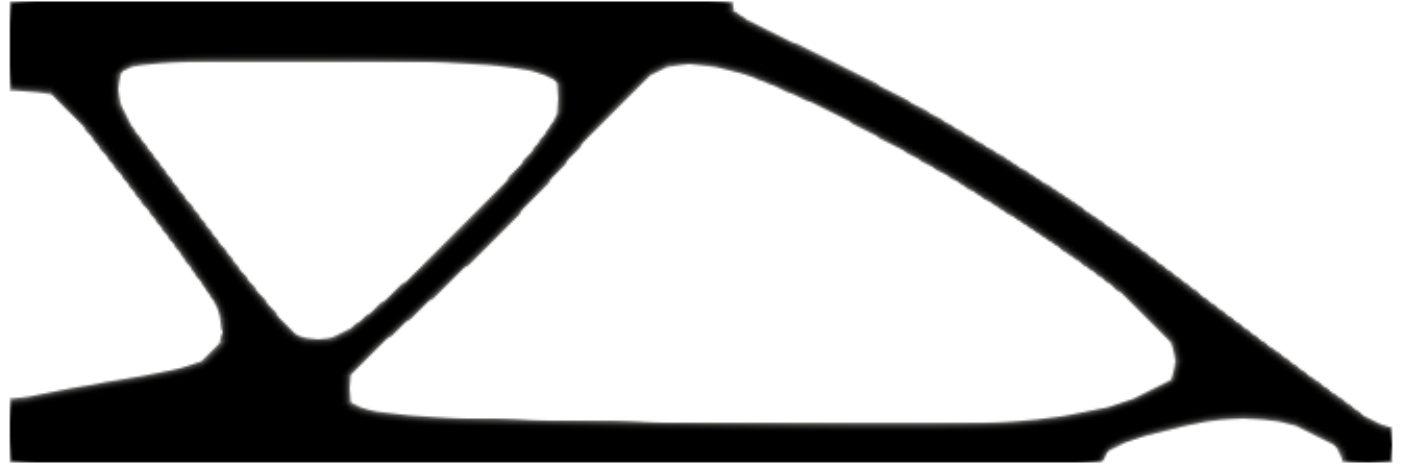}
\end{subfigure} \\
\hline 

\begin{minipage}[c][\height][c]{1.75cm}
  \centering
  \vspace{-1.0cm}
  $N=0.99$
\end{minipage}
 &
\begin{subfigure}[b]{0.25\linewidth}
\vspace{2mm} 
    \includegraphics[width=\linewidth]{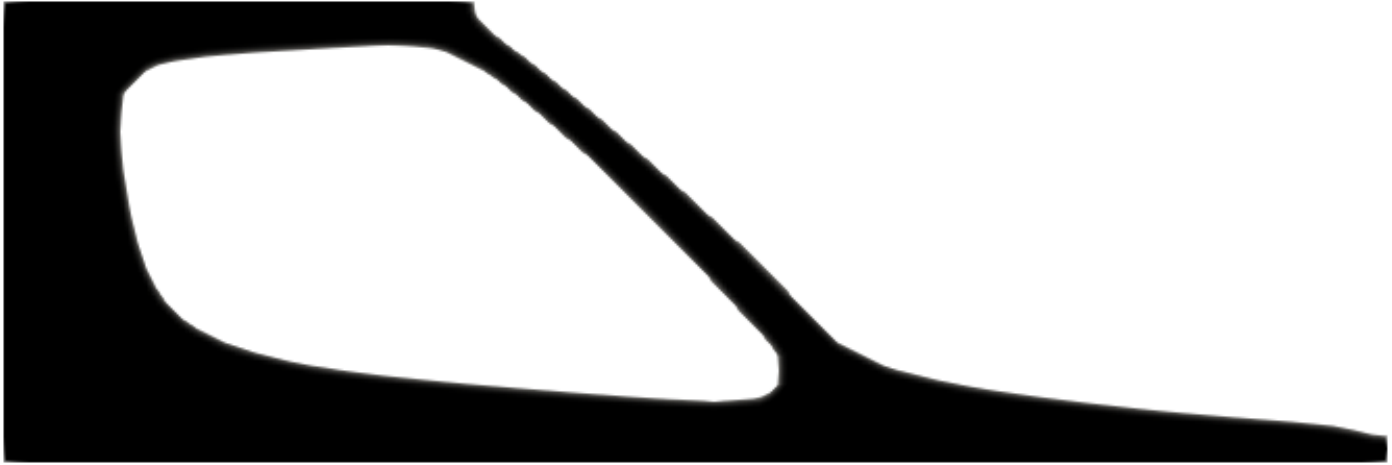}
\end{subfigure} &
\begin{subfigure}[b]{0.25\linewidth}
    \includegraphics[width=\linewidth]{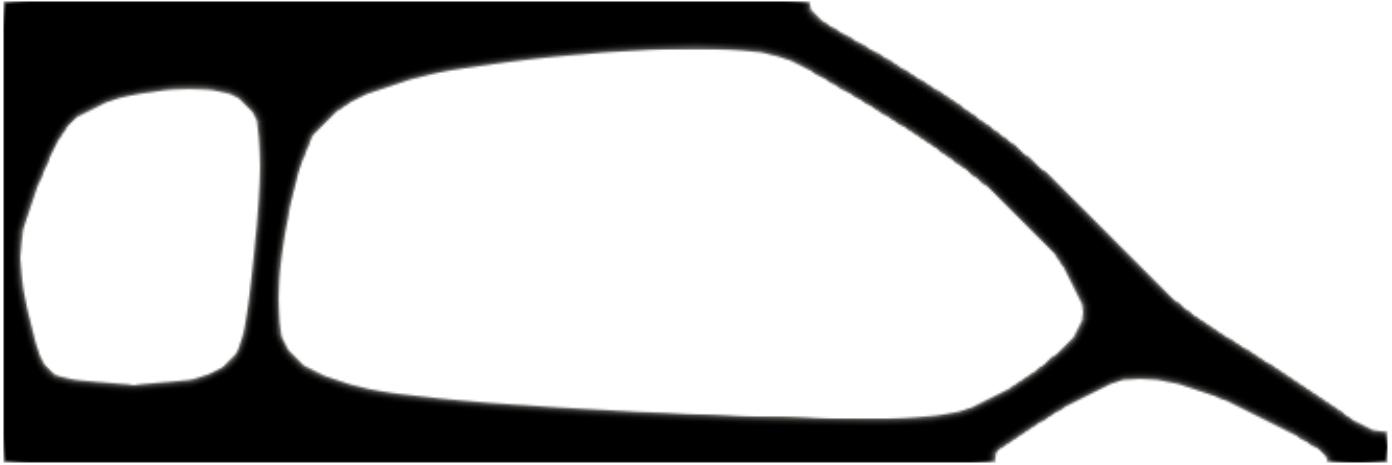}
\end{subfigure} &
\begin{subfigure}[b]{0.25\linewidth}
    \includegraphics[width=\linewidth]{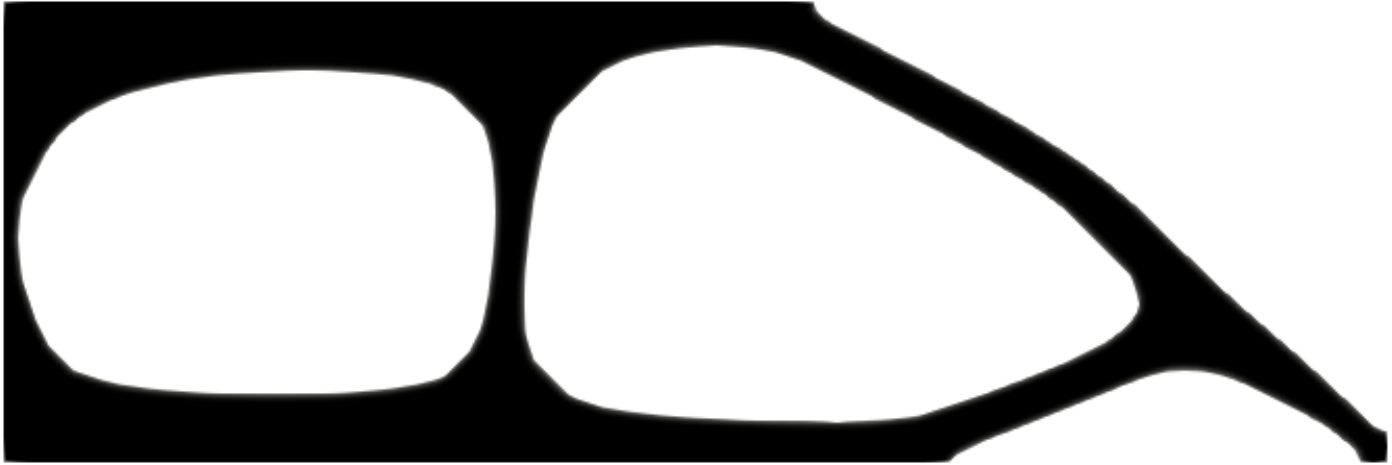}
\end{subfigure} \\
\hline
\end{tabular}
\caption{Topologically optimized designs of the half-MBB beam in Example I with dimensions $H=10\,\text{mm}, \,L=30\,\text{mm}$ for different bending length scale $\lb$ and micropolar coupling number $N$.}
\label{fig:comparisonforHalfMBBBeamNvariedLbVaried}
\end{figure}

To investigate the influence of the micropolar parameters such as the micropolar coupling number, $N$, and the bending length scale, $\lb$, on the mechanical response and the optimized design of the specimen, a parametric study is conducted. The coupling number is varied from $N=0.0$ to $N=0.99$ while maintaining constant values for the specimen size, aspect ratio, and bending length scale $\lb$. As illustrated by the second and third rows for $N=0.5$ and $0.99$ in Fig.~\ref{fig:comparisonforHalfMBBBeamNvariedLbVaried}, an increase in the micropolar coupling number leads to a significant alteration in the optimized topology. For $N = 0.5$, reducing the bending length scale $\lb$ leads the optimized design to converge toward a truss-like structure with slight modifications. In contrast, for $N = 0.99$, the design does not strictly reduce to a truss-like form; instead, the material tends to concentrate along the edges with fewer members even at $\lb = 2\,\text{mm}$. Thus, at a higher micropolar coupling number, the influence of micropolar effects remains significant even for small $\lb$. From these results, it can be deduced that for lower values of the micropolar parameters, \textit{i.e.}, smaller $N$ and $\lb$, the optimized structure adopts a predominantly truss-like structure. In particular, when $N=0.0$, which corresponds to the classical topology optimization result without any micropolar effects, the structure remains truss-shaped regardless of variations in $\lb$. The result is expected, since at $N=0.0$, the micropolar constitutive relations do not depend on the bending length scale $\lb$, and no independent microrotation occurs. The micropolar coupling number $N$ quantifies the extent of micro-rotation in the material: $N = 0.0$ corresponds to no micro-rotation (classical continuum behavior), while $N = 0.99$ represents micro-rotations that can reach magnitudes comparable to the macroscopic rotation. The result is expected, since at $N=0.0$, the micropolar constitutive relations do not depend on the bending length scale $\lb$, and no independent microrotation occurs. However, as the degree of micropolarity increases, the optimized topology transitions from a truss-like to a frame-like structure. This is because bending is rarely used as an efficient load-carrying mechanism in classical Cauchy continuum-based topology optimization. Since bending resistance arises only through shear stresses generated from the symmetric stress tensor, the specimen cannot develop internal torque to resist the macro-rotation. As a result, the optimization algorithm naturally avoids bending paths and focuses on purely axial truss-shaped members to minimize the objective for a given volume of material. However, when a beam within the structure bends, there are substantial spatial changes in the microscopic rotations of the material points (\textit{i.e}, gradient of microrotation, $\nabla \thetab$). This is particularly significant when the internal bending length scale $\lb$ becomes comparable to the dimensions of the specimen.  

\begin{figure}[H]
    \centering
    \begin{subfigure}[b]{0.49\linewidth}
    \centering
        \includegraphics[width=\linewidth]{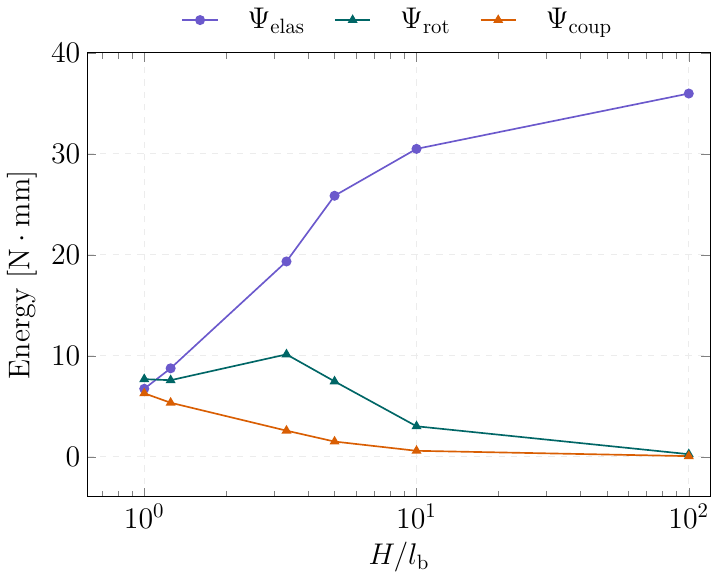}
    \end{subfigure}
\caption{Elastic strain energy $(\Psi_{\text{elas}})$, rotational strain energy $(\Psi_{\text{rot}})$, and coupled strain energy $(\Psi_{\text{coup}})$  are plotted against the change in $H/\lb$ ratio from $0.1$ to $100$ for micropolar coupling number $N = 0.5$ in Example I for the half-MBB beam with dimensions $H = 10$ mm, $L = 30$ mm.}
\label{fig:difflbStrainEnergyRatioValuesN0.5}
\end{figure}

\begin{figure}[H]
    \centering
    \begin{subfigure}[t]{0.49\linewidth}
    \centering
        \caption{For micropolar coupling number $N = 0.5$}        \includegraphics[width=0.98\linewidth]{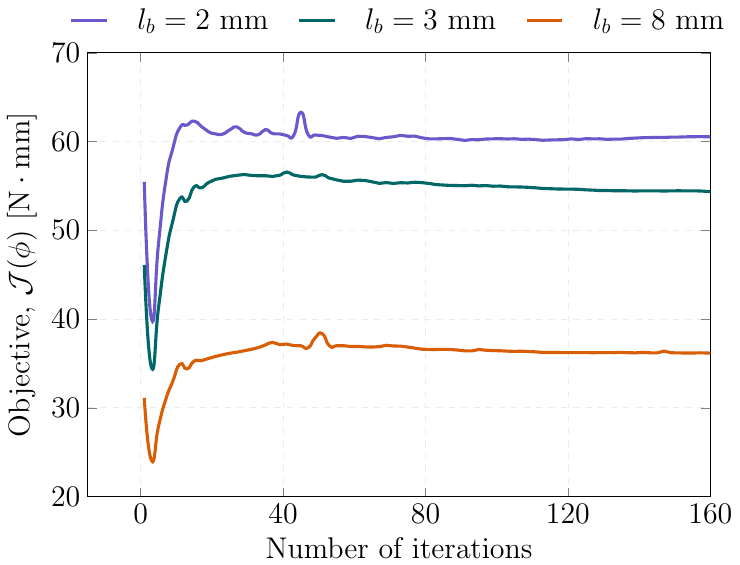}
        \label{fig:ObjFuncforDiffLbN0.5}
    \end{subfigure}
    \begin{subfigure}[t]{0.49\linewidth}
    \centering
        \caption{For bending length scale $\lb = 3$ mm.}        \includegraphics[width=\linewidth]{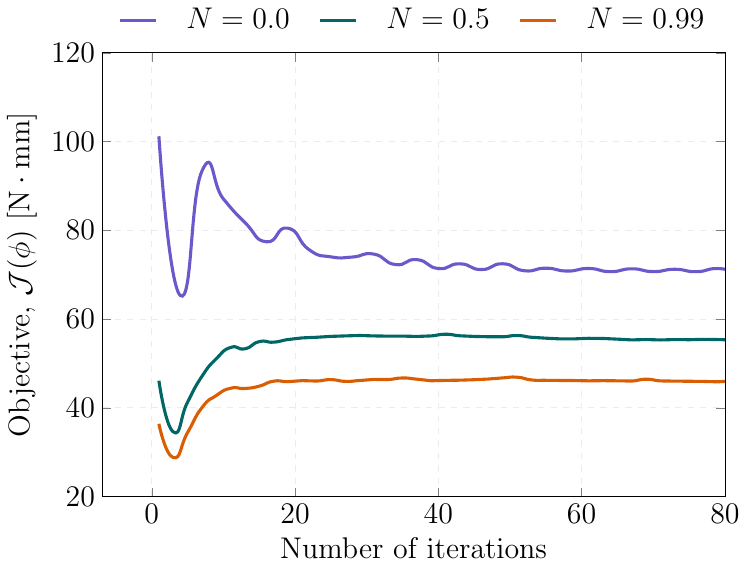}
        \label{fig:ObjFuncforDiffNLb3}
    \end{subfigure}
    \caption{The objective is plotted for the half-MBB beam with dimensions $H = 10$ mm, $L = 30$ mm. In (a), the objective is plotted for different values of bending length scale $\lb$ as $2,~3,~8$ mm for the micropolar coupling number $N=0.5$. In  (b), the objective is plotted for different values of micropolar coupling number $N = 0.0,~0.5,~0.99$ for bending length scale $\lb=3$ mm.}
\label{fig:HalfMBBN0.5LbVary}
\end{figure}

\begin{figure}[H]
    \centering
    \begin{subfigure}[b]{0.49\linewidth}
    \centering
        \includegraphics[width=\linewidth]{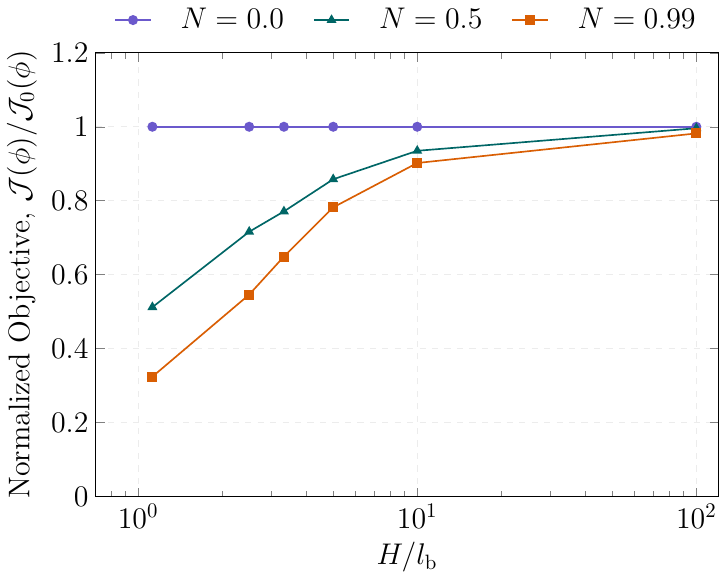}
    \end{subfigure}
\caption{Normalized objective is plotted against the change in $H/\lb$ ratio from $0.1$ to $100$ for different values of micropolar coupling number $N = 0.0,~0.5,~0.99$ in Example I for the half-MBB beam (see Fig.~\ref{fig:HalfMBBDesignDomain}) with dimension $H = 10~\mathrm{mm}, L= 30~\mathrm{mm}$.}
\label{fig:domaindifferentlbValues}
\end{figure}

The effect of micropolarity is also observed in Fig.~\ref{fig:difflbStrainEnergyRatioValuesN0.5}, which compares the elastic strain energy $(\Psi_{\mathrm{elas}})$, the rotational energy $(\Psi_{\mathrm{rot}})$ (associated with $\nabla \thetab$), and the couple stress energy $(\Psi_{\mathrm{coup}})$. As $\lb$ increases, the contribution of rotational energy also rises, leading to the development of couple stresses that directly oppose the microrotation gradient, while the asymmetric stress produces a non-zero net torque. This offers an additional efficient means of carrying moments through the structure. Therefore, when the optimizer minimizes total compliance, it naturally places the material in regions where it can provide the highest resistance to local micro-rotation gradients, ensuring that the structure efficiently resists both axial forces and bending moments. As a result, the optimized topology evolves towards frame-like shapes, as these forms are energetically optimal for transmitting both forces and moments in a micropolar continuum. As the degree of micropolarity increases, the compliance of the optimized structure is noticeably reduced, indicating a global enhancement of structural stiffness due to the additional bending resistance provided by the micropolar effect. This trend is evident in Fig.~\ref{fig:HalfMBBN0.5LbVary}, where compliance decreases both with increasing bending length scale, $\lb$ (see Fig.~\ref{fig:HalfMBBN0.5LbVary} (a)), and with higher values of the micropolar coupling number, $N$ (see Fig.~\ref{fig:HalfMBBN0.5LbVary} (b)). Furthermore, Fig.~\ref{fig:domaindifferentlbValues} shows the normalized objective, defined as the objective value of the optimized structure divided by that of the nonpolar case, $\mathcal{J}(\phi)/\mathcal{J}_0(\phi)$, which directly highlights the compliance variation with respect to $\lb$ and $N$. It can be observed that for very small values of $\lb$ (\textit{i.e.}, when the ratio $H / \lb$ is large, with $H$ being the specimen size), the predictions of the micropolar model converge to those of the classical continuum model, indicating negligible size effects in this regime. Note that as the material bending length scale, $\lb$,  approaches the value of the specimen size, it results in pronounced micro-rotational effects and, consequently, increased stiffness.

Fig.~\ref{fig:DifferentRotationForN0,0.5,0.99} illustrates the variation of macro-rotation $\thetab_\mathrm{m}$, micro-rotation $\thetab$, and relative rotation $\thetab_{\mathrm{r}}$ for different values of the micropolar coupling number $N$. In the non-polar case ($N=0.0$), the micro-rotation $\thetab$ remains zero for all bending length scales $\lb$, while the macro-rotation $\thetab_{\mathrm{m}}$ is significantly high and nearly constant, making the relative rotation $\thetab_{\mathrm{r}} = \thetab_{\mathrm{m}} - \thetab$ identical to the macro-rotation. For $N=0.5$, the macro-rotation decreases compared to the non-polar case and reduces monotonically with increasing $\lb$. The micro-rotation is smaller than the macro-rotation, and the relative rotation takes values between the two. When $N=0.99$, all the rotations are further reduced, with the macro- and micro-rotations becoming nearly equal, which drives the relative rotation close to zero. Overall, the results indicate that as the micropolar coupling number $N$ increases, both macro- and micro-rotation components decrease. In the non-polar case, the gap between macro- and micro-rotation is substantial, whereas for higher $N$ values the two converge, leading to minimal relative rotation. Across all cases, however, the relative rotation remains nearly constant with respect to the bending length, $\lb$.  
\begin{figure}[H]
    \centering
    \begin{subfigure}[t]{0.7\linewidth}
    \centering
        \includegraphics[width=0.4\linewidth]{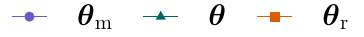}
    \end{subfigure}
    \begin{subfigure}[t]{0.49\linewidth}
    \centering
            \caption{$N = 0.0$.}
     \includegraphics[width=\linewidth]{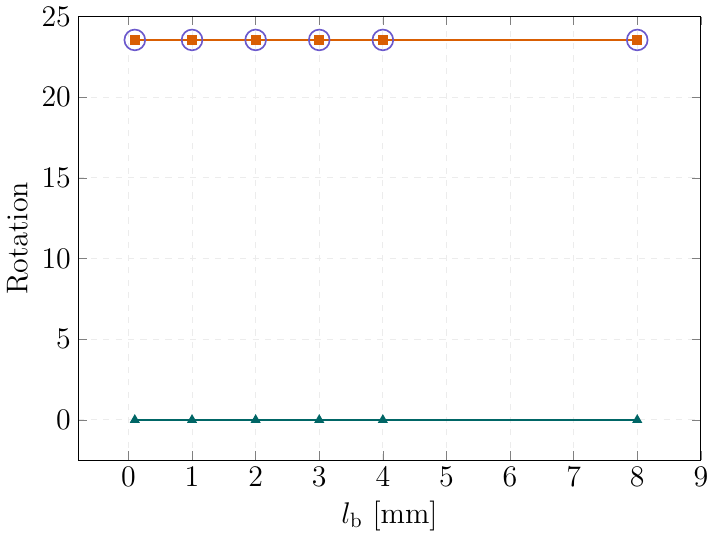}
    \end{subfigure}
     \begin{subfigure}[t]{0.49\linewidth}
    \centering
           \caption{$N = 0.5$.}
     \includegraphics[width=\linewidth]{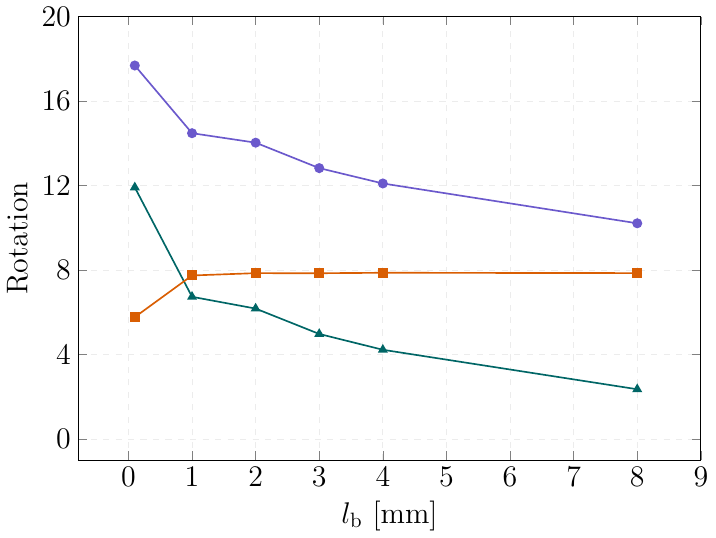}
    \end{subfigure}
     \begin{subfigure}[t]{0.49\linewidth}
    \centering
\caption{$N = 0.99$.}
\includegraphics[width=\linewidth]{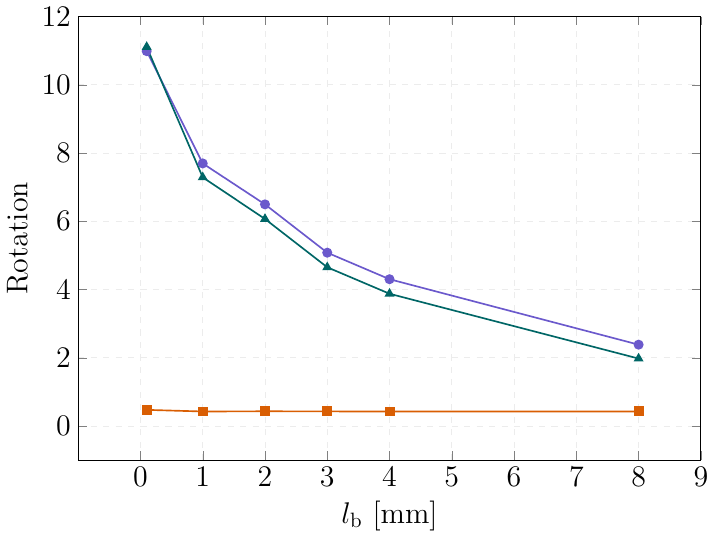}
    \end{subfigure}
    \hfill
    \caption{Macro-rotation $\thetab_{\mathrm{m}}$, micro-rotation $\thetab$, and relative-rotation $\thetab_{\mathrm{r}}$ value is plotted against the change in bending length scale value $\lb$ from $0.1$ to $8$ for micropolar coupling number $N$ varies from $0.0$ to $0.99$ in Example I for the half-MBB beam (see Fig.~\ref{fig:HalfMBBDesignDomain}) with dimensions $H = 10$ mm, $L = 30$ mm.} 
\label{fig:DifferentRotationForN0,0.5,0.99}
\end{figure}

\subsubsection{Influence of specimen dimension}
\begin{figure}[H]
\renewcommand{\arraystretch}{2.5} 
\setlength{\tabcolsep}{6pt}     

\begin{tabular}{>{\centering\arraybackslash}m{1.75cm} | c | c | c}
\hline
 & $H=10$\,mm & $H=20$\,mm & $H=500$\,mm \\
 \hline
 \begin{minipage}[c][\height][c]{1.75cm}
  \centering
  \vspace{-1.0cm}
  $N=0.0$
\end{minipage} &
\begin{subfigure}[b]{0.25\linewidth}
\vspace{2mm} 
    \includegraphics[width=\linewidth]{Figures/HalfMBB10times30N0.5Lb0.1.pdf}
\end{subfigure} &
\begin{subfigure}[b]{0.25\linewidth}
    \includegraphics[width=\linewidth]{Figures/HalfMBB10times30N0.5Lb0.1.pdf}
\end{subfigure} &
\begin{subfigure}[b]{0.25\linewidth}
    \includegraphics[width=\linewidth]{Figures/HalfMBB10times30N0.5Lb0.1.pdf}
\end{subfigure} \\

\hline 

\begin{minipage}[c][\height][c]{1.75cm}
  \centering
  \vspace{-1.0cm}
  $N=0.5$
\end{minipage} &
\begin{subfigure}[b]{0.25\linewidth}
\vspace{2mm} 
    \includegraphics[width=\linewidth]{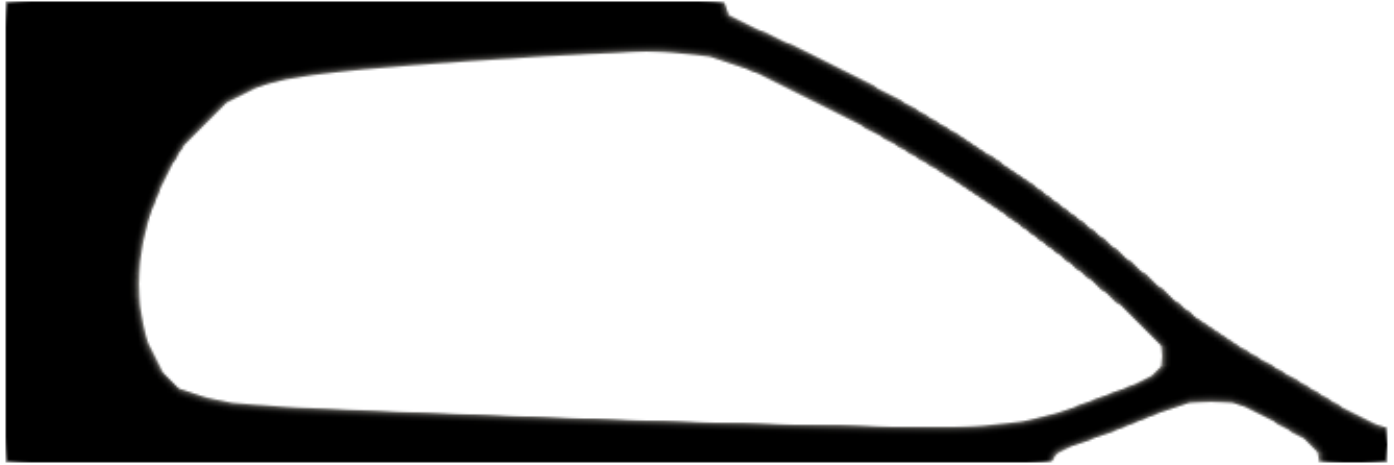}
\end{subfigure} &
\begin{subfigure}[b]{0.25\linewidth}
    \includegraphics[width=\linewidth]{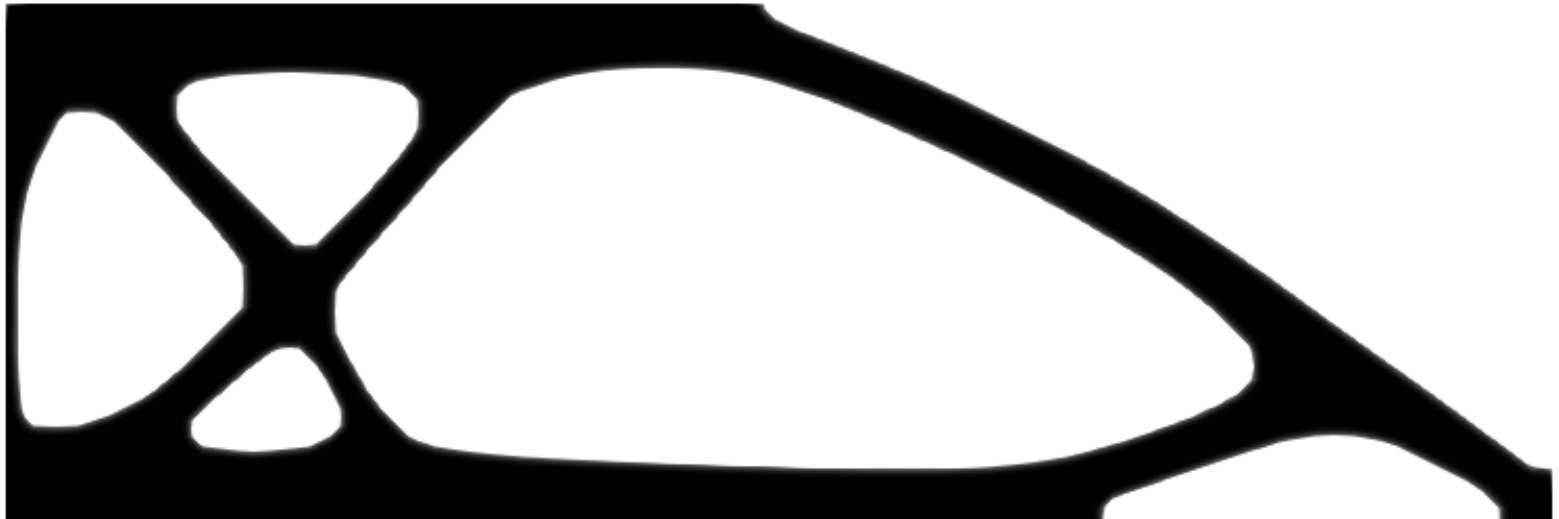}
\end{subfigure} &
\begin{subfigure}[b]{0.25\linewidth}
    \includegraphics[width=\linewidth]{Figures/HalfMBB10times30N0.5Lb0.1.pdf}
\end{subfigure} \\

\hline 

\begin{minipage}[c][\height][c]{1.75cm}
  \centering
  \vspace{-1.0cm}
  $N=0.99$
\end{minipage}
 &
\begin{subfigure}[b]{0.25\linewidth}
\vspace{2mm} 
    \includegraphics[width=\linewidth]{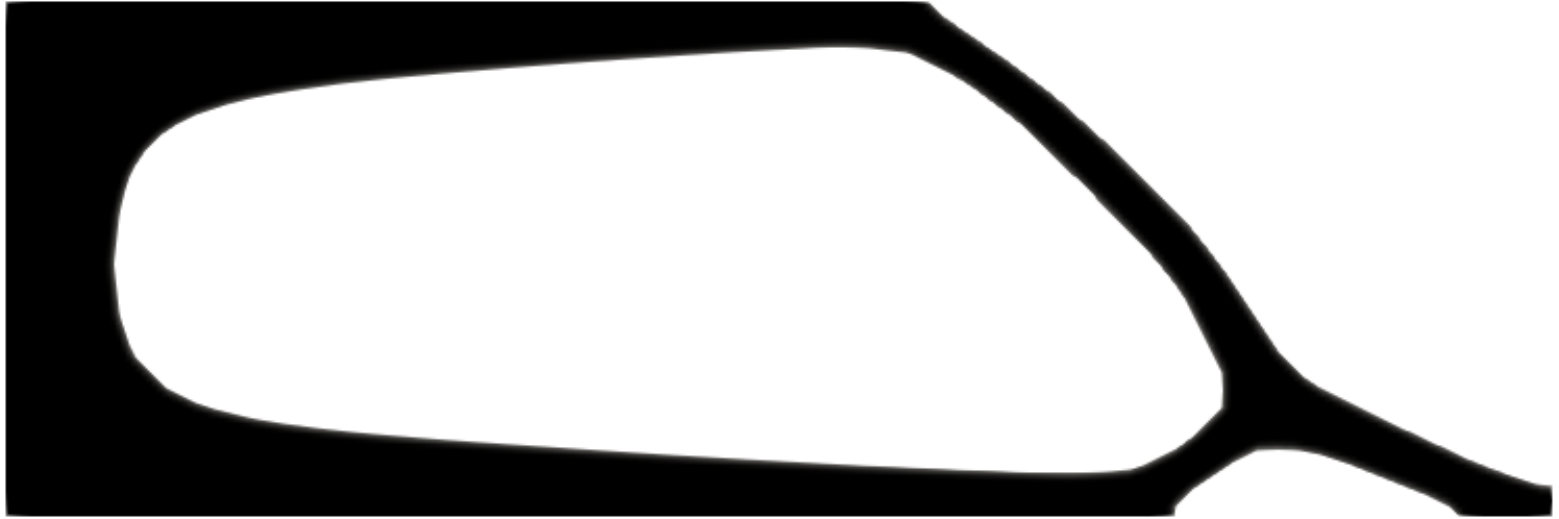}
\end{subfigure} &
\begin{subfigure}[b]{0.25\linewidth}
    \includegraphics[width=\linewidth]{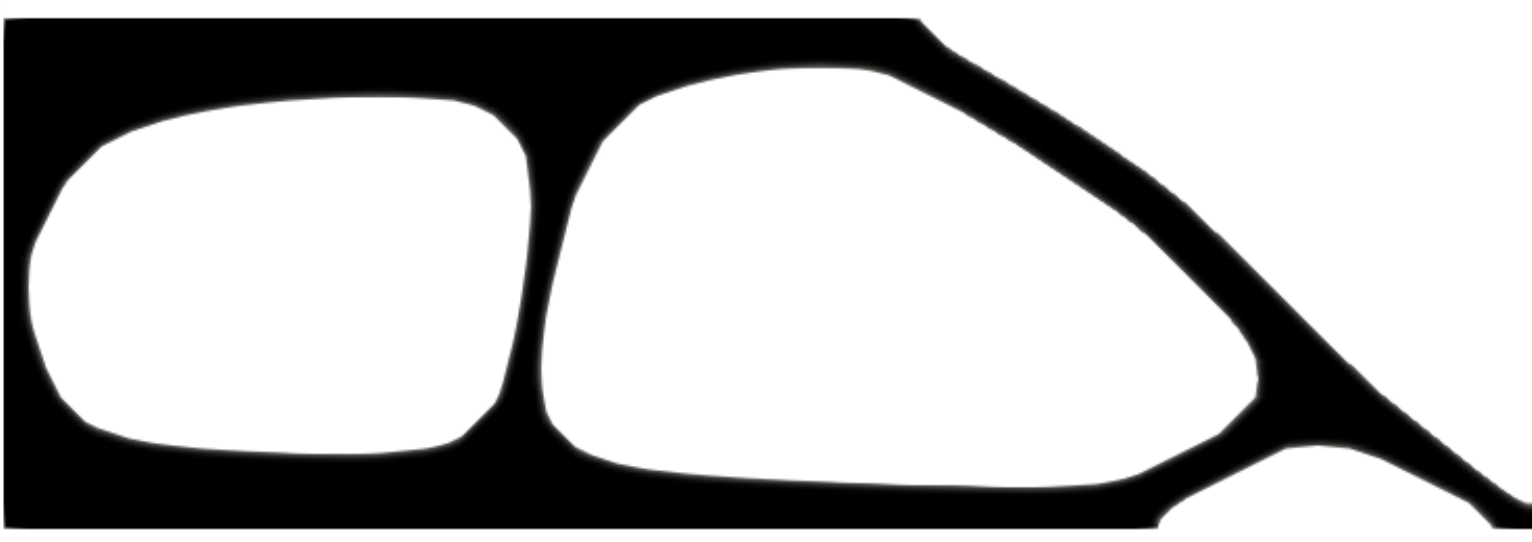}
\end{subfigure} &
\begin{subfigure}[b]{0.25\linewidth}
    \includegraphics[width=\linewidth]{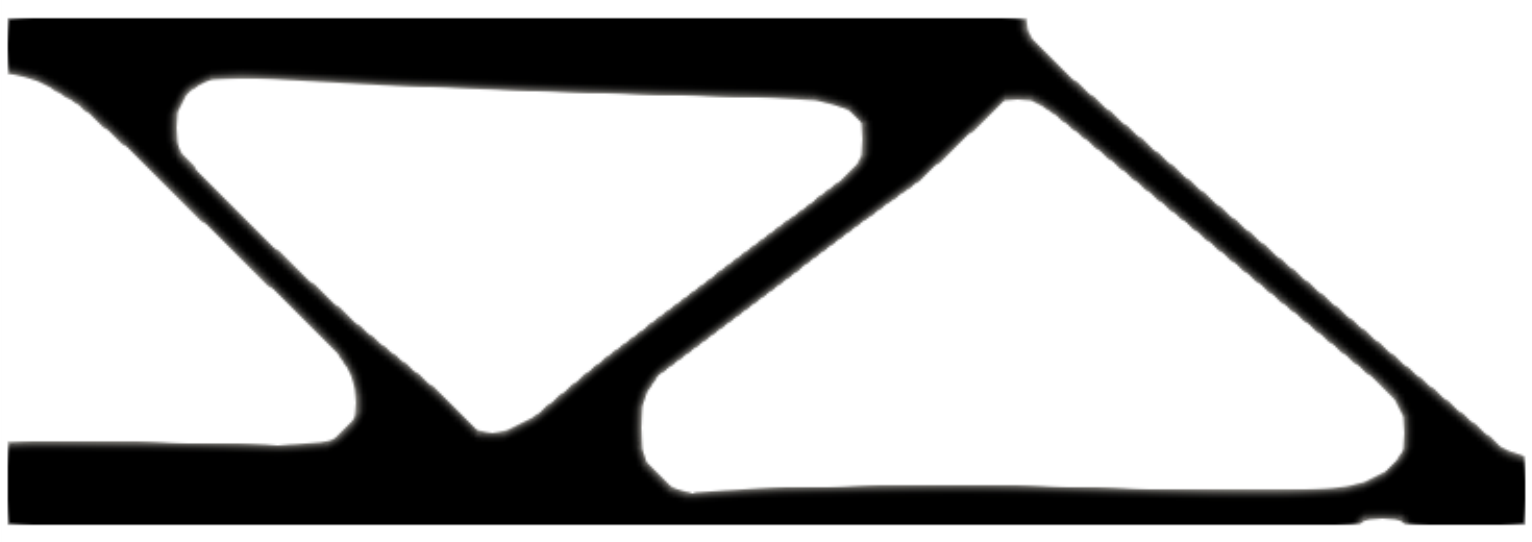}
\end{subfigure} \\
\hline
\end{tabular}
\caption{Topologically optimized design obtained in Example I for the half-MBB beam (see Fig.~\ref{fig:HalfMBBDesignDomain}) are presented against variations in its dimensions and the micropolar coupling number $N$, with the bending length scale $\lb = 5$ mm.}
\label{fig:CompareHalfMBBBeamNvariedLbVaried}
\end{figure}

\begin{figure}[H]
    \centering
    \begin{subfigure}[t]{0.49\linewidth}
    \centering
         \caption{Area normalized objective is plotted with varying dimension}
       \includegraphics[width=\linewidth]{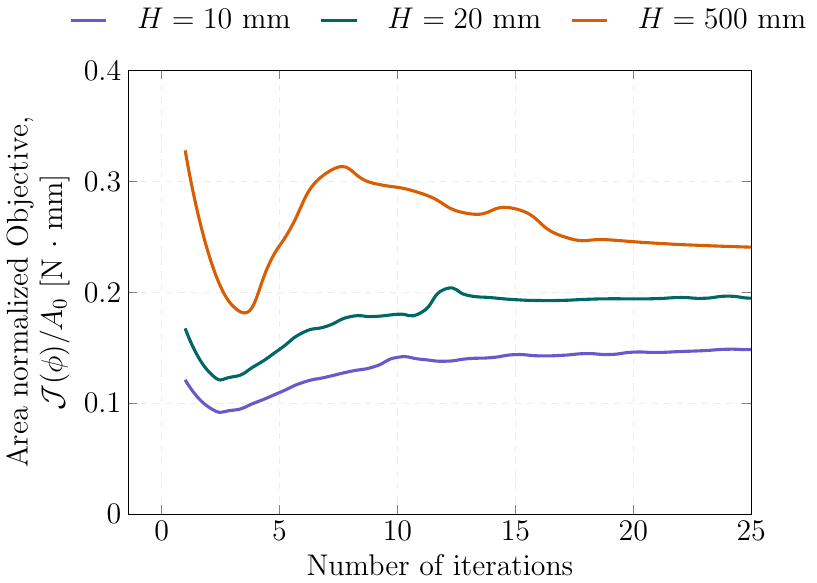}
    \end{subfigure}
    \begin{subfigure}[t]{0.49\linewidth}
    \centering
        \caption{Objective is plotted with varying $N$}        \includegraphics[width=0.95\linewidth]{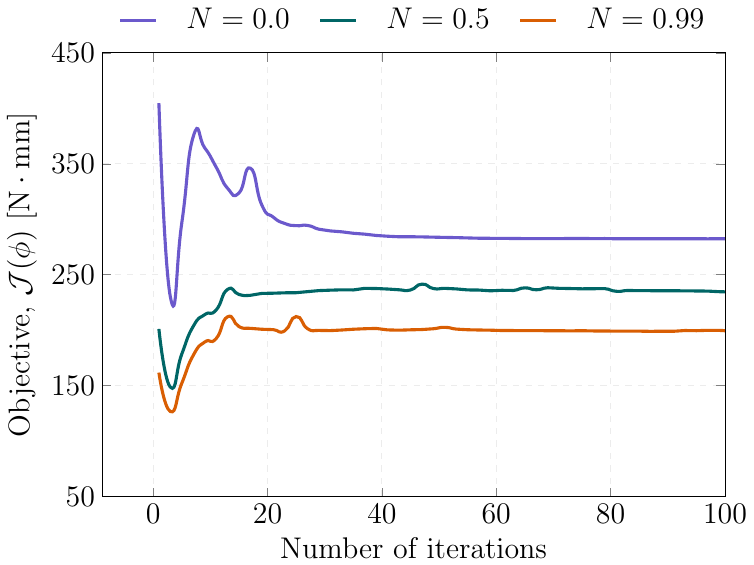}
    \end{subfigure}
   \caption{For the half-MBB beam in Example I, (a) shows the area normalized objective, obtained by dividing the objective $\mathcal{J}(\phi)$ by the specimen area $A_0$, for $N = 0.5$ and $\lb = 5.0~\mathrm{mm}$ while varying the specimen dimensions, and (b) shows the objective for $\lb = 5.0~\mathrm{mm}$ while varying the micropolar coupling number $N$ for a beam of dimensions $L = 60~\mathrm{mm}$ and $H = 20~\mathrm{mm}$.}
\label{fig:HalfMBBDimension study}
\end{figure}

\begin{figure}[H]
    \centering
    \begin{subfigure}[b]{0.49\linewidth}
    \centering
        \includegraphics[width=\linewidth]{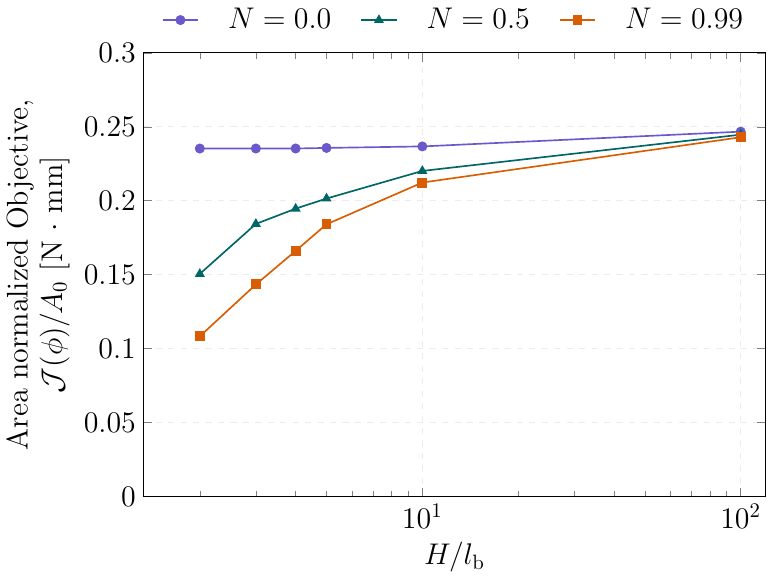}
    \end{subfigure}
\caption{Area normalized objective, defined as the objective $\mathcal{J}(\phi)$ divided by the specimen area $A_0$, for the half-MBB beam in Example I. Results are plotted for varying beam dimensions and micropolar coupling number $N$, while keeping the bending length scale $\lb = 3~\text{mm}$.}
\label{fig:domaindifferentlbNValues}
\end{figure}

The material bending length scale, $\lb$, is an intrinsic property of a material and ideally remains constant for a given material. A more practical approach to evaluating size effects is to vary the specimen dimensions while keeping the bending length scale $\lb$ fixed. A parametric study is conducted by varying the dimensions of the beam while maintaining a constant aspect ratio, as depicted in Fig.~\ref{fig:HalfMBBDesignDomain}. The dimensions used are $10$ mm$\times 30$ mm, $20$ mm$\times 60$ mm, and $500$ mm$\times 1500$ mm. The bending length scale $\lb = 5$ mm, and the micropolar coupling number $N\in\{0,0.5,\,\mathrm{and}\, 0.99\}$. The specimen dimensions are varied such that the ratio of Height-to-bending length scale, $H/\lb\in\{2,4,100\}$. From the first row of Fig.~\ref{fig:CompareHalfMBBBeamNvariedLbVaried}, it is apparent that increasing the specimen size results in nearly identical optimized designs under conventional optimization ($N=0$). This indicates that classical topology optimization fails to capture design modifications necessary when size effects manifest through enhanced micropolar rotations and consequent stiffness variations. In contrast, for higher micropolarity levels ($N=0.5$ and $N=0.99$, shown in the second and third rows of Fig.~\ref{fig:CompareHalfMBBBeamNvariedLbVaried}), pronounced changes occur in smaller specimens (\textit{i.e.,} low $H/\lb$ values). Specifically, the topology transitions from a truss-like to a frame-like structure as size effects become significant. However, as the specimen size increases up to $50$ times, the optimized designs from both conventional and micropolar frameworks converge, reflecting the diminishing influence of micropolar effects at larger scales.
Furthermore, Fig.~\ref{fig:HalfMBBDimension study} (a) shows that as the beam dimension increases, the area normalized objective (objective divided by specimen area) also increases, and convergence is achieved in fewer iterations. Fig.~\ref{fig:HalfMBBDimension study} (b) illustrates that for a specimen with dimensions $H = 20~\text{mm}$ and $L = 60~\text{mm}$, an increase in the micropolar coupling number $N$ from $0$ to $0.99$ results in a significant reduction in compliance, indicating enhanced stiffness of the optimized structure under a given load due to the additional rotational stiffness introduced by micropolar effects. To provide a concise summary, Fig.~\ref{fig:domaindifferentlbNValues} plots the normalized compliance on a logarithmic scale for various $H/\lb$ ratios ranging from $2$ to $100$. The plot clearly demonstrates that as $H/\lb$ approaches $100$, the influence of the micropolar coupling number on compliance becomes negligible. Conversely, for small ratios, such as $H/\lb = 2$, compliance values for $N=0.5$ and $0.99$ are markedly lower than for $N = 0.0$. This confirms that size-dependent effects dominate when the specimen dimension is comparable to the bending length scale $\lb$. In these scenarios, conventional topology optimization methods are insufficient to accurately capture the underlying size-dependent phenomena.

\subsection{Example II: Fixed beam under thermo-elastic Loading}
\label{subsec:FixedBeam}
\begin{figure}[H]
    \begin{subfigure}[t]{0.49\linewidth}
        \centering
        \caption{}
        \includegraphics[width=\linewidth]{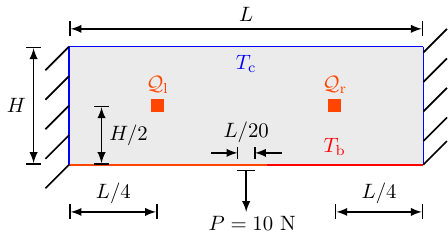}
    \end{subfigure}
    \begin{subfigure}[t]{0.49\linewidth}
        \centering
        \caption{}
        \vspace{8mm}
\includegraphics[width=0.75\linewidth]{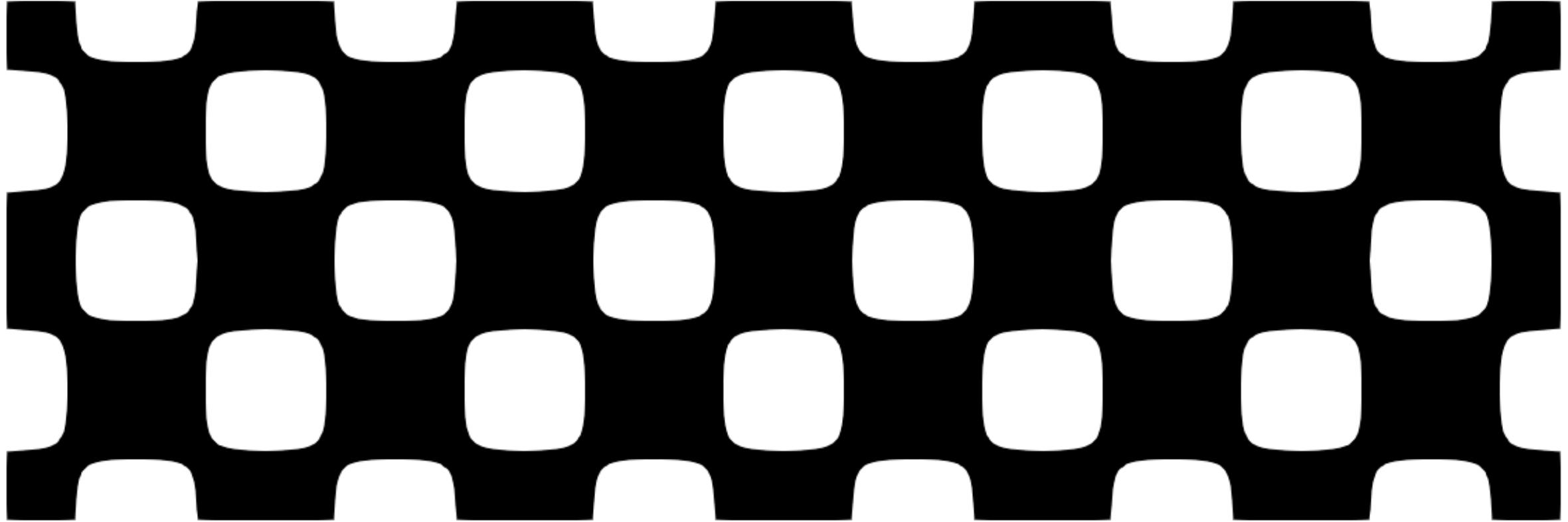}
    \end{subfigure}
\caption{In (a), the design domain and boundary conditions are shown for a fixed beam in Example II under thermo-mechanical loading. The beam dimensions are $H = 10~\text{mm}$ and $L = 30~\text{mm}$. The top, left, and right faces are maintained at a constant temperature $T_\mathrm{c} = 0\degree$C, while the bottom face is kept at a variable temperature $\Tb$. Two internal heat-sources, $\mathcal{Q}_{\mathrm{l}}$ and $\mathcal{Q}_{\mathrm{r}}$, located at a distance $L/4$ from the corresponding fixed edges and $H/2$ from the bottom face, are also prescribed at the variable temperature $\Tb$. A load of $P = 10~\text{N}$ is applied at the midpoint of the bottom face over a width of $L/20$, and (b) shows the initial design.}
\label{fig:domainForThermMech_FixeBeam}
\end{figure}

In this section, we present a numerical example of a fixed beam to demonstrate the proposed micropolar thermo-elastic topology optimization method under steady-state heat conduction conditions. An extensive study is conducted to investigate the effect of thermal and mechanical loads, micropolar coupling number $N$, and bending length scale $\lb$. The topology optimization problem for the considered thermo-elastic test case is formulated as  
\begin{equation}
\begin{aligned}
\min_{\phi} \quad & 
\mathcal{J}(\phi) := \int_{\mathcal{D}} \Ie\, \boldsymbol{\sigma^B}(\mathbf{u},T) : \epsElasSym(\mathbf{u},T) \, \mathrm{d} V, \\
\text{subjected to} \quad & \mathcal{C}(\phi) \leq 0, \\
& a_{\text{MP}}\big((\mathbf{u},\boldsymbol{\theta},T),(\mathbf{w},\boldsymbol{\vartheta},\tau),\phi \big) 
   = b_{\text{MP}}\big((\mathbf{w},\boldsymbol{\vartheta},\tau),\phi\big), 
\end{aligned}
\label{eq:TopOptTM}
\end{equation}
where $\mathcal{J}(\phi)$ is the elastic strain energy and $\mathcal{C}(\phi)$ is computed using Eq.~\eqref{eq:VolConst}. The forms $a_{\text{MP}}$ and $b_{\text{MP}}$ denote the bilinear and linear weak formulations of the thermo-elastic micropolar governing PDEs, respectively, as given in Eq.~\eqref{eq:cc}. Table~\ref{tab:ThermoElasticMatProp} summarizes the material properties employed in the thermo-elastic problems of Examples~II and~III as discussed in Subsections~\ref{subsec:FixedBeam} and \ref{subsec:HalfMBBTm}.
\begin{table}[H]
\centering
\caption{Material properties used in the thermo-elastic analyses for Examples II and  III.}
\label{tab:ThermoElasticMatProp}
\renewcommand{\arraystretch}{1.3} 
\begin{tabular}{@{} l @{\hspace{3.8em}} c @{\hspace{3.8em}} c @{}}
\toprule
\textbf{Property} & \textbf{Value} & \textbf{Units} \\
\midrule
Young's modulus (solid), $E$ & $30.0 \times 10^{3}$ & $\mathrm{N/mm^{2}}$ \\
Void stiffness parameter, $\varepsilon_{\mathrm{elas}}$ & $1.0 \times 10^{-9}$ & -- \\
Thermal conductivity (solid), $k$ & $1.0$ & $\mathrm{W/(mm~\!\degree C)}$ \\
Void thermal conductivity parameter, $\varepsilon_{\mathrm{therm}}$ & $0.03$ & -- \\
Thermal expansion coefficient, $\alpha_t$ & $12\times 10^{-6}$ & $\mathrm{\degree C^{-1}}$ \\
Poisson's ratio, $\nu$ & $0.30$ & -- \\
\bottomrule
\end{tabular}
\end{table}
A fixed beam subjected to thermo-mechanical loading under plane-stress conditions is considered to analyze the influence of different micropolar parameters on the optimal topology. The geometric configuration, along with the mechanical and thermal boundary conditions, is illustrated in Fig.~\ref{fig:domainForThermMech_FixeBeam}. The beam has a height of $H = 10~\mathrm{mm}$, a length of $L = 30~\mathrm{mm}$, and a thickness of $t = 1~\mathrm{mm}$, giving an aspect ratio of $L/H = 3$. The design domain is discretized into $225 \times 75$ bilinear quadrilateral finite elements. As shown in the Fig.\,\ref{fig:domainForThermMech_FixeBeam}, a mechanical load of $P = 10~\mathrm{N}$ is applied on the center of the bottom face of the beam over a width of $L/20$. All other faces, except the bottom, are maintained at a constant temperature of $0\degree$C. At the bottom face and at both internal heat-sources, $\mathcal{Q}_{\mathrm{l}}$ and $\mathcal{Q}_{\mathrm{r}}$, the temperature is maintained at $\Tb$. The initial temperature $(T_0)$ of the beam is assumed to be $0\degree$C. The prescribed volume fraction $V_f$ for the designs is $50\%$ of the design domain. The initial holes are generated using the parameters $\zeta = 12/L$ and $b = 0.1$, as defined in Eq.~\eqref{eq:LSF}. The material properties used herein are provided in Table~\ref{tab:ThermoElasticMatProp}.

\subsubsection{Influence of thermal gradient}
\begin{figure}[H]
    \centering
    \begin{subfigure}[t]{\linewidth}
    \centering
        \includegraphics[width=0.6\linewidth]{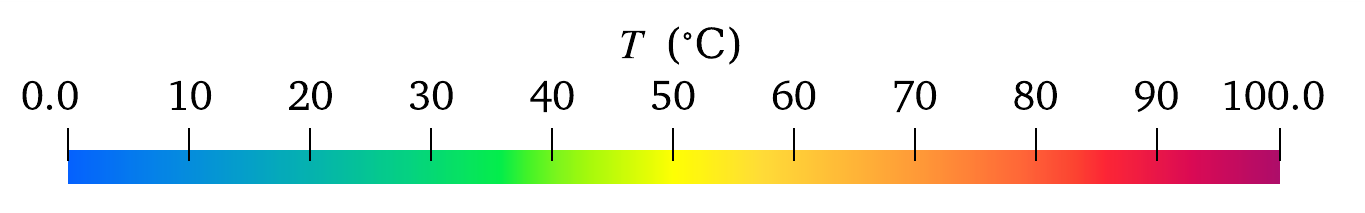}
    \end{subfigure}
    \\[-0.5cm]
    \begin{subfigure}[t]{0.3\linewidth}
    \centering
            \caption{$\Tb = 1\degree$C}
\includegraphics[width=\linewidth]{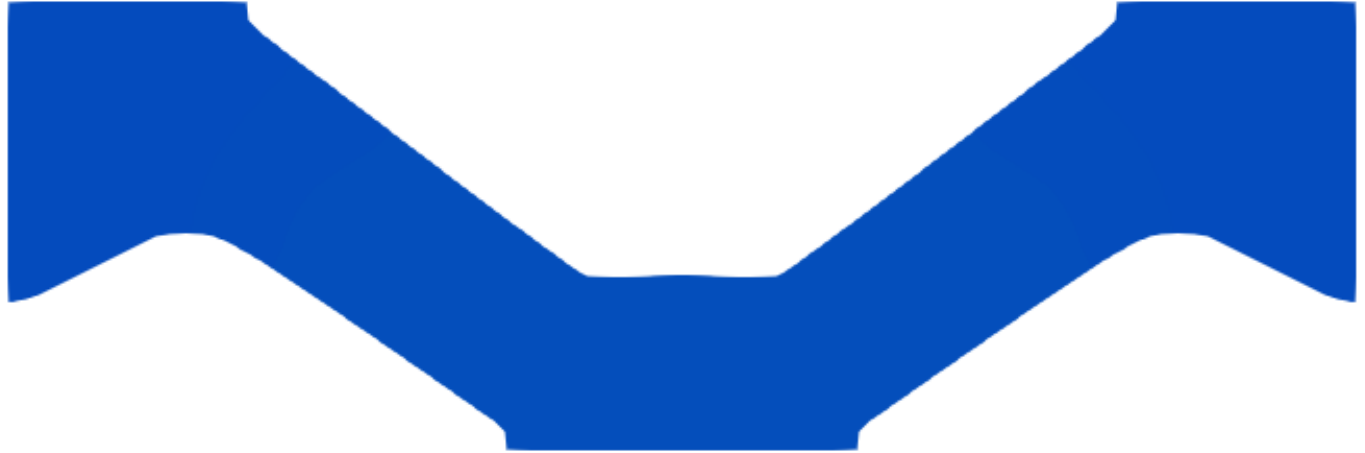}
    \label{fig:Tb0}
    \end{subfigure}
    \hspace{0.2cm}
    \begin{subfigure}[t]{0.3\linewidth}
    \centering
         \caption{$\Tb = 50\degree$C}
   \includegraphics[width=\linewidth]{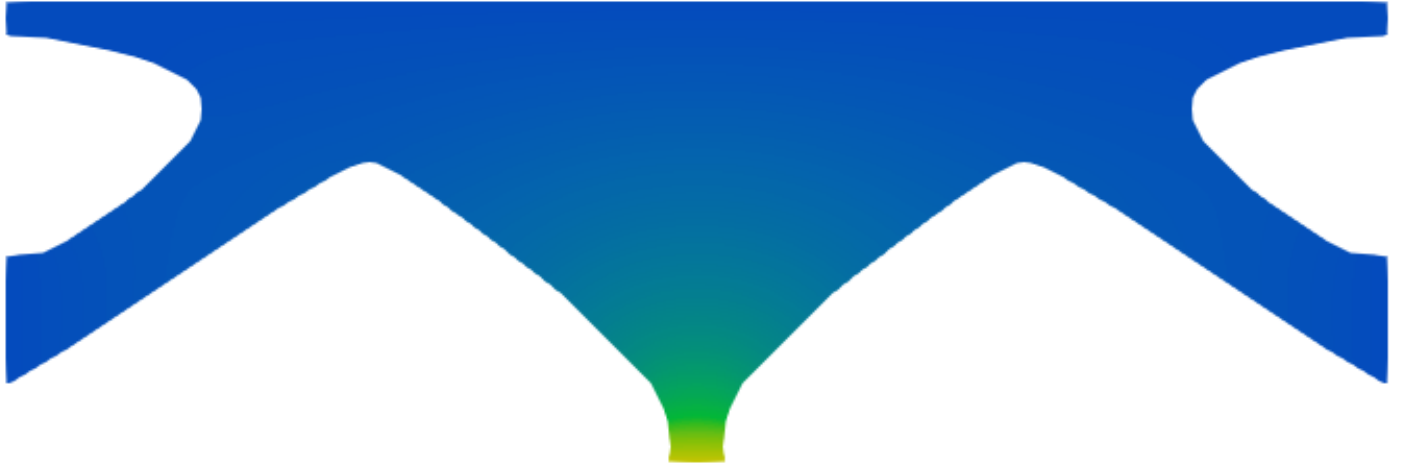}
    \end{subfigure}
    \hspace{0.2cm}
    \begin{subfigure}[t]{0.3\linewidth}
    \centering
        \caption{$\Tb = 100\degree$C}
    \includegraphics[width=\linewidth]{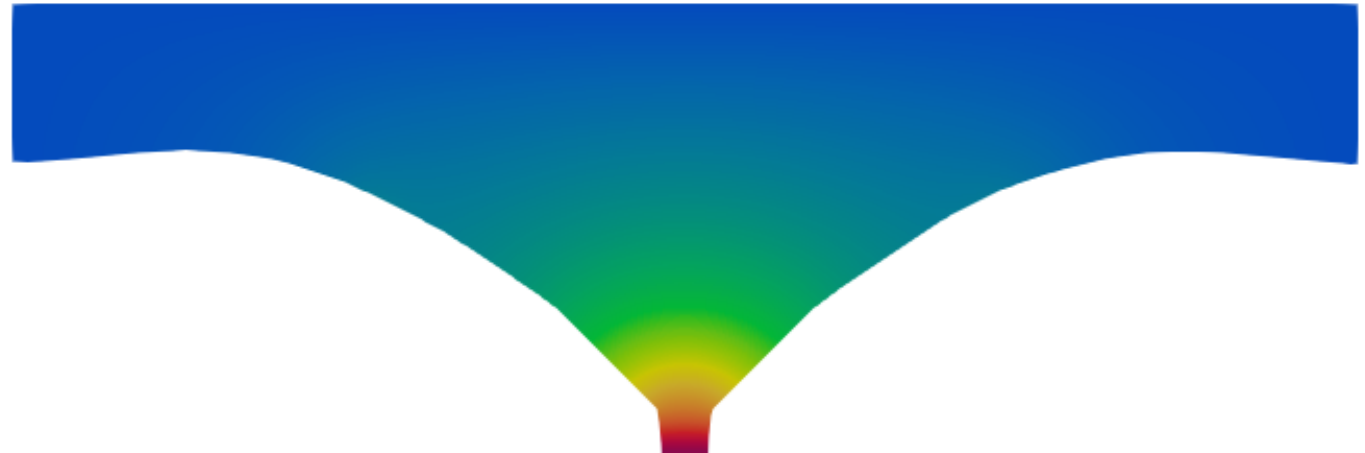}
    \end{subfigure}
\caption{Topologically optimized designs for different applied bottom-face, and two internal heat-sources temperature (a) $\Tb = 1\degree$C, (b) $\Tb = 50\degree$C, and (c) $\Tb = 100\degree$C  for the fixed beam in Example II.}
\label{fig:comparDifferBottomFaceTemp_FixedBeam}
\end{figure}

\begin{figure}[H]
    \centering
    \begin{subfigure}[t]{\linewidth}
    \centering
        \includegraphics[width=0.6\linewidth]{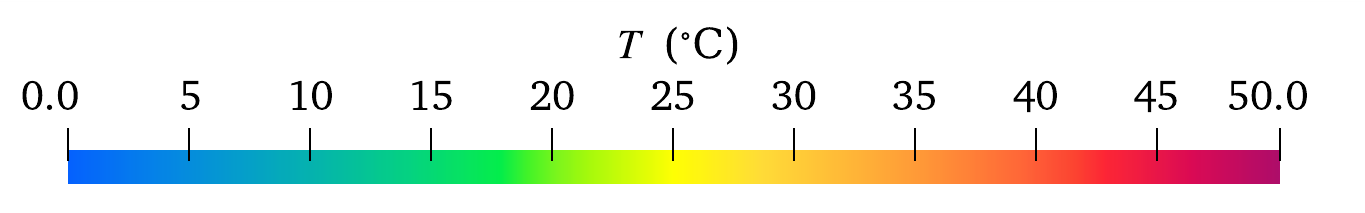}
    \end{subfigure}
    \\[-0.5cm]
    \begin{subfigure}[t]{0.4\linewidth}
    \centering
    \caption{Temperature distribution in $\Omega$}
        \includegraphics[width=\linewidth]{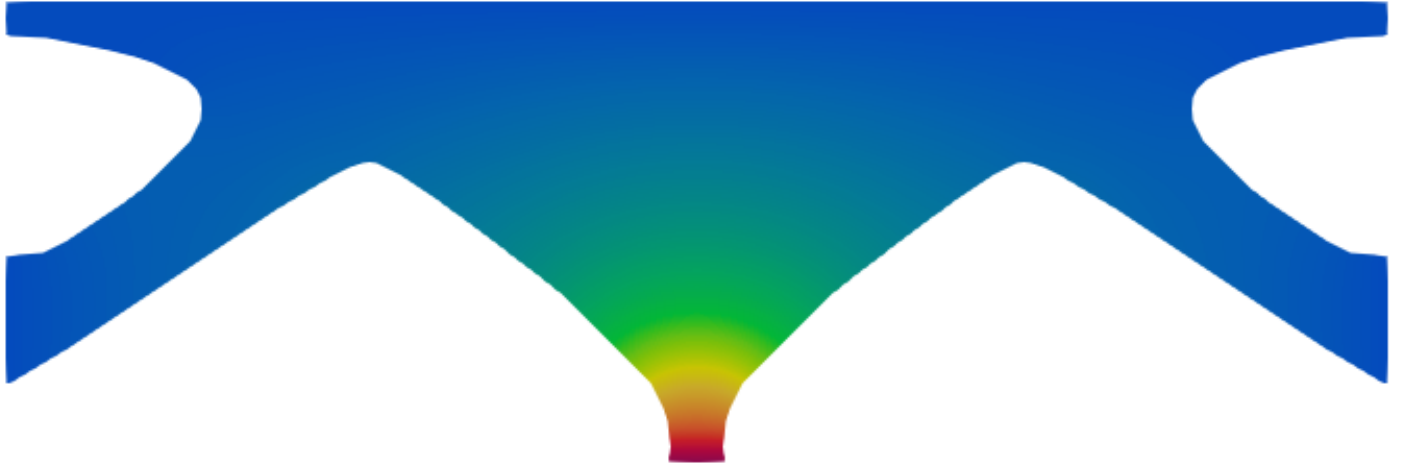}
    
    \label{fig:Fixedbeam25TempDistri}
    \end{subfigure}
    \hspace{0.9cm}
    \begin{subfigure}[t]{0.4\linewidth}
    \centering
    \caption{Temperature distribution within the domain $\mathcal{D}$}
        \includegraphics[width=\linewidth]{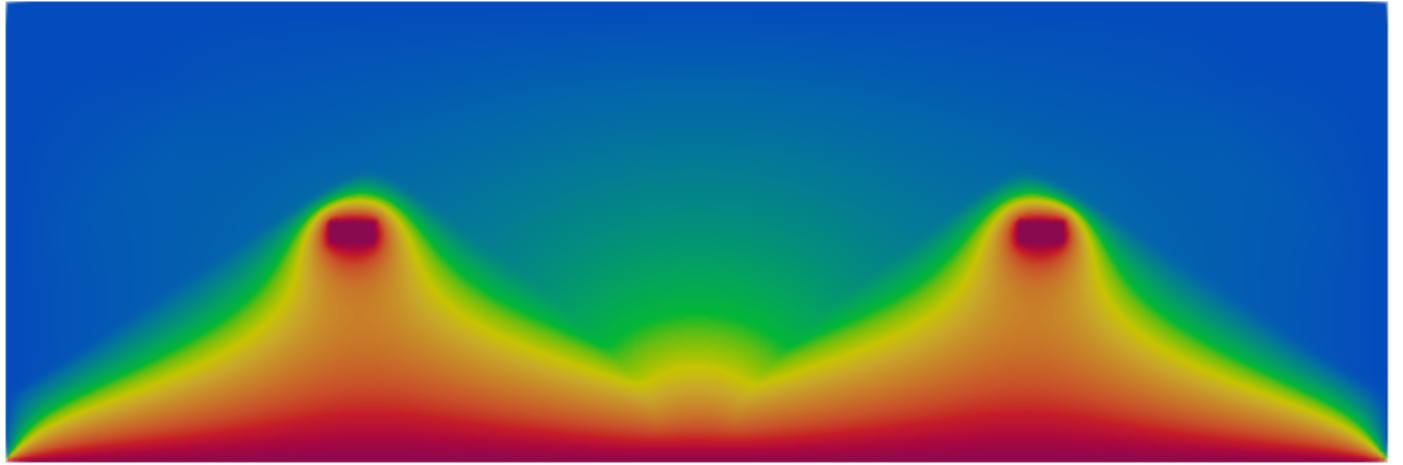}
    
    \label{fig:Fixedbeam25TempDistriFullDomain}
    \end{subfigure}
    \caption{In (a), topologically optimized design and the corresponding temperature distribution in the domain $\Omega$ are shown, while in (b), the temperature distribution in the whole design domain $\mathcal{D}$ is shown for a fixed beam (see Fig.~\ref{fig:domainForThermMech_FixeBeam}) in Example II, 
    subjected to $\Tb = 50^{\circ}\text{C}$.}
\label{fig:BottomFaceTemp25_FixedBeam}
\end{figure}

In this section, the influence of the thermal gradient on the optimized topology is investigated using the fixed beam configuration shown in Fig.~\ref{fig:domainForThermMech_FixeBeam}. To isolate the effect of the thermal gradient, a non-polar model with $N = 0.0$ is employed, while the temperature of the bottom face and the two internal heat-sources, $\Tb$, is varied from $1\,\degree$C to $100\,\degree$C.
 As illustrated in Fig.~\ref{fig:comparDifferBottomFaceTemp_FixedBeam}, increasing the $\Tb$ drives material redistribution toward the cooler top face to reduce the thermal strains. For small thermal gradients (\textit{e.g.}, $\Tb = 1\degree$C), the optimized topology resembles a truss-like structure which efficiently distributes the applied mechanical load primarily through diagonal members. As the thermal gradient rises with $\Tb = 50\degree$C, significant changes in the topology are observed. The optimization favors material placement near the cooler region, leading to a structure that departs from the initial truss-like configuration. Then, at the highest temperature gradients studied ($\Tb = 100\degree$C), the load-carrying members along the beam height diminish, with material primarily concentrated along the beam’s central axis and top boundary to further accommodate the thermal-induced stresses. Hence, it can be inferred that with an increase in the thermal gradient, the optimized structure evolves to minimize thermal strain energy by redistributing material toward cooler regions.
Fig.~\ref{fig:BottomFaceTemp25_FixedBeam} (a) depicts the optimized design $\Omega$ and corresponding distribution for $T_{\mathrm{b}} = 50\degree\mathrm{C}$, while Fig.~\ref{fig:BottomFaceTemp25_FixedBeam} (b) illustrates the temperature field across the entire design domain $D$. The optimized topology strategically positions material to reduce direct exposure to high thermal gradients, with hollow regions functioning as thermal buffers that dissipate heat and enhance cooling efficiency. In the solid regions, the temperature variation is minimal, indicating effective thermal insulation, whereas the hollow zones exhibit significant heat dissipation, contributing to overall efficient thermal management of the structure.
\subsubsection{Influence of micropolar coupling number $N$}
\begin{figure}[H]
\centering
\includegraphics[width=0.6\linewidth]{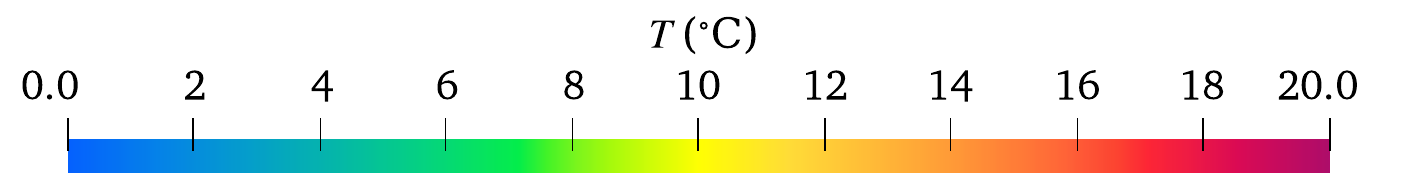}
    \vspace{0.3cm}
\renewcommand{\arraystretch}{2.5} 
\setlength{\tabcolsep}{6pt}     

\begin{tabular}{>{\centering\arraybackslash}m{1.75cm} | c | c | c}
\hline
 & {$\Tb = 1\degree$C} & {$\Tb = 8\degree$C} & {$\Tb = 20\degree$C} \\
 \hline
 \begin{minipage}[c][\height][c]{1.75cm}
  \centering
  \vspace{-1.0cm}
  $N=0.0$
\end{minipage} &
\begin{subfigure}[b]{0.25\linewidth}
\vspace{2mm} 
    \includegraphics[width=\linewidth]{Figures/FixedBeam10times30TMTb1P10N0.pdf}
\end{subfigure} &
\begin{subfigure}[b]{0.25\linewidth}
    \includegraphics[width=\linewidth]{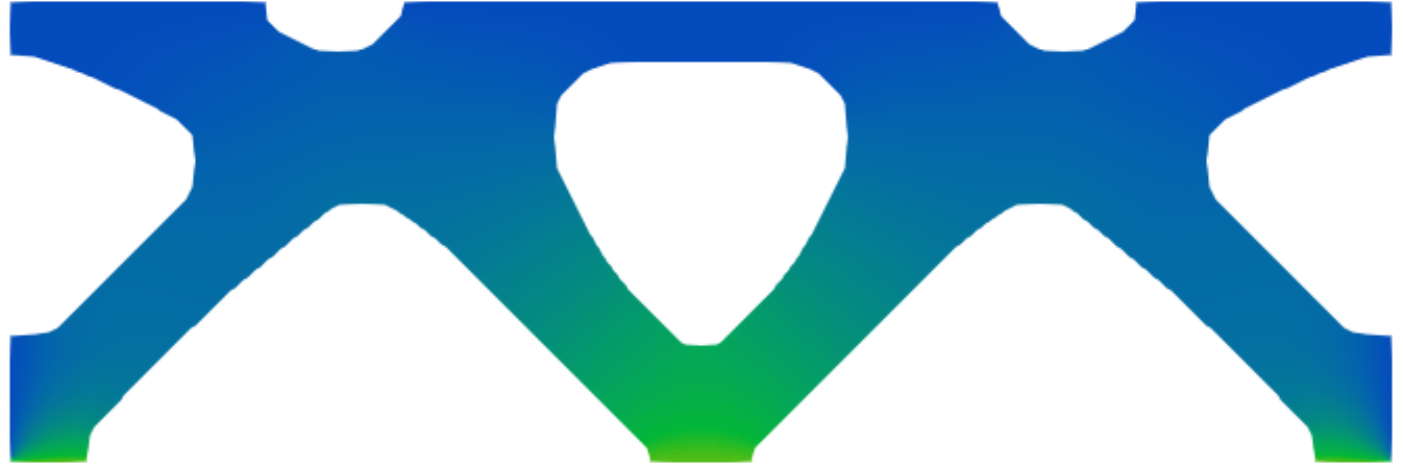}
\end{subfigure} &
\begin{subfigure}[b]{0.25\linewidth}
    \includegraphics[width=\linewidth]{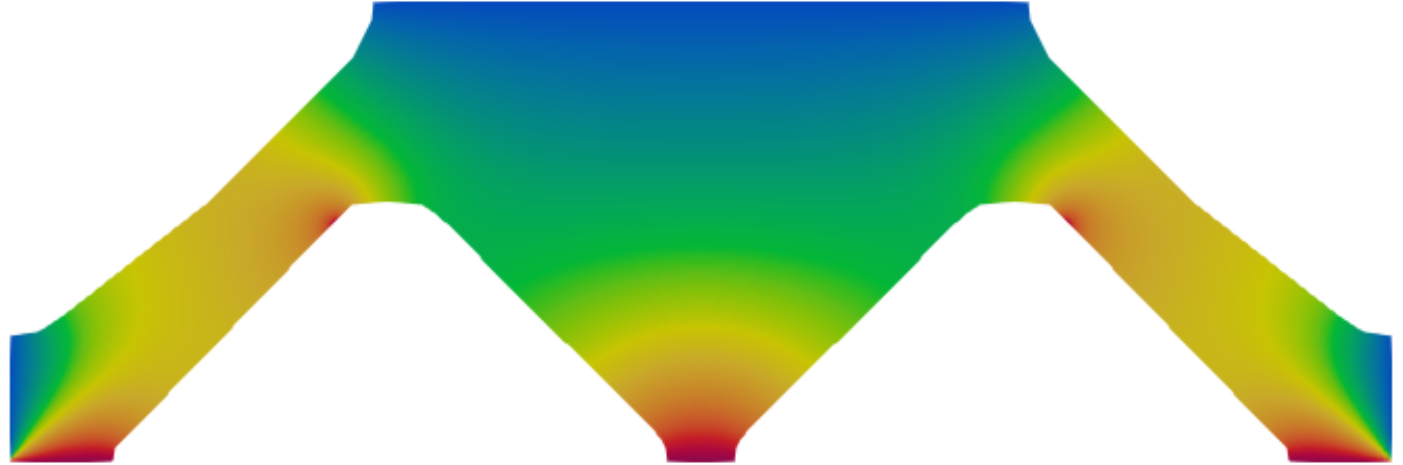}
\end{subfigure} \\

\hline 

\begin{minipage}[c][\height][c]{1.75cm}
  \centering
  \vspace{-1.0cm}
  $N=0.5$
\end{minipage} &
\begin{subfigure}[b]{0.25\linewidth}
\vspace{2mm} 
    \includegraphics[width=\linewidth]{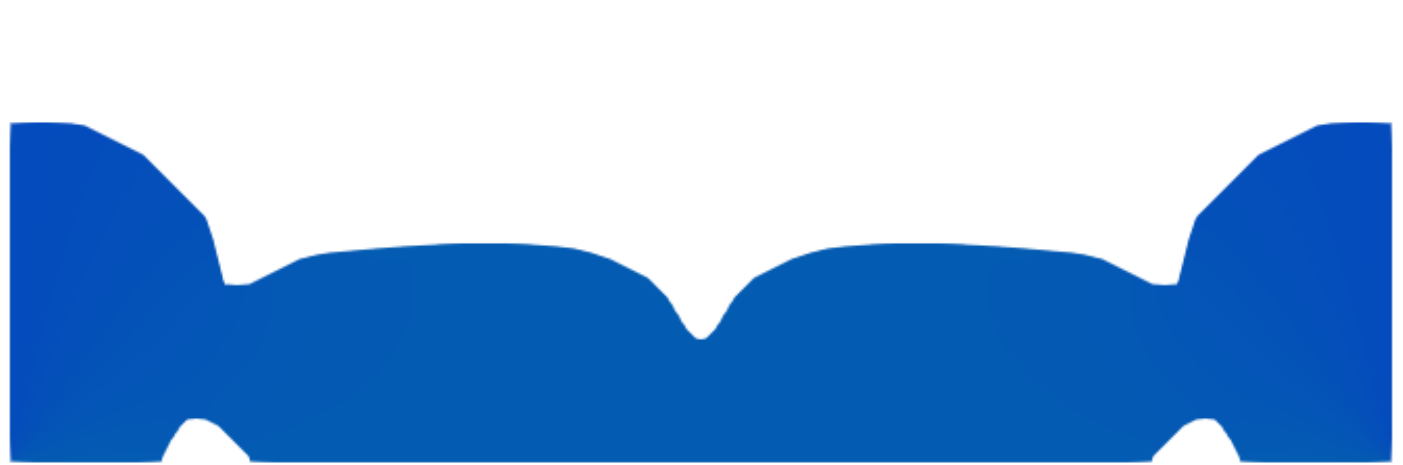}
\end{subfigure} &
\begin{subfigure}[b]{0.25\linewidth}
    \includegraphics[width=\linewidth]{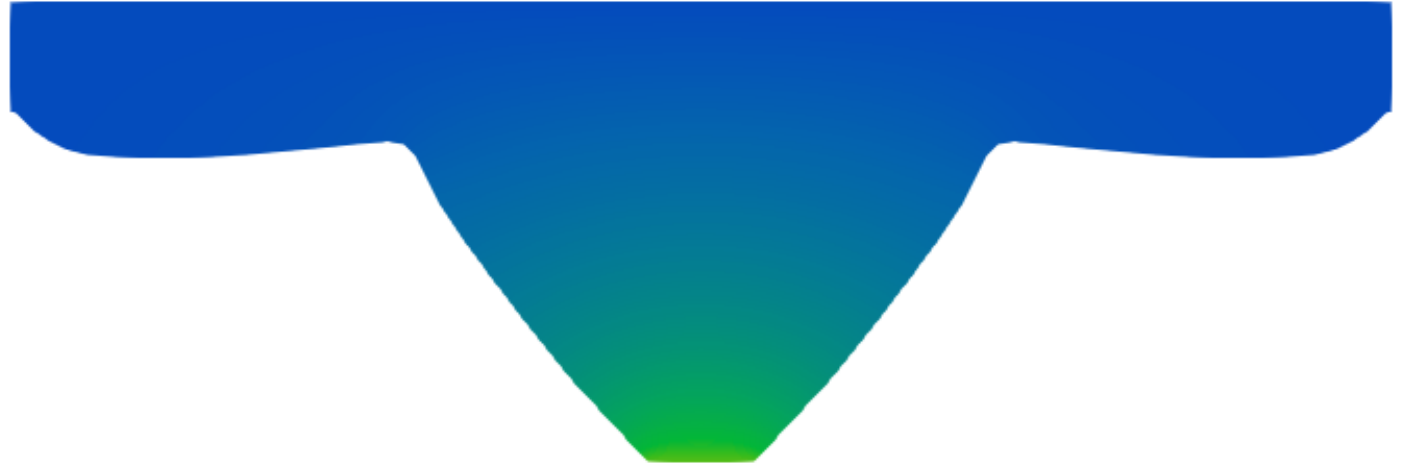}
\end{subfigure} &
\begin{subfigure}[b]{0.25\linewidth}
    \includegraphics[width=\linewidth]{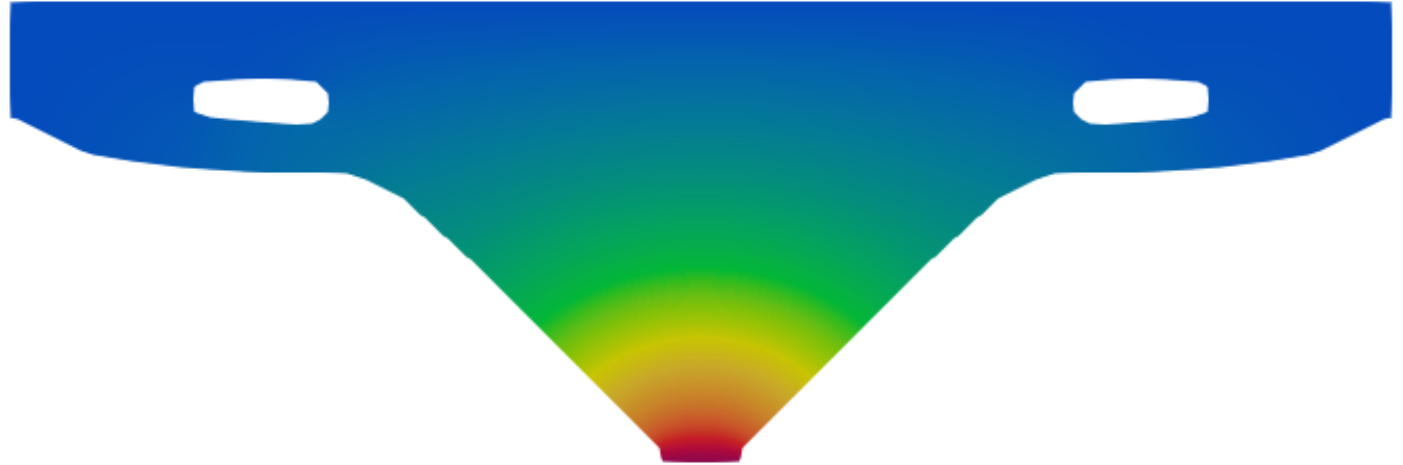}
\end{subfigure} \\

\hline 

\begin{minipage}[c][\height][c]{1.75cm}
  \centering
  \vspace{-1.0cm}
  $N=0.99$
\end{minipage}
 &
\begin{subfigure}[b]{0.25\linewidth}
\vspace{2mm} 
    \includegraphics[width=\linewidth]{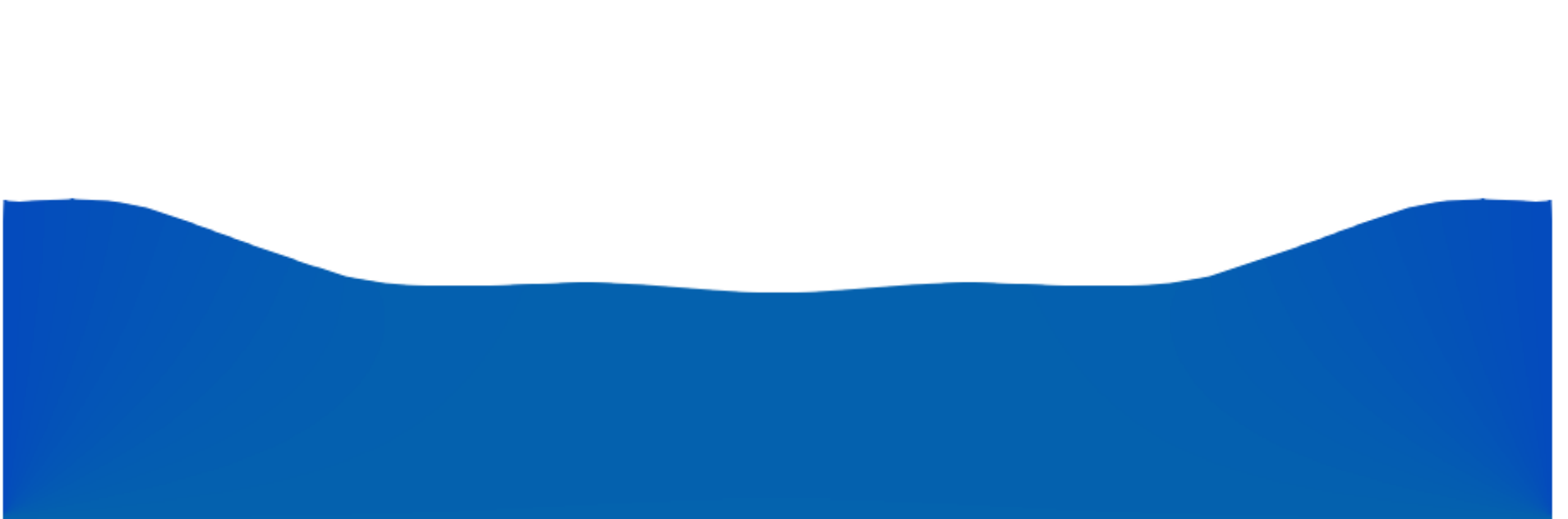}
\end{subfigure} &
\begin{subfigure}[b]{0.25\linewidth}
    \includegraphics[width=\linewidth]{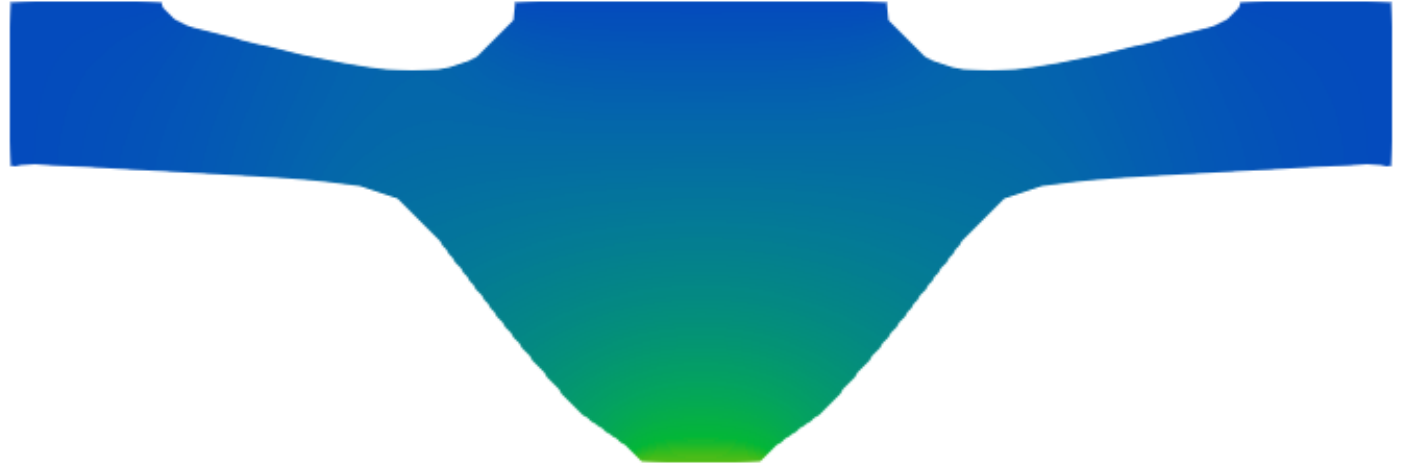}
\end{subfigure} &
\begin{subfigure}[b]{0.25\linewidth}
    \includegraphics[width=\linewidth]{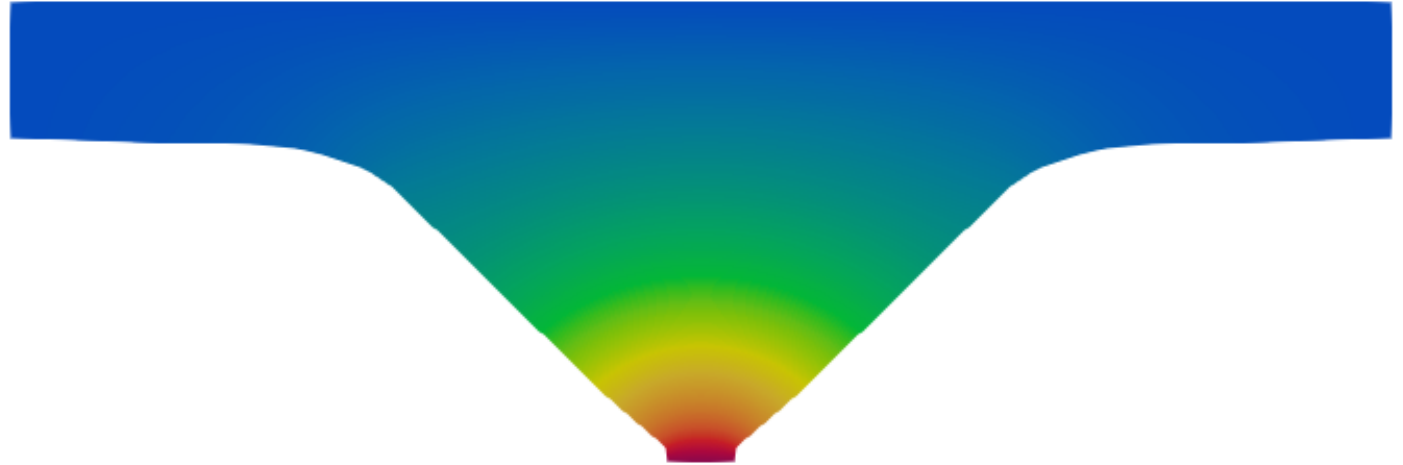}
\end{subfigure} \\
\hline
\end{tabular}
\caption{Topologically optimized designs for the fixed beam in Example II  for different temperature values $\Tb$ and micropolar coupling number $N$ with bending length scale $\lb = H/5 = 2$ mm.}
\label{fig:comparisonforTempValuesFixed_BeamNvariedLbfixed}
\end{figure}

In this section, the effect of the micropolar coupling number $N$ on the topologically optimized design under varying thermal gradients, $\Tb$, is examined for the fixed beam shown in Fig.~\ref{fig:domainForThermMech_FixeBeam}. Fig.~\ref{fig:comparisonforTempValuesFixed_BeamNvariedLbfixed} shows the optimized topologies of the beam under combined thermo-mechanical loading, with a downward load of $P = 10$~N applied at the bottom face. The bending length scale $\lb = H/5 = 2$~mm, and all other parameters remain constant. The applied bottom-face and the two internal heat-sources temperatures $\Tb$ increase across the columns from $1^\circ$C to $20^\circ$C, and the micropolar coupling number $N$ increases down the rows from $0.0$ to $0.99$. For the non-polar model ($N = 0.0$), increasing $\Tb$ causes material to shift upward toward the cooler regions. The diagonal members connected to the top (visible at $\Tb = 1^\circ$C) gradually disappear, while the members connected to the bottom (prominent at $\Tb = 8^\circ$C) thicken. At $\Tb = 20^\circ$C, these bottom-connected members further thicken, with additional material concentrated along the central axis, making the design strongly temperature-driven, removing material from the bottom face with higher temperature. As the micropolar coupling number $N$ increases, the material distribution changes notably. At $\Tb = 1^\circ$C, higher $N$ shifts material toward the bottom where the load is applied. At $\Tb = 8^\circ$C, the bottom-connected members vanish, and material accumulates along the central axis. By $\Tb = 20^\circ$C, the members that were connected to the bottom at $N = 0.0$ shift toward the cooler top region with increasing micropolarity, while additional material continues to concentrate along the beam’s central axis.

\begin{figure}[H]
    \centering
    \begin{subfigure}[t]{0.49\linewidth}
    \centering
    \caption{Mechanical and thermal strain energy with varying $\Tb$.}
    \includegraphics[width=\linewidth]{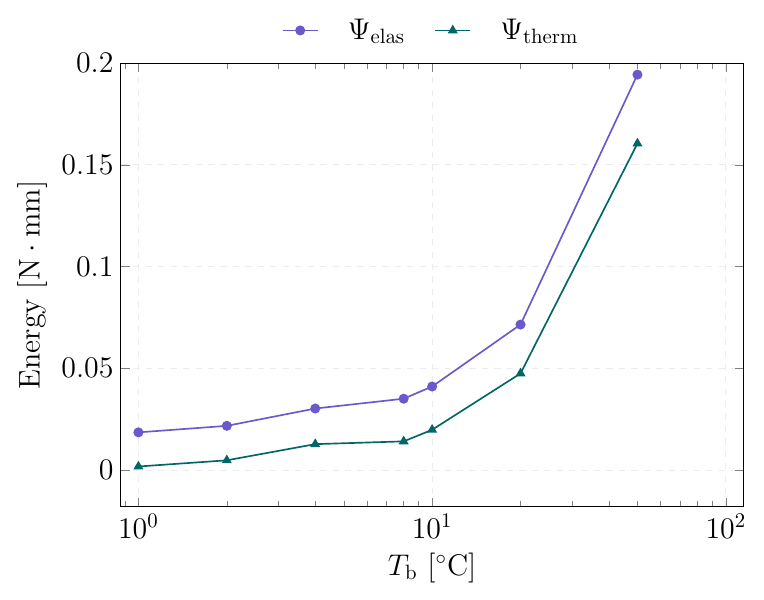}
    \end{subfigure}
    \begin{subfigure}[t]{0.49\linewidth}
    \centering
    \caption{Normalized strain energy contributions with varying $\Tb$.}
\includegraphics[width=\linewidth]{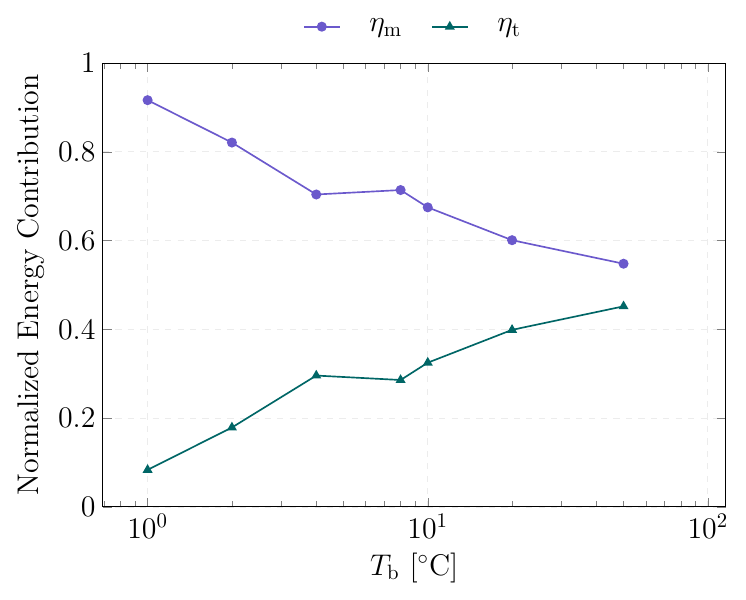}

\end{subfigure}
\caption{Strain energies of the optimized structures in Example II for the fixed beam (see Fig.~\ref{fig:domainForThermMech_FixeBeam}) are plotted. In (a), mechanical strain energy $\Psi_{\mathrm{elas}}$ and thermal strain energy $\Psi_{\mathrm{therm}}$ are shown as the applied bottom-face temperature $\Tb$ increases from $1^\circ$C to $50^\circ$C, for bending length scale $\lb = H/5 = 2$~mm and micropolar coupling number $N = 0.5$, and in (b), normalized mechanical strain energy $\eta_{\mathrm{m}} = \Psi_{\mathrm{elas}}/(\Psi_{\mathrm{elas}}+\Psi_{\mathrm{therm}})$ and normalized thermal strain energy $\eta_{\mathrm{t}} = \Psi_{\mathrm{therm}}/(\Psi_{\mathrm{elas}}+\Psi_{\mathrm{therm}})$ are plotted.}
\label{fig:FixedBeamdomaindifferentTbN0.5StrainEnergy}
\end{figure}

\begin{figure}[H]
    \centering
    \begin{subfigure}[b]{0.49\linewidth}
    \centering        \includegraphics[width=\linewidth]{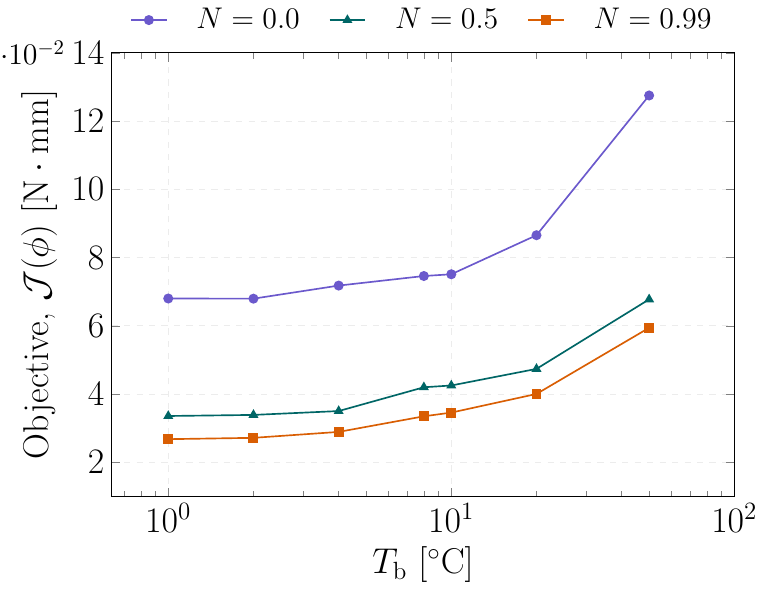}
    \end{subfigure}
\caption{Variation of elastic strain energy \textit{i.e.}, the objective with applied bottom-face temperature $\Tb$ ranging from $1\degree$C to $50\degree$C, for micropolar coupling number $N \in \{0.0, 0.5, 0.99\}$, and a bending length scale of $\lb = H/5 = 2$~mm in Example~II.}
\label{fig:FixedBeamdomaindifferentNValues}
\end{figure}

The material distribution pattern is further illustrated in Fig.~\ref{fig:FixedBeamdomaindifferentTbN0.5StrainEnergy} for a micropolar coupling number $N = 0.5$. In Fig.~\ref{fig:FixedBeamdomaindifferentTbN0.5StrainEnergy} (a), both the mechanical strain energy $\Psi_{\mathrm{elas}}$ and the thermal strain energy $\Psi_{\mathrm{therm}}$ increase monotonically as the applied bottom-face temperature $\Tb$ rises. Fig.~\ref{fig:FixedBeamdomaindifferentTbN0.5StrainEnergy} (b) presents the normalized mechanical strain energy, $\eta_{\mathrm{m}} = \Psi_{\mathrm{elas}} / (\Psi_{\mathrm{elas}} + \Psi_{\mathrm{therm}})$, and the normalized thermal strain energy, $\eta_{\mathrm{t}} = \Psi_{\mathrm{therm}} / (\Psi_{\mathrm{elas}} + \Psi_{\mathrm{therm}})$. As $\Tb$ increases, $\eta_{\mathrm{t}}$ rises while $\eta_{\mathrm{m}}$ correspondingly decreases. When $\Tb = 20^\circ$C, $\eta_{\mathrm{t}}$ accounts for nearly $40\%$ of the total thermo-mechanical energy, becoming significant enough to drive a major shape transformation toward the cooler top face. Beyond this point, the change in the optimized topology remains minor, as $\eta_{\mathrm{t}}$ saturates at around $50\%$ of the total energy. This explains the upward shift of the side legs for $N = 0.5$, indicating that the structure is less sensitive to temperature effects compared to the non-polar case ($N = 0.0$). Overall, these results demonstrate that stronger micropolar effects, corresponding to higher values of $N$, mitigate thermally induced material redistribution and enhance structural stability by providing resistance to micro-rotation $\thetab$ under combined thermal and mechanical loading.
In Fig.~\ref{fig:FixedBeamdomaindifferentNValues}, the elastic strain energy \textit{i.e.}, the objective, is plotted against the bottom-face and two internal heat-sources temperature $\Tb$ ranging from $1\degree$C to  $50\degree$C, for micropolar coupling number $N \in \{0.0,0.5,0.99\}$, and a bending length scale of $\lb = H/5 = 2$ mm. For all values of $N$, the elastic strain energy increases monotonically with $\Tb$. A marked reduction in the objective is observed when $N$ increases from $0.0$ to $0.5$, whereas further increasing $N$ from $0.5$ to $0.99$ produces no significant changes. The observed trend is consistent with the optimized design in Fig.~\ref{fig:comparisonforTempValuesFixed_BeamNvariedLbfixed}, which shows minimal variation between $N = 0.5$ and $0.99$. These results indicate that moderate micropolar coupling is sufficient to capture the primary size-dependent effects under thermo-mechanical loading, with limited additional benefit at higher $N$ values. Therefore, the micropolar theory plays a significant role in thermo-mechanical loading scenarios. It demonstrates that as size effects become prominent, structures modeled using micropolar theory exhibit improved efficiency and higher stiffness compared to those designed under the conventional Cauchy model.

\subsubsection{Influence of bending length scale $\lb$}
\begin{figure}[H]
\centering
    \includegraphics[width=0.6\linewidth]{Figures/ColorBarFixedBeamDifferentNT20.pdf}
    \vspace{0.2cm}
\renewcommand{\arraystretch}{2.5} 
\setlength{\tabcolsep}{6pt}     

\begin{tabular}{>{\centering\arraybackslash}m{1.75cm} | c | c | c}
\hline
 & {$\Tb = 1\degree$}C & {$\Tb = 8\degree$C} & {$\Tb = 20\degree$C} \\
 \hline
 \begin{minipage}[c][\height][c]{1.75cm}
  \centering
  \vspace{-1.0cm}
  $\lb = H/5$
\end{minipage} &
\begin{subfigure}[b]{0.25\linewidth}
\vspace{2mm} 
    \includegraphics[width=\linewidth]{Figures/FixedBeam10times30TMTb1P10N0.5Lb2.pdf}
\end{subfigure} &
\begin{subfigure}[b]{0.25\linewidth}
    \includegraphics[width=\linewidth]{Figures/FixedBeam10times30TMTb8P10N0.5Lb2.pdf}
\end{subfigure} &
\begin{subfigure}[b]{0.25\linewidth}
    \includegraphics[width=\linewidth]{Figures/FixedBeam10times30TMTb20P10N0.5Lb2.pdf}
\end{subfigure} \\

\hline 

\begin{minipage}[c][\height][c]{1.75cm}
  \centering
  \vspace{-1.0cm}
  $\lb = H/4$
\end{minipage} &
\begin{subfigure}[b]{0.25\linewidth}
\vspace{2mm} 
    \includegraphics[width=\linewidth]{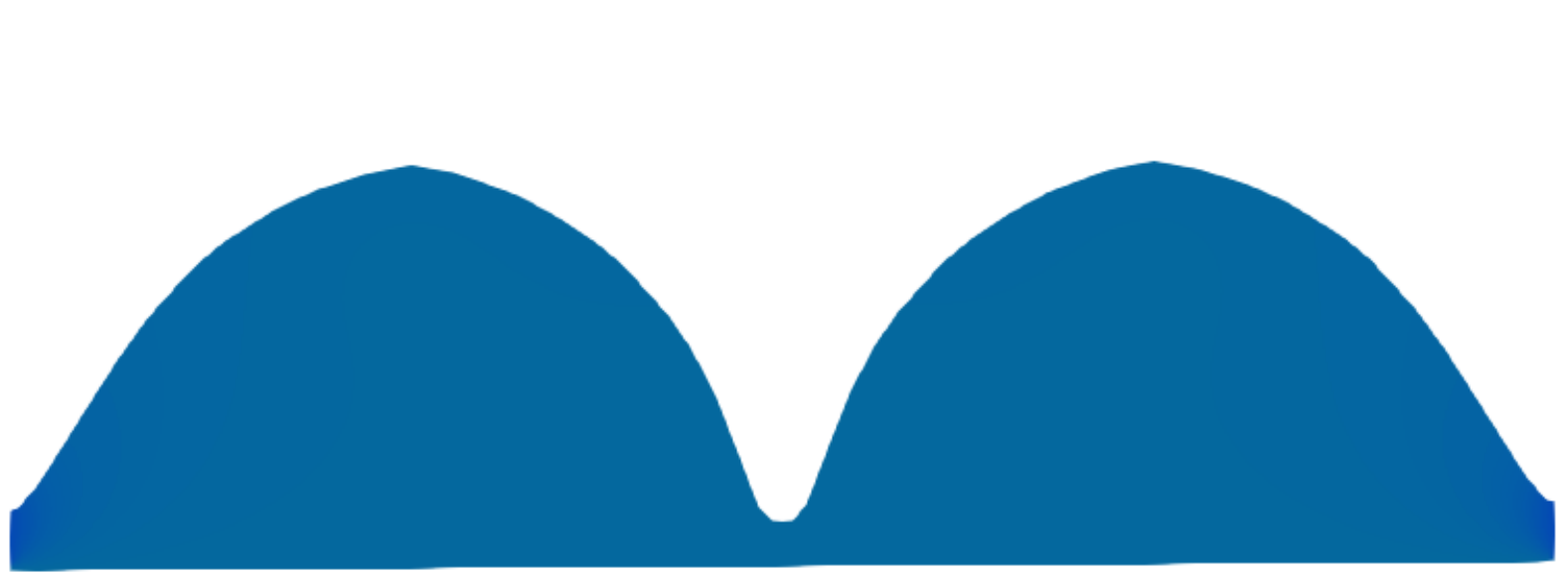}
\end{subfigure} &
\begin{subfigure}[b]{0.25\linewidth}
    \includegraphics[width=\linewidth]{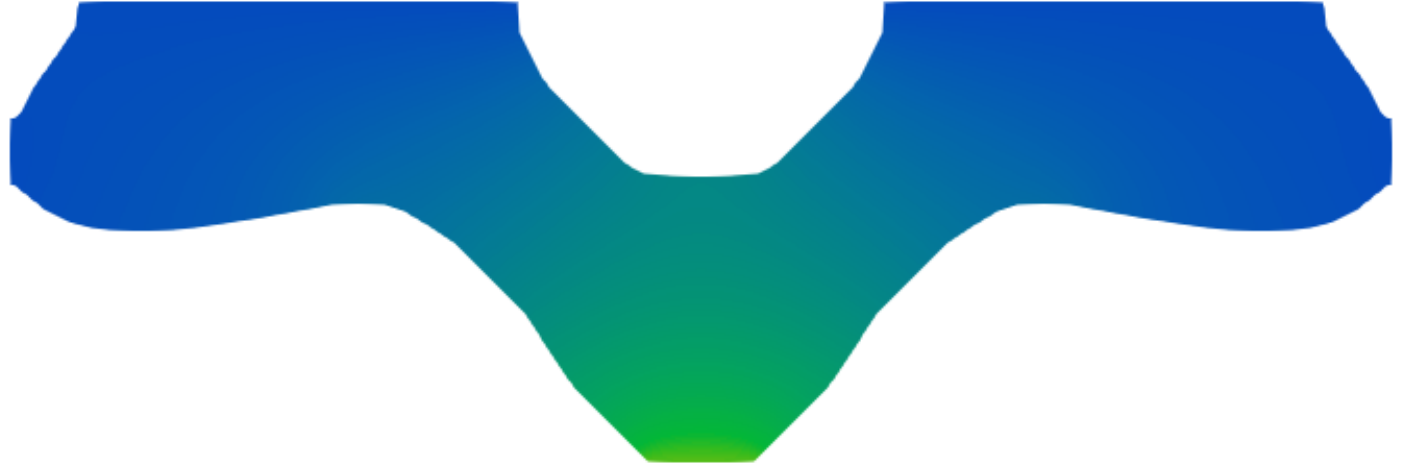}
\end{subfigure} &
\begin{subfigure}[b]{0.25\linewidth}
    \includegraphics[width=\linewidth]{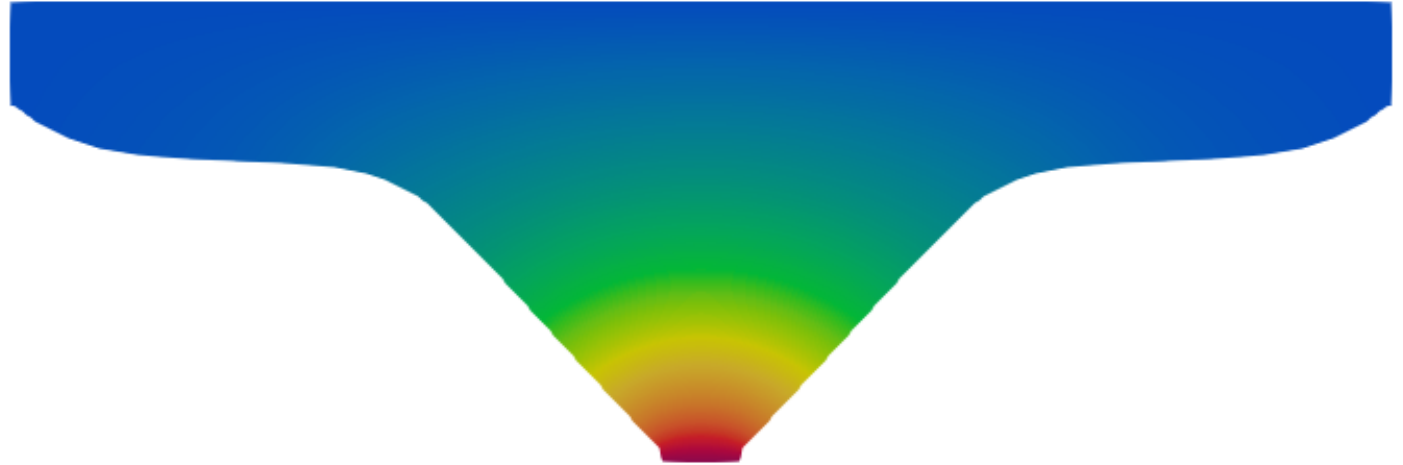}
\end{subfigure} \\

\hline 

\begin{minipage}[c][\height][c]{1.75cm}
  \centering
  \vspace{-1.0cm}
  $\lb = H/2$
\end{minipage}
 &
\begin{subfigure}[b]{0.25\linewidth}
\vspace{2mm} 
    \includegraphics[width=\linewidth]{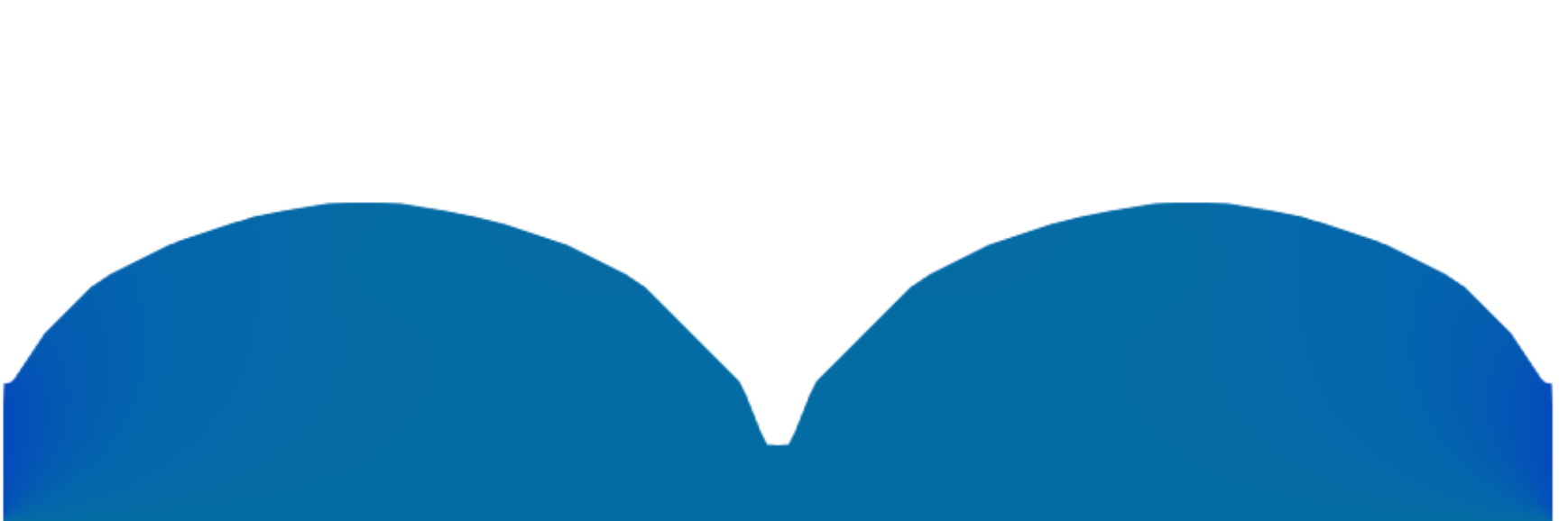}
\end{subfigure} &
\begin{subfigure}[b]{0.25\linewidth}
    \includegraphics[width=\linewidth]{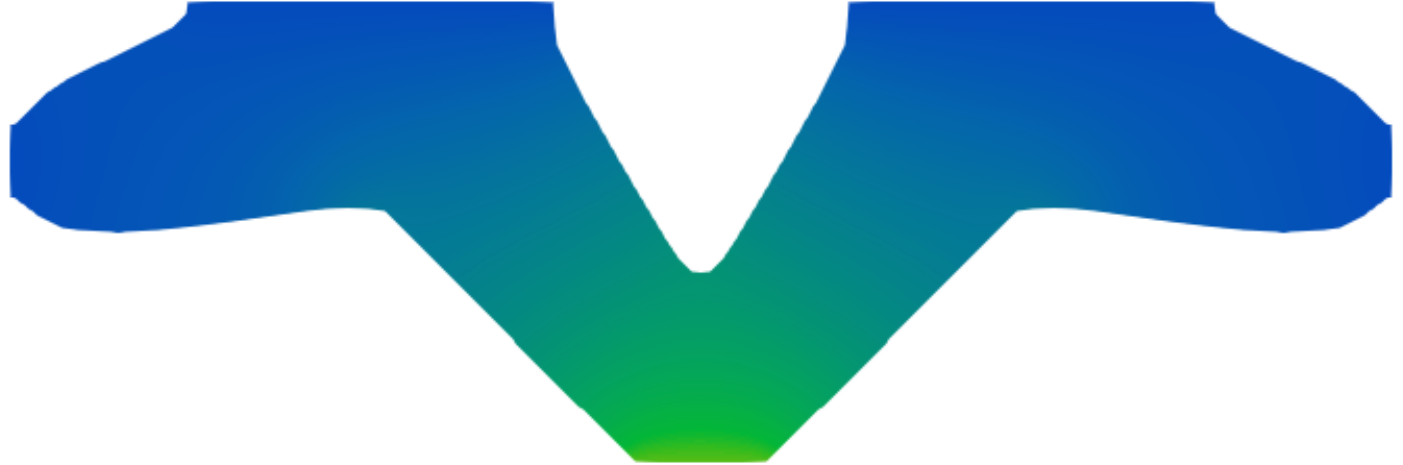}
\end{subfigure} &
\begin{subfigure}[b]{0.25\linewidth}
    \includegraphics[width=\linewidth]{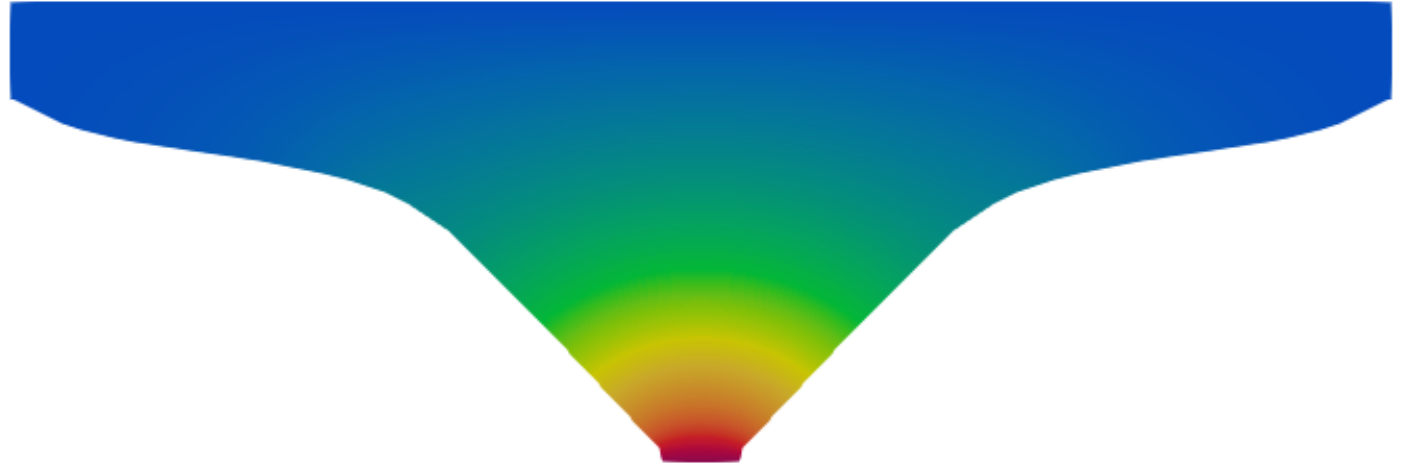}
\end{subfigure} \\
\hline
\end{tabular}
\caption{Topologically optimized designs in Example II for the fixed beam for different applied temperature $\Tb$ and bending length scale $\lb$ with the micropolar coupling number $N=0.5$.}
\label{fig:comparisonforTempValuesFixed_BeamLbvariedNfixed}
\end{figure}

\begin{figure}[H]
    \centering
    \begin{subfigure}[t]{0.49\linewidth}
    \centering
    \caption{Objective versus $\Tb$ and $\lb$ at $N = 0.5$.}
    \includegraphics[width=\linewidth]{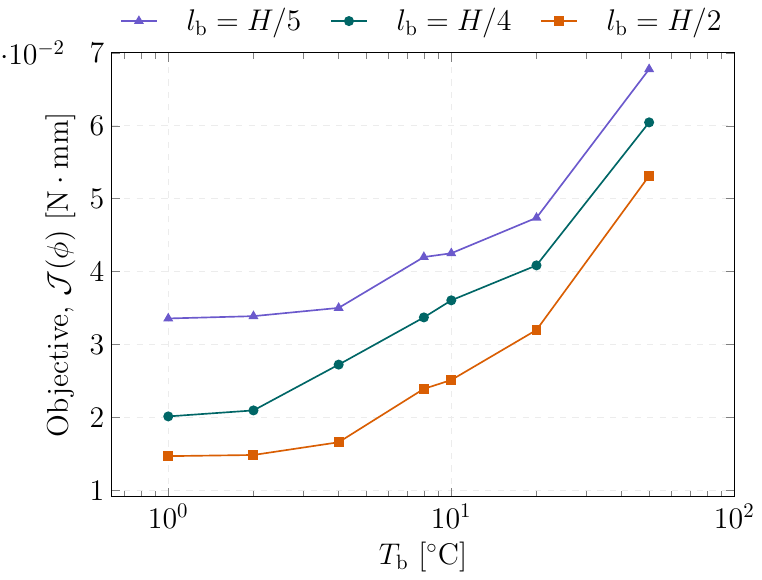}
    \label{fig:FixedBeamdomaindifferentLbValues}
    \end{subfigure}
    \begin{subfigure}[t]{0.49\linewidth}
    \centering
    \caption{Normalized objective versus $N$ and $\lb$ at $\Tb = 8\degree$C.}
    \includegraphics[width=0.95\linewidth]{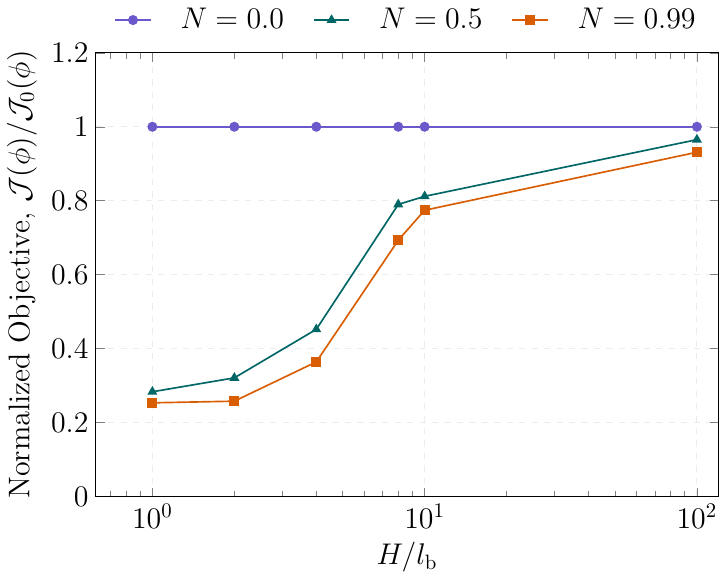}
    \label{fig:FixedBeamdomaindifferentN,LbValues}
\end{subfigure}
\caption{Elastic strain energy, \textit{i.e.}, the objective, is plotted. In (a), its variation with the thermal gradient $\Tb$ is shown for different bending length scales $\lb$, with micropolar coupling number $N = 0.5$, and in (b) the normalized objective is compared across micropolar coupling number $N$ and height-to-bending length scale ratios $(H/\lb)$ at applied temperature $\Tb = 8^\circ$C, in Example II for a fixed beam (see Fig.~\ref{fig:domainForThermMech_FixeBeam}).}
\label{fig:FixedBeam}
\end{figure}

In this section, the influence of the bending length scale $\lb$ on the topology-optimized configuration of the fixed beam in Fig.~\ref{fig:domainForThermMech_FixeBeam} subjected to varying $\Tb$ is investigated. Fig.~\ref{fig:comparisonforTempValuesFixed_BeamLbvariedNfixed} illustrates the optimized topologies under combined thermo-mechanical loading for micropolar coupling number $N = 0.5$, and all other parameters held constant. The applied bottom-face temperature $\Tb$ increases across the columns from $1\degree$C to $20\degree$C, while the bending length scale $\lb$ varies down the rows from $\lb = H/5 = 2$~mm to $\lb = H/2 = 5$~mm.  At a lower temperature ($\Tb = 1\degree$C), the optimized design concentrates material near the beam’s lower region, with reduced thickness around the loading point for all $\lb$ values. For a smaller bending length scale ($\lb = H/5$), the material distribution is relatively uniform, whereas for $\lb = H/4$, and $H/2$, the structure becomes more consolidated, forming broader, arch-like zones across the domain—an effect attributed to stronger micro-rotational influences in the micropolar framework. As the temperature rises to $\Tb = 8\degree$C, the material shifts toward the cooler upper region while remaining connected to the bottom loading point. For $\lb = H/5$, the topology becomes more centralized, with material concentrated near the beam’s upper midsection and connected to the side boundaries. With larger $\lb$ values ($H/4$ and $H/2$), the structure transitions into a V-shaped configuration, efficiently redistributing loads under combined thermal and mechanical effects as micropolar effects intensify. At the highest temperature gradient ($\Tb = 20\degree$C), the design for $\lb = H/5$ exhibits partially hollow zones, with material concentrated along the central axis. Further increases in $\lb$ to $H/4$ and $H/2$ result in minimal topological changes, indicating convergence toward an efficient configuration under these thermal conditions. Overall, the designs reveal that both thermal gradients and micropolar bending length scales significantly influence the optimized topology. Higher thermal loads and larger bending length scale promote centralized and compact designs, enhancing the structure’s ability to withstand coupled thermo-mechanical effects.
In Fig.~\ref{fig:FixedBeam} (a), the elastic strain energy \textit{i.e.}, the objective, is plotted against varying the applied temperature $\Tb$, for different bending length scales $\lb \in \{H/2,~H/4,~H/5\}$, and a micropolar coupling number $N = 0.5$. Fig.~\ref{fig:FixedBeam} (a) shows that as $\Tb$ increases, elastic strain energy increases for all the bending length scale $\lb$ values. For the smallest $\lb = H/5 = 2$ mm, the elastic strain energy increases monotonically with the thermal gradient with slight deviation in the low-temperature range. Importantly, increasing $\lb$ consistently reduces the elastic strain energy across the entire temperature range. This reduction can be attributed to the enhanced rotational stiffness introduced by larger bending length scales in the micropolar formulation, which promotes a stiffer structure with materials concentrated near the central region and thereby limits strain energy accumulation under combined thermal and mechanical loads. Whereas, in Fig.~\ref{fig:FixedBeam} (b), the normalized objective defined as the ratio of the elastic strain energy of the optimized design to that of the corresponding elastic strain energy of the non-polar case $(N = 0.0)$, is plotted for $\Tb = 8\degree$C as a function of the specimen height-to-bending length scale ratio, $H/\lb$. 
For $(N = 0.0)$, the proposed method reproduces the behavior of the conventional model, as can be seen from Fig.~\ref{fig:FixedBeam} (b), the normalized objective remains essentially constant across all $H/\lb$ values, indicating no sensitivity to the bending length scale $\lb$ value. However, for higher micropolar coupling number (\textit{i.e.}, $N = 0.5,~0.99$), the normalized strain energy decreases markedly, with the largest reduction observed at $H/\lb = 1$. As $H/\lb$ increases and approaches $100$, the values converge for all $N$, reflecting diminishing micropolar effects at lower bending length scales compared to the dimension of the specimen. It is also evident that the strain energy reduction is substantial when moving from $N = 0.0$ to $0.5$, but the change from $N = 0.5$ to $0.99$ is comparatively minor. The observed trend indicates that $N = 0.5$ captures most of the stiffening effect, and further increases in $N$ yield only marginal improvement in stiffness. 

\subsection{Example III: Half-MBB beam under thermo-elastic Loading}
\label{subsec:HalfMBBTm}
\begin{figure}[H]
    \centering
    \begin{subfigure}[t]{0.49\textwidth}
    \centering
    \caption{}
    \includegraphics[width=\linewidth]{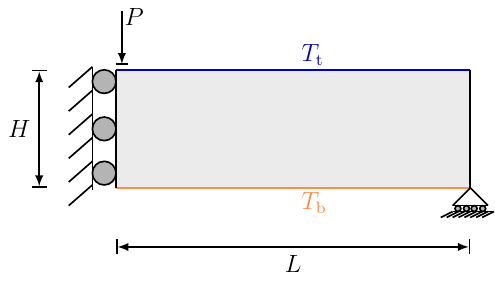}
    \end{subfigure}
    \begin{subfigure}[t]{0.49\textwidth}
    \centering
    \caption{}
    \vspace{11mm}
    \includegraphics[width=0.75\linewidth]{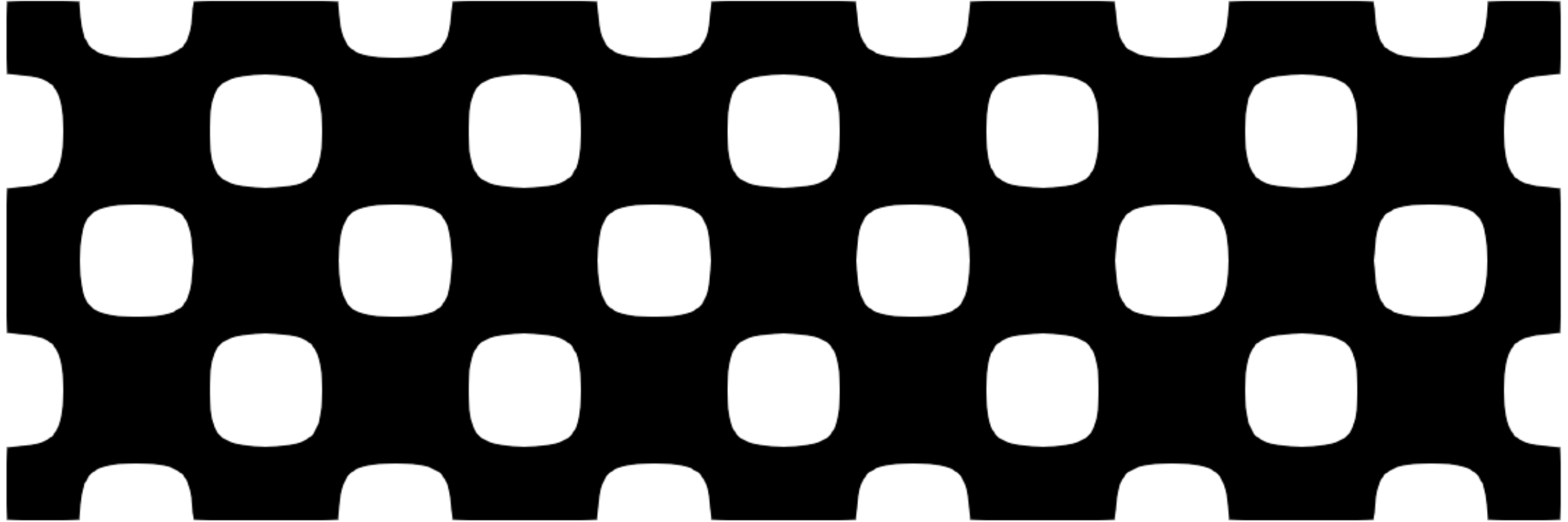}
    \end{subfigure}
    \caption{In (a), the design domain and boundary conditions are shown for a half-MBB beam in Example III under thermo-mechanical loading. The beam has a height of $H = 400~\text{mm}$ and a length of $L = 1200~\text{mm}$. Thermal boundary conditions are imposed on the top face of the beam, denoted as $T_{\text{t}} = 0\degree$C, and on the bottom face as $\Tb$. A load $P$ is applied at the top left corner of the beam in the downward direction, and (b) shows the initial design.}
\label{fig:domainForThermMech_HalfMBBBeam}
\end{figure}

In this section, we present a numerical example of a half-MBB beam under plane-stress conditions to investigate the influence of various parameters using the proposed micropolar topology optimization framework under thermo-mechanical loading. The geometric configuration, along with the mechanical and thermal boundary conditions, is illustrated in Fig.~\ref{fig:domainForThermMech_HalfMBBBeam}. The optimized topology for this example is determined from the finite element solution of Eq.~\eqref{eq:TopOptTM}. The beam has a height of $H = 400$ mm, length of $L = 1200$ mm, and thickness $t = 1$ mm, maintaining an aspect ratio of $L/H = 3$. The design domain is discretized into $300 \times 100$ bilinear quadrilateral finite elements. As shown in the Fig.~\ref{fig:domainForThermMech_HalfMBBBeam}, a mechanical load $P$ is applied on the top-left end of the beam. Due to symmetry about the vertical axis, half of the beam is modeled for finite element implementation. The thermal boundary conditions include a prescribed temperature of $T_{\text{t}} = 0\degree$C on the top face, while the bottom face is maintained at $\Tb$. The initial temperature of the beam is $T_0 = 0\degree$C. The prescribed volume fraction for the design is $40\%$ of the design domain. The initial holes are generated using the parameters $\zeta = 12/L$ and $b = 0.2$, as defined in Eq.~\eqref{eq:LSF}. The material properties used herein are listed in Table~\ref{tab:ThermoElasticMatProp}.

\subsubsection{Influence of mechanical and thermal loads} 
\label{sec:MechandThermLoad}
\begin{figure}[H]
    \centering
    \begin{subfigure}[t]{\linewidth}
        \centering
        \includegraphics[width=0.6\linewidth]{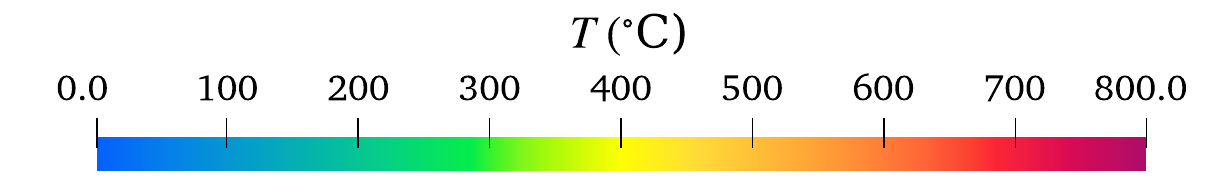}
    \end{subfigure}
    \\[-0.5cm]
    \begin{subfigure}[t]{0.3\linewidth}
        \centering
        \caption{$P = 500~\mathrm{N}$}
        \includegraphics[width=\linewidth]{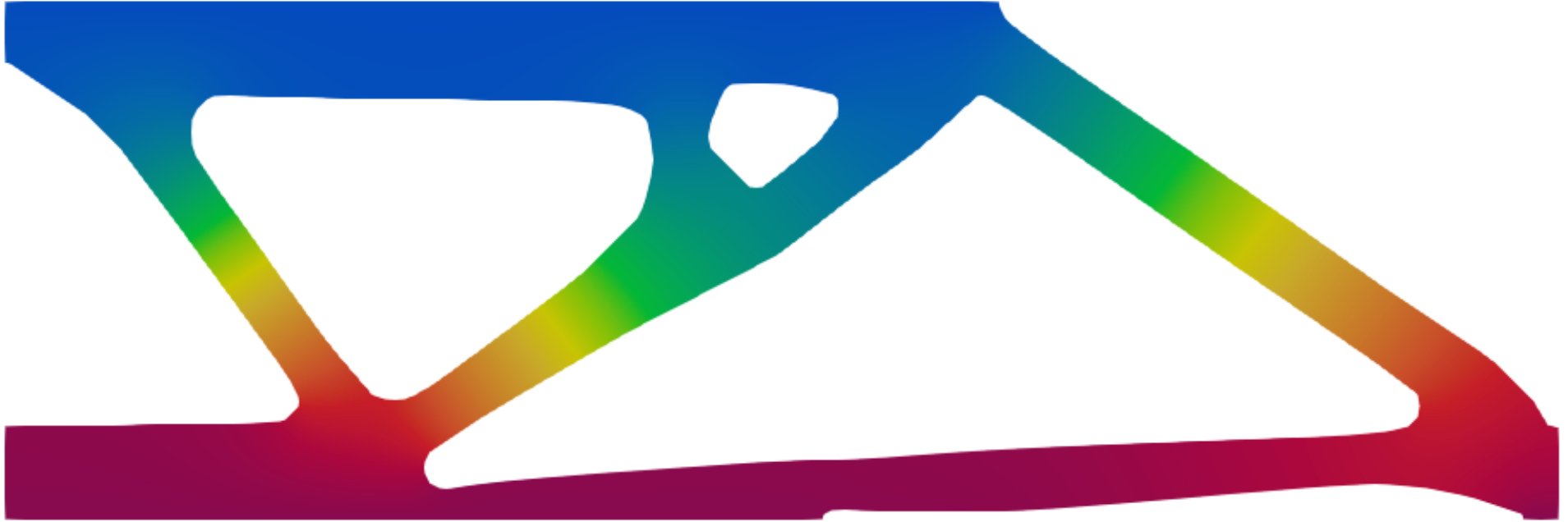}
        
    \end{subfigure}
    \hspace{0.2cm}
    \begin{subfigure}[t]{0.3\linewidth}
    \centering
    \caption{$P = 300~\mathrm{N}$}
    \includegraphics[width=\linewidth]{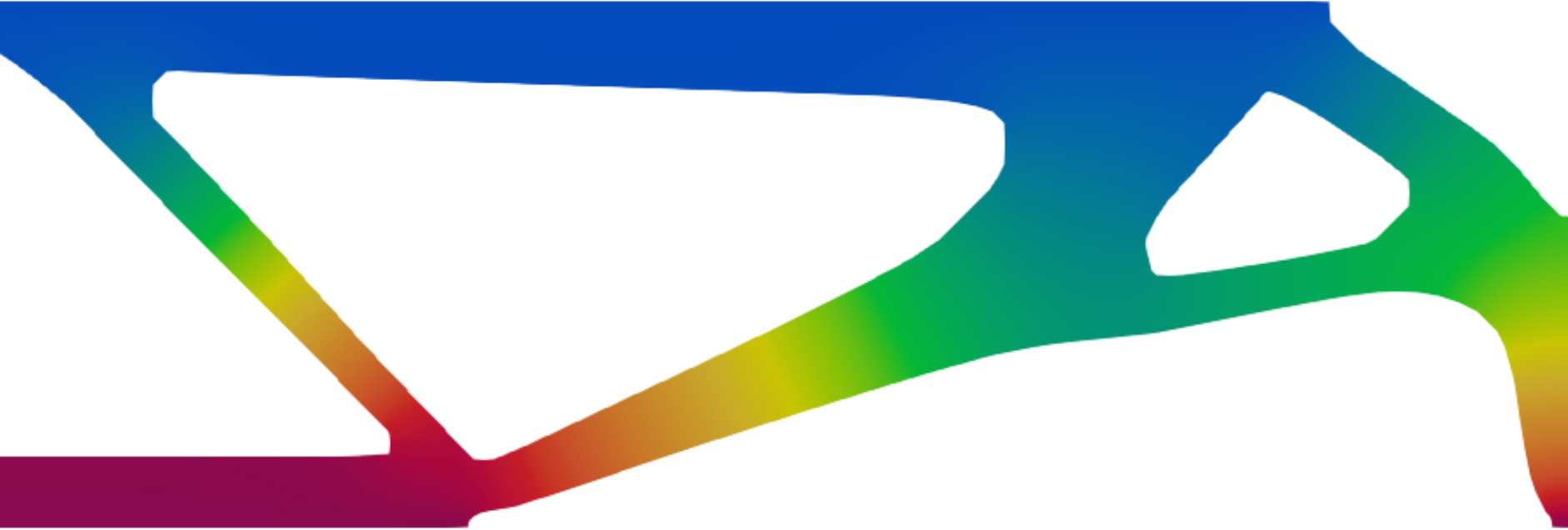}
        
    \end{subfigure}
    \hspace{0.2cm}
    \begin{subfigure}[t]{0.3\linewidth}
    \centering
    \caption{$P = 100~\mathrm{N}$}
    \includegraphics[width=\linewidth]{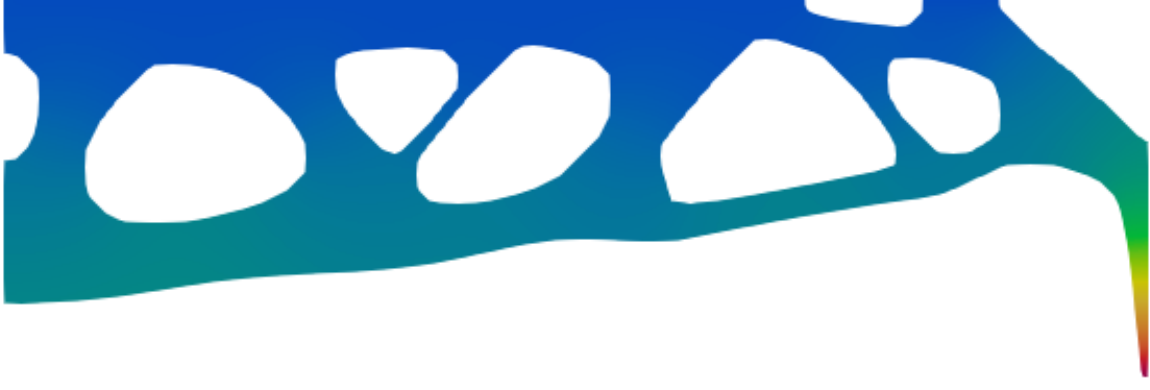}
    \end{subfigure}

    \vspace{0.3cm}

    \begin{subfigure}[t]{0.3\linewidth}
    \centering
    \caption{$\Tb = 200\degree$C}
    \includegraphics[width=\linewidth]{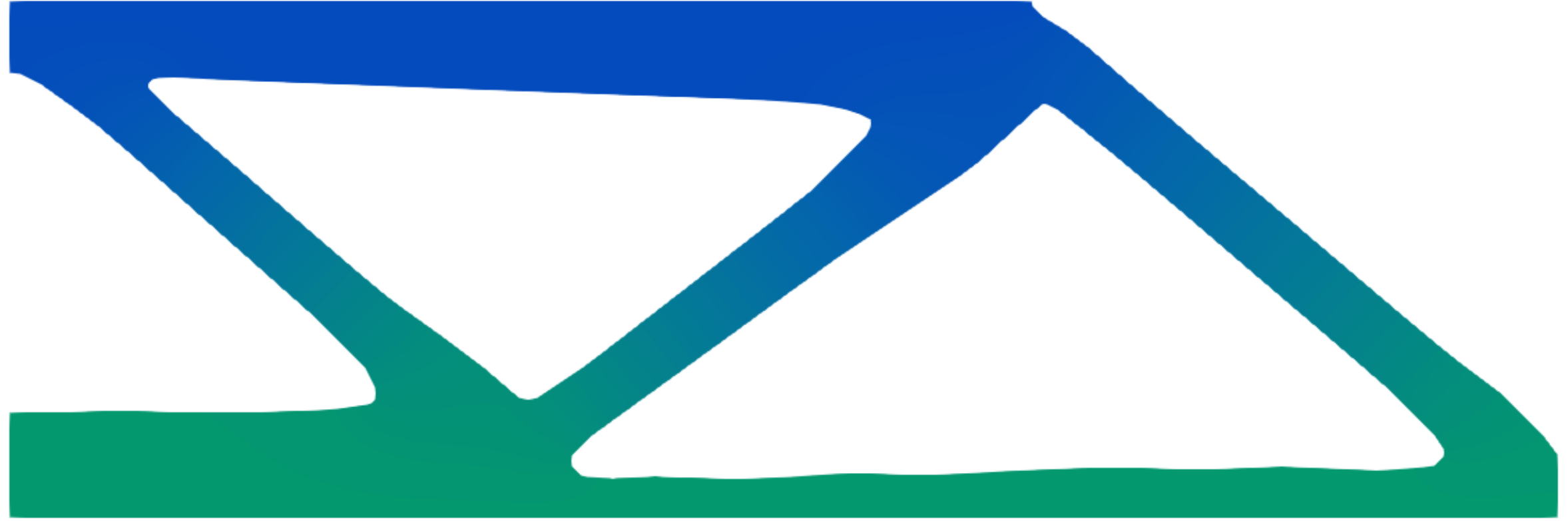}
    \end{subfigure}
    \hspace{0.2cm}
    \begin{subfigure}[t]{0.3\linewidth}
    \centering
    \caption{$\Tb = 600\degree$C}
    \includegraphics[width=\linewidth]{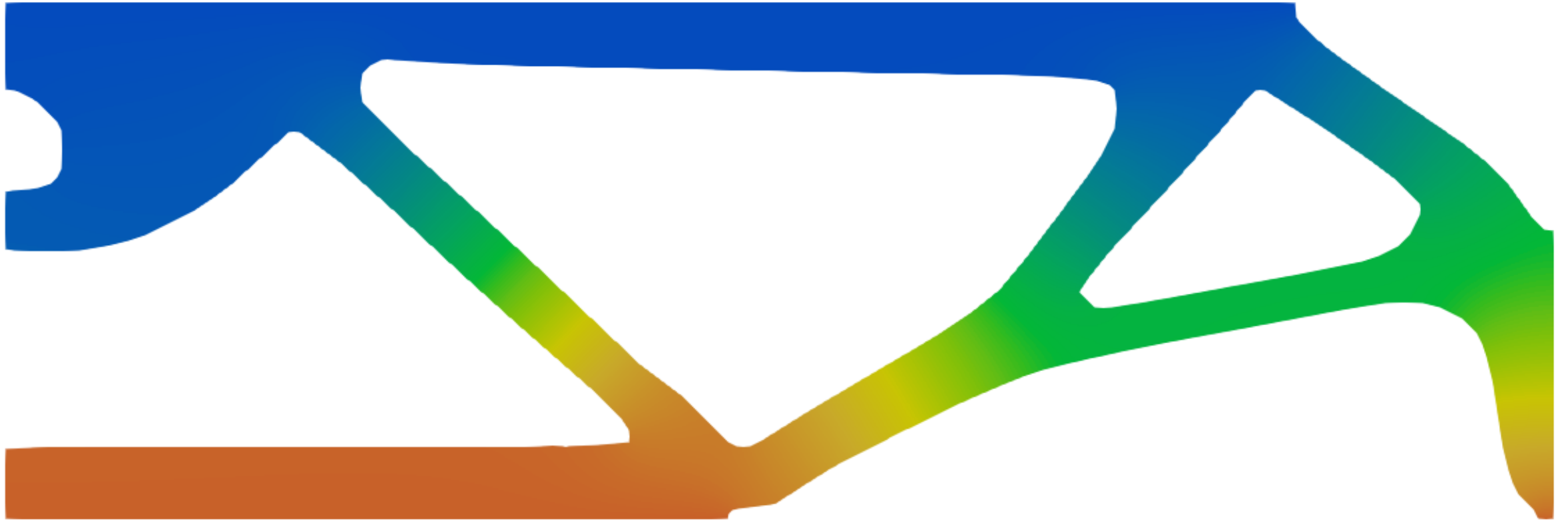}
    \end{subfigure}
    \hspace{0.2cm}
    \begin{subfigure}[t]{0.3\linewidth}
    \centering
    \caption{$\Tb = 800\degree$C}
    \includegraphics[width=\linewidth]{Figures/HAlfMBB_TM_Tb800_P300N.pdf}
    \end{subfigure}
\caption{Topologically optimized designs for the half-MBB beam in Example III under different mechanical and thermal loading conditions, with micropolar coupling number $N = 0.0$. {Row 1:} demonstrates the influence of load magnitude $P$ with $\Tb = 800\degree$C and $T_{\text{t}} = 0\degree$C. {Row 2:} demonstrates the influence of applied bottom-face temperature $\Tb$ while the load $P = 300$ N and $T_{\mathrm{t}} = 0\degree$C.}
\label{fig:LoadTempParamStudyHalfMBB}
\end{figure}

For the half-MBB beam shown in Fig.~\ref{fig:domainForThermMech_HalfMBBBeam}, a parametric study is carried out under the non-polar case to investigate the influence of thermo-mechanical loading conditions on the optimized topology. Two key parameters are considered: \textit{(i)} the applied mechanical load $P \in \{500,300,100\}~\mathrm{N}$ while maintaining the top and applied bottom-face temperatures at $T_{\text{t}} = 0\degree$C and $\Tb = 800\degree$C, respectively; and \textit{(ii)} the applied bottom-face temperature $\Tb\in \{200,600,800\}~\degree \mathrm{C}$ while fixing the top-face temperature at $T_{\text{t}} = 0\degree$C and the applied load at $P = 300$ N. The topologically optimized designs from both the parametric studies are presented in Fig.~\ref{fig:LoadTempParamStudyHalfMBB}. The first row illustrates the effect of varying the applied mechanical load $P$ while fixing the $\Tb = 800\degree$C. As the load decreases, the optimized material distribution shifts upward, transitioning from a truss-like structure with slender members to a more cellular configuration containing multiple internal voids. This evolution not only enhances structural safety through improved load redistribution but also aids thermal management, as the porous design facilitates heat dissipation and accommodates thermal expansion-induced deformations. The corresponding temperature fields reveal that regions with stronger thermal influence exhibit comparatively lower temperatures within the optimized domain. In the second row of Fig.~\ref{fig:LoadTempParamStudyHalfMBB}, the effect of varying the applied bottom-face temperature $\Tb$ is illustrated. As $\Tb$ increases, the thermal contribution to the strain energy within the domain becomes more pronounced, causing the optimized material distribution to shift toward the cooler zone. The topology evolves from a truss-like structure at $\Tb = 200\degree$C to a cellular arrangement at higher thermal gradients, such as $\Tb = 800\degree$C, while the applied mechanical load $P$ is kept constant at $300$ N. In both cases, when the thermal contribution in the overall elastic strain energy increases, there is a major redistribution of structural material toward the cooler top face. This shift is a direct result of the thermal strain difference between the top and bottom surfaces, which promotes material removal from the hotter regions. Consequently, the lower portion of the beam becomes increasingly hollow, and material near the roller support accumulates into a nearly vertical strut, suggesting a load path dominated by bending resistance in the cooler region.  

\subsubsection{Influence of micropolar coupling number $N$}
\begin{figure}[H]
\centering
\includegraphics[width=0.6\linewidth]{Figures/ColorBar_HalfMBBTM_800.pdf}
    \vspace{0.1cm}
\renewcommand{\arraystretch}{2.5} 
\setlength{\tabcolsep}{6pt}     
\begin{tabular}{>{\centering\arraybackslash}m{1.75cm} | c | c | c}
\hline
 & {$\Tb = 400\degree$C} & {$\Tb = 600\degree$C} & {$\Tb = 800\degree$C} \\
 \hline
 \begin{minipage}[c][\height][c]{1.75cm}
\centering
  \vspace{-1.0cm}
  $N=0.0$
\end{minipage} &
\begin{subfigure}[b]{0.25\linewidth}
\vspace{2mm} 
\includegraphics[width=\linewidth]{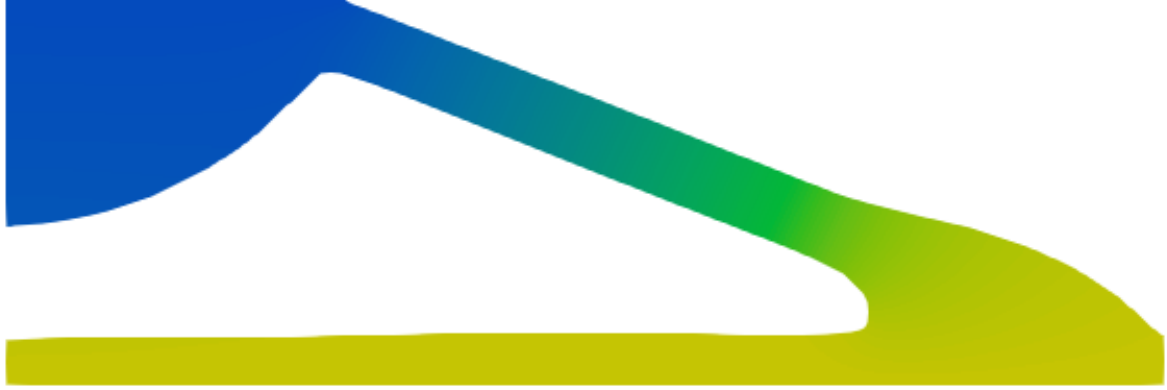}
\end{subfigure} &
\begin{subfigure}[b]{0.25\linewidth}
\includegraphics[width=\linewidth]{Figures/HalfMBBTM_Tb600P300N0.pdf}
\end{subfigure} &
\begin{subfigure}[b]{0.25\linewidth}
\includegraphics[width=\linewidth]{Figures/HAlfMBB_TM_Tb800_P300N.pdf}
\end{subfigure} \\

\hline 

\begin{minipage}[c][\height][c]{1.75cm}
  \centering
  \vspace{-1.0cm}
  $N=0.5$
\end{minipage} &
\begin{subfigure}[b]{0.25\linewidth}
\vspace{2mm} 
\includegraphics[width=\linewidth]{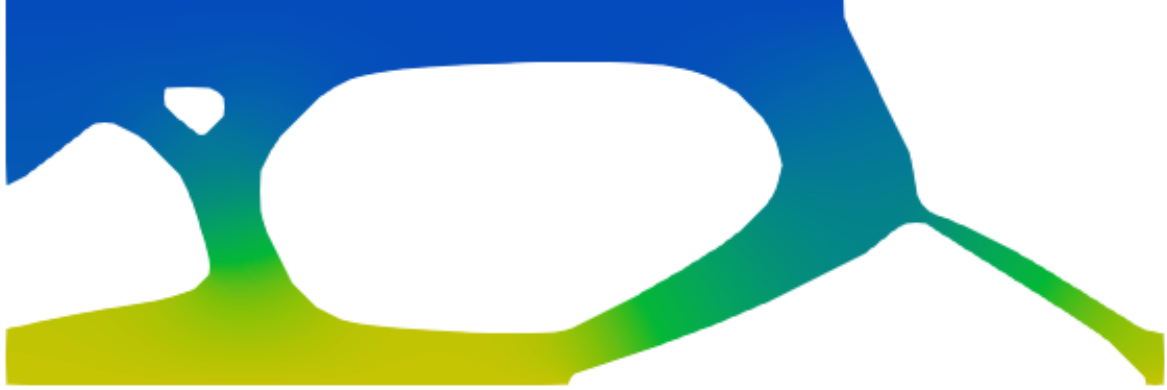}
\end{subfigure} &
\begin{subfigure}[b]{0.25\linewidth}
\includegraphics[width=\linewidth]{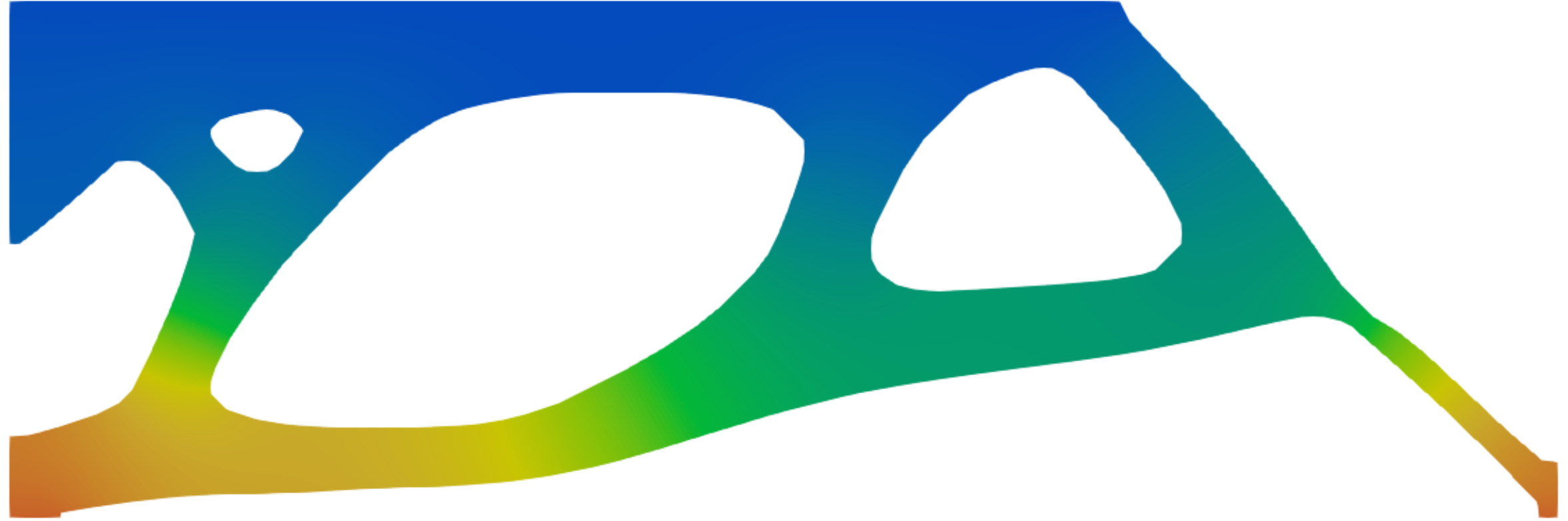}
\end{subfigure} &
\begin{subfigure}[b]{0.25\linewidth}
\includegraphics[width=\linewidth]{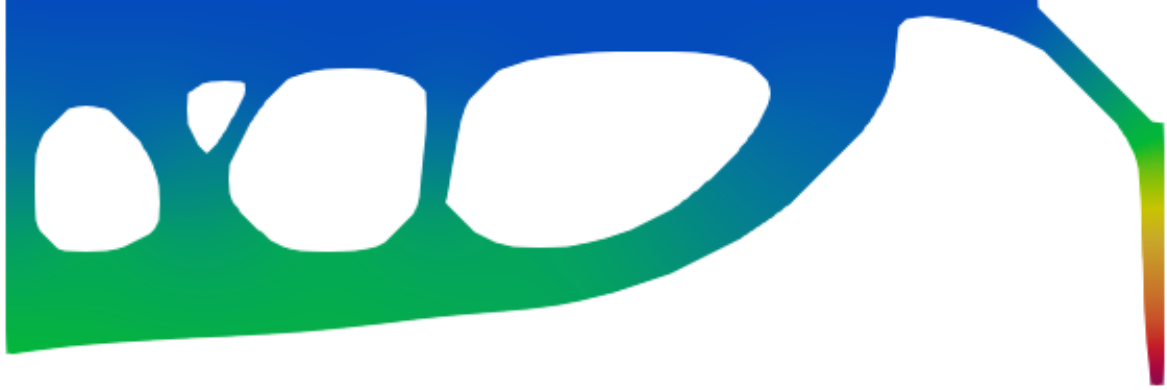}
\end{subfigure} \\
\hline 

\begin{minipage}[c][\height][c]{1.75cm}
  \centering
  \vspace{-1.0cm}
  $N=0.99$
\end{minipage}
 &
\begin{subfigure}[b]{0.25\linewidth}
\vspace{2mm} 
\includegraphics[width=\linewidth]{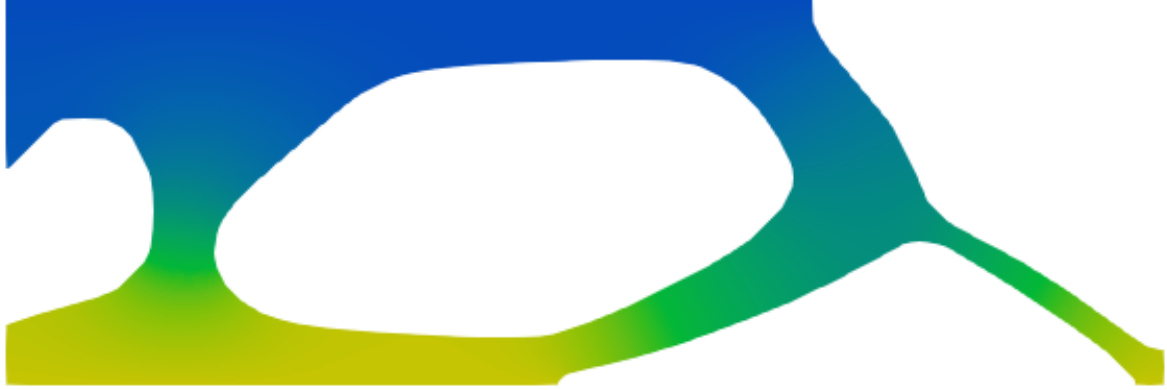}
\end{subfigure} &
\begin{subfigure}[b]{0.25\linewidth}
\includegraphics[width=\linewidth]{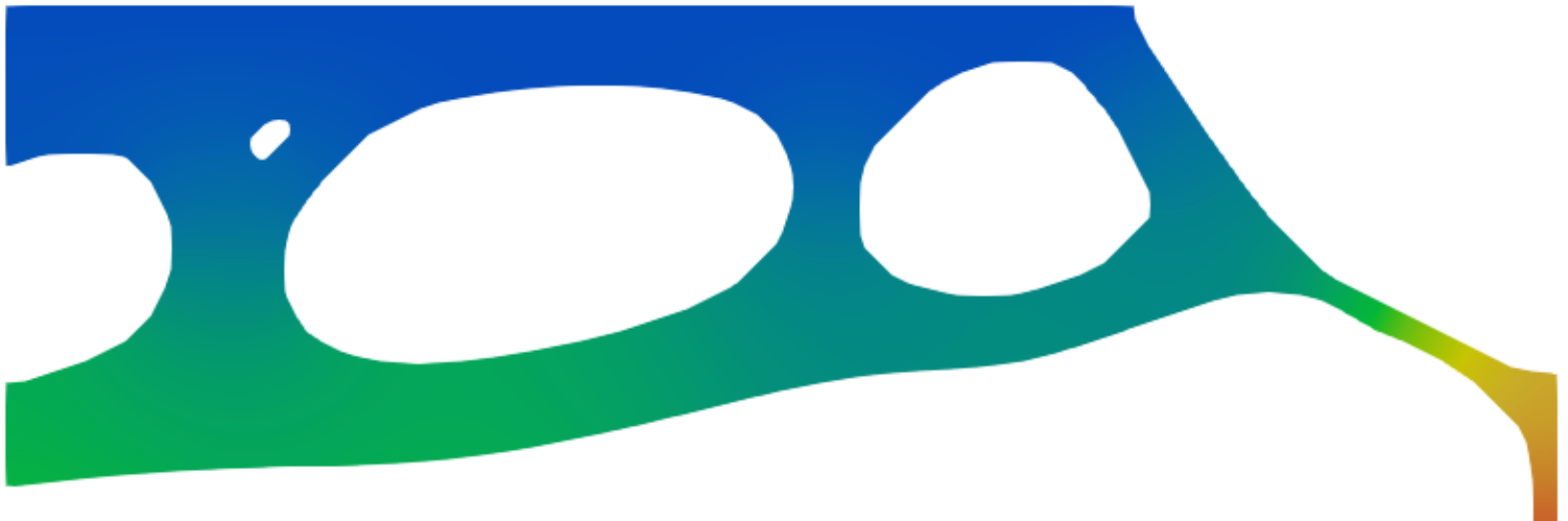}
\end{subfigure} &
\begin{subfigure}[b]{0.25\linewidth}
\includegraphics[width=\linewidth]{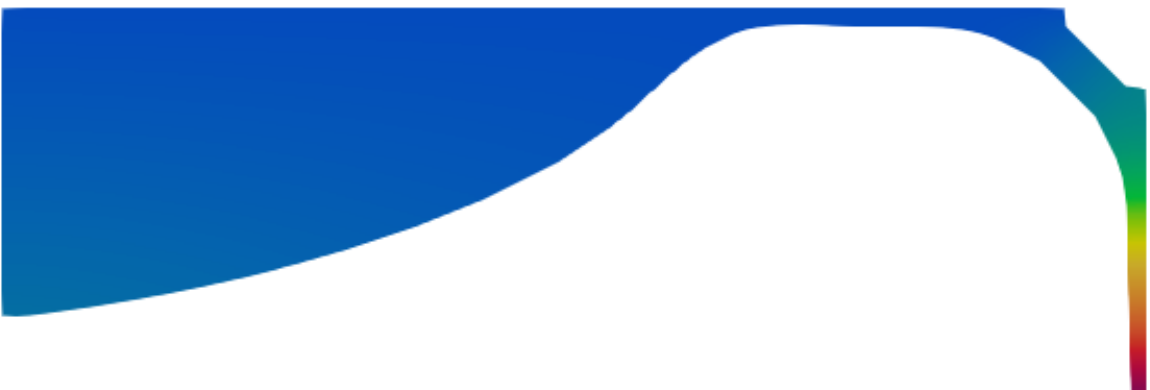}
\end{subfigure} \\
\hline
\end{tabular}
\caption{Topologically optimized designs for the fixed beam in Example III against different applied bottom-face temperature $\Tb$ and micropolar coupling number $N$ with the bending length scale $\lb = H/10 = 40$ mm.}
\label{fig:comparisonforTempValuesHalf-MBB_BeamN0.5}
\end{figure}

\begin{figure}[H]
    \centering
    \begin{subfigure}[b]{0.49\linewidth}
    \centering
        \includegraphics[width=\linewidth]{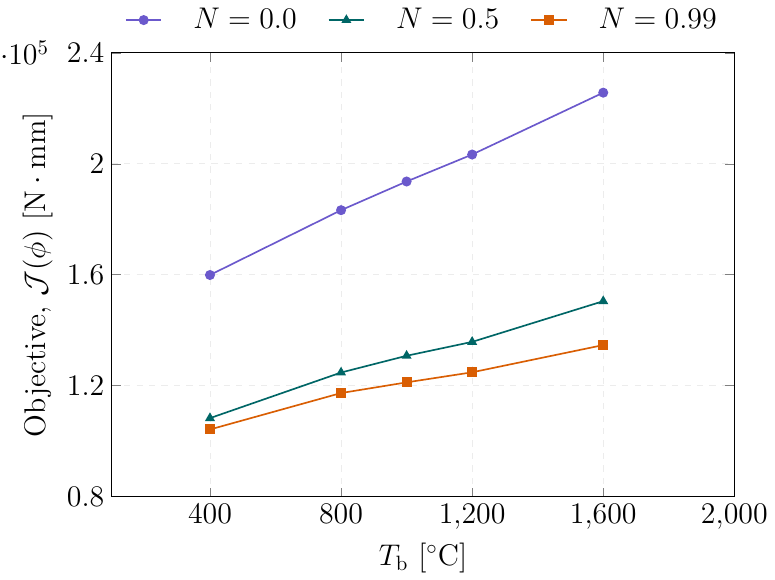}
    \end{subfigure}
    \caption{Elastic strain energy (\textit{i.e.,} the objective) is plotted against varying micropolar coupling number $N$, and applied bottom-face temperature $\Tb$ in Example III for the fixed beam (see Fig.~\ref{fig:domainForThermMech_HalfMBBBeam}) with the bending length scale $\lb = H/10 = 40$ mm.}
\label{fig:HalfMBBdifferentNValues}
\end{figure}

To investigate the influence of the micropolar coupling number $N$ under varying bottom-face temperature $\Tb$, the top surface of the beam is kept at $T_{\text{t}} = 0\degree$C, while the bottom-face temperature $\Tb \in \{400, 600, 800\}$$\degree$C and subjected to a mechanical load of $P = 300$~N. The bending length scale is considered to be $\lb = H/10 = 40$~mm, with all other parameters remaining constant. Fig.~\ref{fig:comparisonforTempValuesHalf-MBB_BeamN0.5} presents the optimized design with a change in $\Tb$ arranged across the columns, while the micropolar coupling number $N$ varies down the rows with values $N = 0.0$, $0.5$, and $0.99$. In the non-polar solids $(N = 0.0)$, the topologically optimized results in the first row of Fig.~\ref{fig:comparisonforTempValuesHalf-MBB_BeamN0.5} show a strong influence of the thermal gradient $\Tb$, as discussed in detail in Section~\ref{sec:MechandThermLoad}. When the micropolar coupling number $N$ is increased to $0.5$, the optimized structure undergoes a notable transformation, exhibiting a more interconnected and distributed material layout. While the effect of $\Tb$ remains evident, the lower part of the beam develops larger void regions, and the material shifts upward toward the load application point. This redistribution is driven by micro-rotations $\thetab$, which generate a strut-like member connecting the main body of the beam to the roller support.  As shown in Fig.~\ref{fig:HalfMBBdifferentNValues}, the inclusion of micropolar effects at $N = 0.5$ enhances structural stiffness by lowering the elastic strain energy of the optimized design and providing greater resistance to thermally induced deformations. The improvement in stiffness arises from the couple stresses and micro-rotations inherent in the micropolar framework. Furthermore, Fig.~\ref{fig:comparisonforTempValuesHalf-MBB_BeamN0.5} highlights that the internal temperature within the optimized domain decreases for $N = 0.5$ compared to the non-polar case, accompanied by a reduction in hole size. The temperature reduction and material redistribution effects become more pronounced at higher thermal gradients, highlighting the stabilizing role of micropolarity under thermo-elastic loading. For a higher value of micropolar coupling number $(N = 0.99)$, the optimized topology evolves further, with thick and continuous members dominating the structure. Compared to the $N = 0.0$ and $N = 0.5$ cases, the hole sizes shrink further, the internal temperature within the domain decreases more noticeably, and the length of the strut-like member increases, reflecting the stronger influence of couple stresses. At higher thermal gradients, the structure demonstrates markedly improved resistance to thermally induced distortions,  making it significantly safer and more stable in maintaining its mechanical performance under combined thermo-elastic loading. Overall, these results establish a clear trend: As $N$ increases from $0.0$ to $0.99$, the half-MBB beam transitions from a thermally sensitive, hollow-dominated design to a dense and thermally stable configuration, highlighting the stabilizing influence of micropolarity. Fig.~\ref{fig:HalfMBBdifferentNValues} further supports this trend by showing the variation of elastic strain energy with increasing applied bottom-face temperature $\Tb$ for different micropolar coupling numbers $N$. While the strain energy consistently rises with $\Tb$, the introduction of micropolarity markedly reduces it—the most notable reduction occurring as $N$ increases from $0.0$ to $0.5$, beyond which ($N = 0.99$) only marginal improvement is observed. This behavior aligns with the minimal geometric differences seen at lower thermal gradients ($\Tb = 400\degree$C), emphasizing the potential of micropolarity in enhancing both mechanical performance and thermal reliability.

\subsubsection{Influence of bending length scale $\lb$}
\label{subsec:lb_and_N_effects}
\begin{figure}[H]
\centering
    \includegraphics[width=0.6\linewidth]{Figures/ColorBar_HalfMBBTM_800.pdf}
    \vspace{0.2cm}
\renewcommand{\arraystretch}{2.5} 
\setlength{\tabcolsep}{6pt}     

\begin{tabular}{>{\centering\arraybackslash}m{1.85cm} | c | c | c}
\hline
 & {$\Tb = 400\degree$C} & {$\Tb = 600\degree$C} & {$\Tb = 800\degree$C} \\
 \hline
 \begin{minipage}[c][\height][c]{1.85cm}
  \centering
  \vspace{-1.0cm}
  $\lb = H/20$
\end{minipage} &
\begin{subfigure}[b]{0.25\linewidth}
\vspace{2mm} 
    \includegraphics[width=\linewidth]{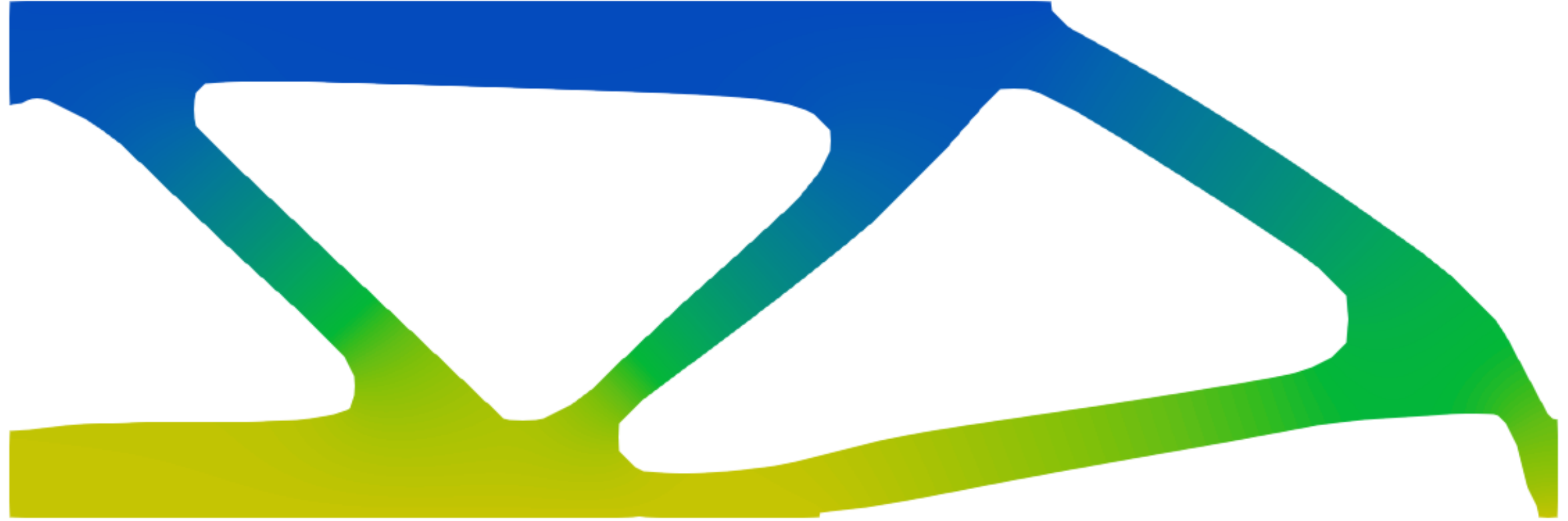}
\end{subfigure} &
\begin{subfigure}[b]{0.25\linewidth}
    \includegraphics[width=\linewidth]{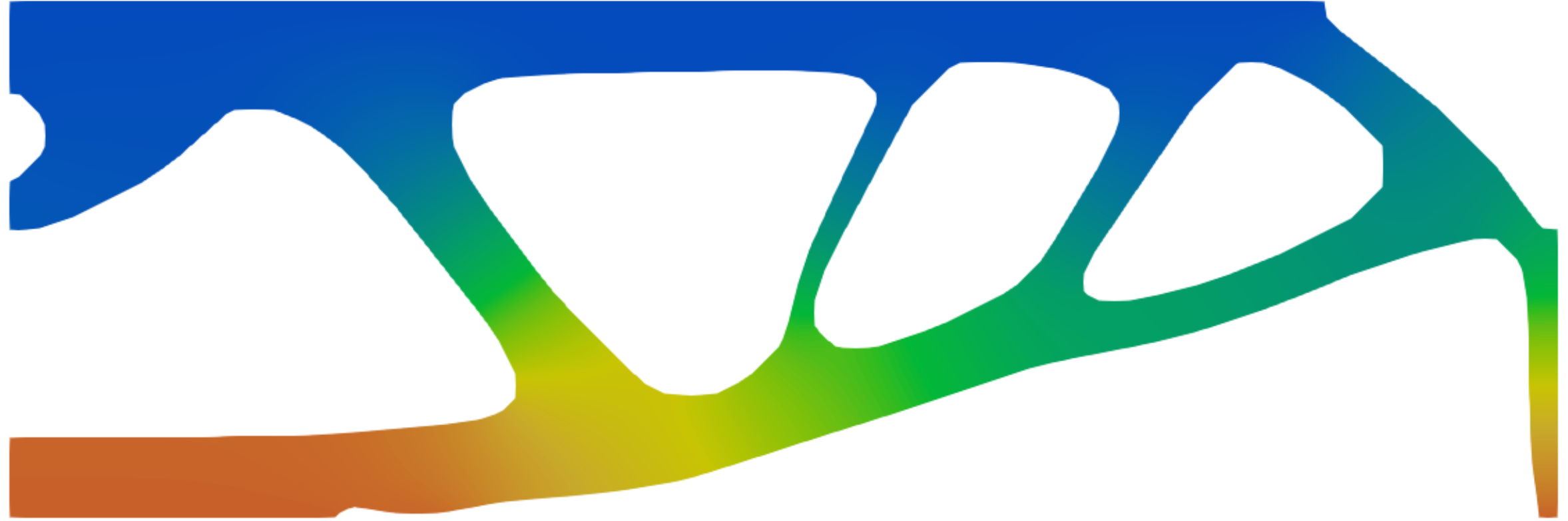}
\end{subfigure} &
\begin{subfigure}[b]{0.25\linewidth}
    \includegraphics[width=\linewidth]{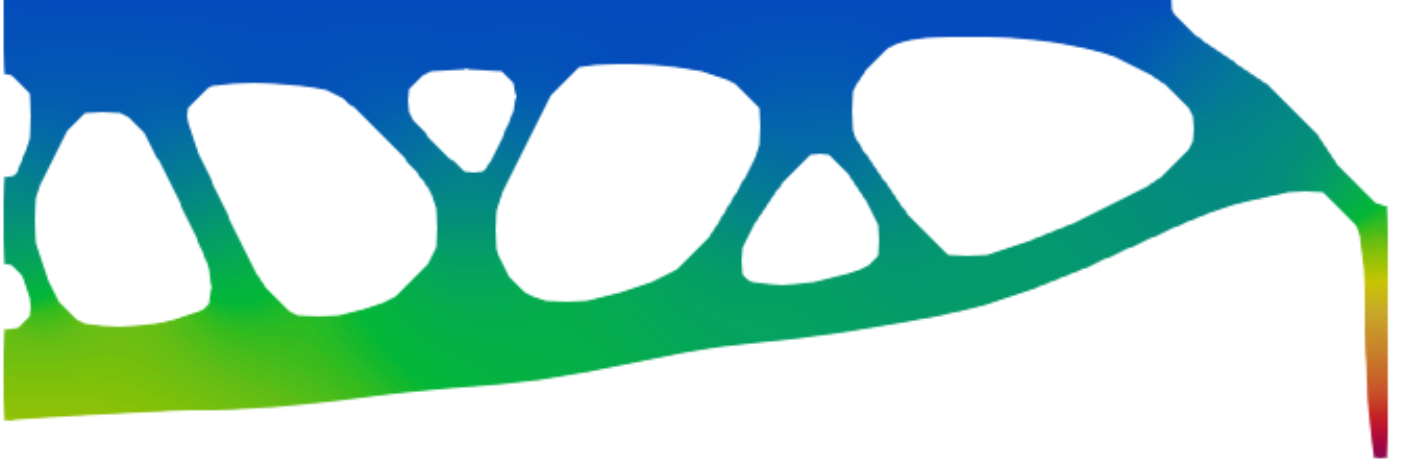}
\end{subfigure} \\

\hline 

\begin{minipage}[c][\height][c]{1.85cm}
  \centering
  \vspace{-1.0cm}
  $\lb = H/10$
\end{minipage} &
\begin{subfigure}[b]{0.25\linewidth}
\vspace{2mm} 
    \includegraphics[width=\linewidth]{Figures/HalfMBBTM_Tb400P300N0.5.pdf}
\end{subfigure} &
\begin{subfigure}[b]{0.25\linewidth}
    \includegraphics[width=\linewidth]{Figures/HalfMBBTM_T600P300N0.5Lb40.pdf}
\end{subfigure} &
\begin{subfigure}[b]{0.25\linewidth}
    \includegraphics[width=\linewidth]{Figures/HalfMBBTM_Tb800P300N0.5Lb40.pdf}
\end{subfigure} \\

\hline 

\begin{minipage}[c][\height][c]{1.85cm}
  \centering
  \vspace{-1.0cm}
  $\lb = H/8$
\end{minipage}
 &
\begin{subfigure}[b]{0.25\linewidth}
\vspace{2mm} 
    \includegraphics[width=\linewidth]{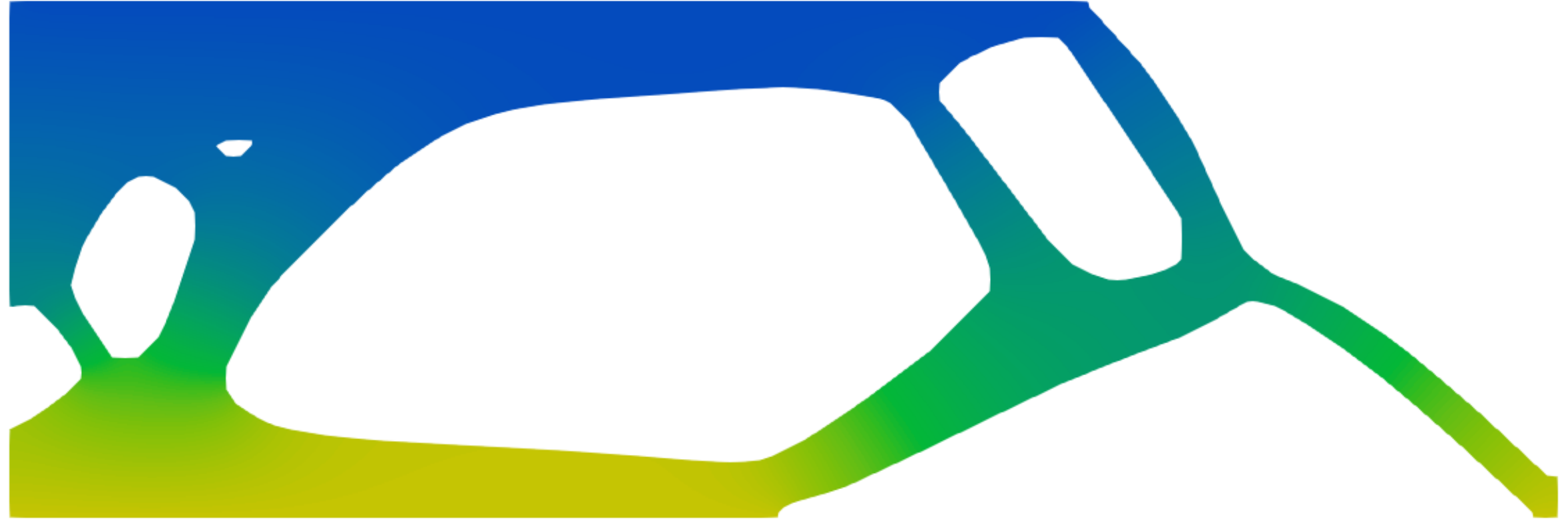}
\end{subfigure} &
\begin{subfigure}[b]{0.25\linewidth}
    \includegraphics[width=\linewidth]{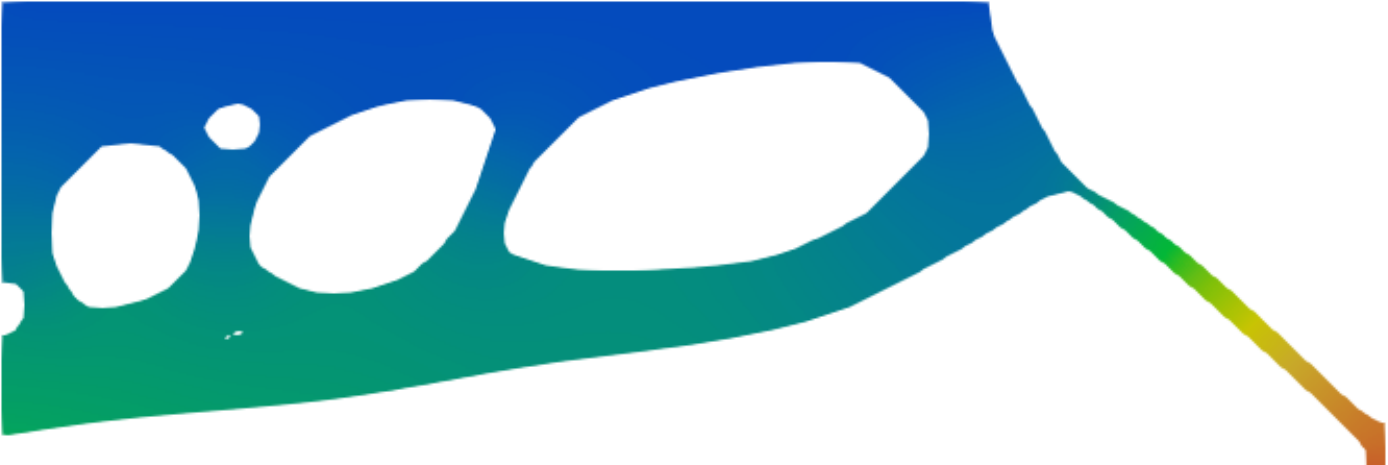}
\end{subfigure} &
\begin{subfigure}[b]{0.25\linewidth}
    \includegraphics[width=\linewidth]{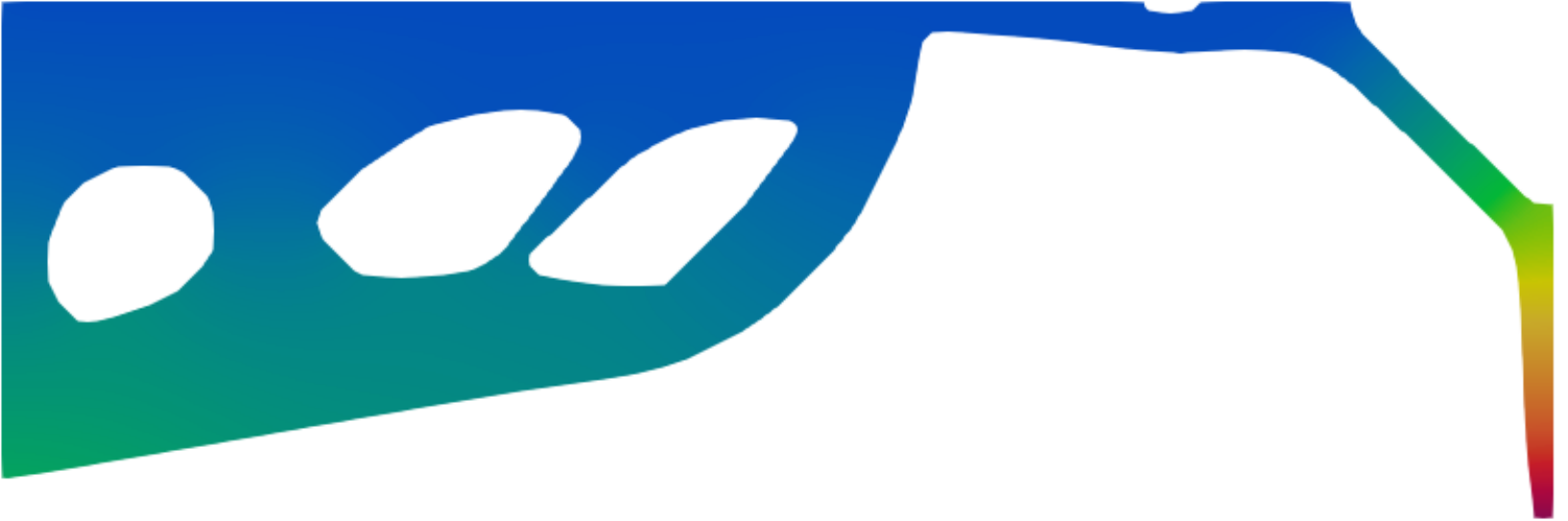}
\end{subfigure} \\
\hline
\end{tabular}
\caption{Topologically optimized designs for the half-MBB beam in Example III against varying applied bottom-face temperature $\Tb$ and bending length scale $\lb$ values with the micropolar coupling number $N = 0.5$.}
\label{fig:comparisonforTempValuesHalf-MBB_BeamN0.5Lbisvaried}
\end{figure}

\begin{figure}[H]
    \centering
    \begin{subfigure}[t]{0.49\linewidth}
    \centering
    \caption{}
    \includegraphics[width=\linewidth]{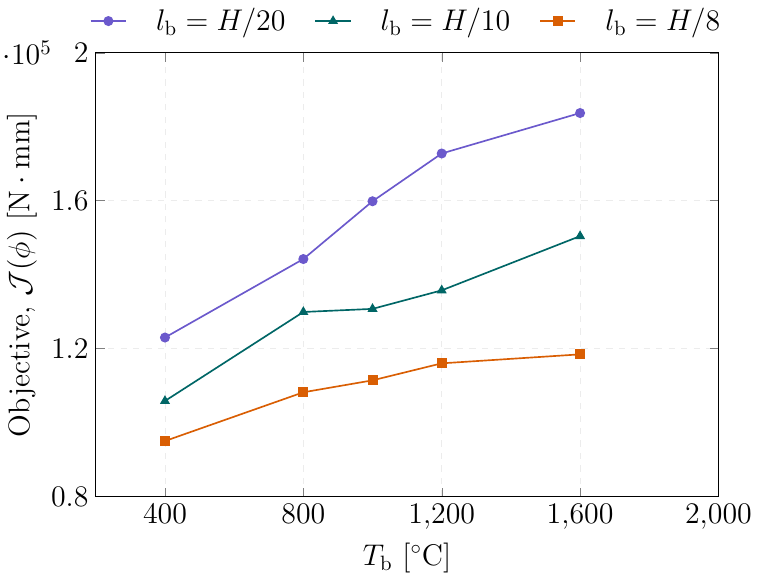}
        
    \label{fig:HalfMBBdifferentLbValues}
    \end{subfigure}
    \begin{subfigure}[t]{0.49\linewidth}
    \centering
    \caption{}
    \includegraphics[width=0.95\linewidth]{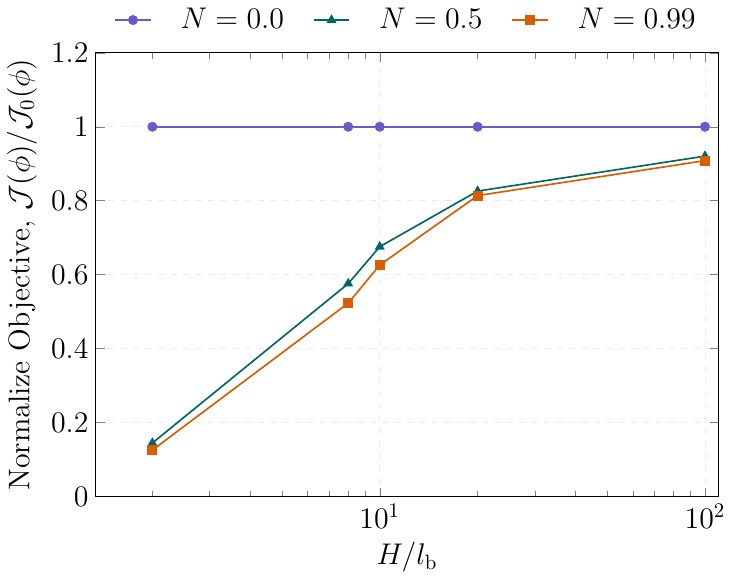}
    \label{fig:HalfMBBdifferentN,LbValuesForTb1000}
    \end{subfigure}
   \caption{Elastic strain energy (\textit{i.e.}, objective) response of the half-MBB beam in Example III is plotted. In (a), it is plotted with the variation of bending length scales at micropolar coupling number $N = 0.5$, and (b) shows the normalized response of the objective for different micropolar parameters $N$, and height-to-bending length scale $(H/\lb)$ while the applied bottom-face temperature $\Tb = 1000\degree$C.}
\label{fig:HalfMBBeam}
\end{figure}

To investigate the influence of the bending length scale $\lb$, we consider a beam subjected to combined thermo-mechanical loading with a concentrated downward point load of $P = 300$ N at the top-left edge under varying applied bottom-face temperature $\Tb$. The top surface is maintained at a constant temperature of $T_{\text{t}} = 0\degree$C, while the applied bottom-face temperature,  $\Tb \in \{ 400\degree$C, $600\degree$C, and $800\degree$C$\}$, and  $\lb \in \{H/20,~H/10,\mathrm{and}~H/8\}$. The micropolar coupling number $N = 0.5$, and all other parameters remain unchanged. Fig.~\ref{fig:comparisonforTempValuesHalf-MBB_BeamN0.5Lbisvaried} presents the resulting topologically optimized designs, with $\Tb$ varying across the columns and the bending length scale $\lb$ decreasing down the rows. For the smallest bending length scale ($\lb = H/20$), the designs retain a pattern similar to conventional non-polar solutions at lower temperatures (see Fig.~\ref{fig:comparisonforTempValuesHalf-MBB_BeamN0.5Lbisvaried}). As the $\Tb$ value increases, the lower edges of the beam undergo significant material removal due to thermal softening, leading to a redistribution of material toward the cooler top region. The resulting material rearrangement reduces the overall temperature within the optimized design. Moreover, as the $\Tb$ value increases, the optimized design changes from a truss-like structure to a cellular shape with more holes which makes the structure more thermally efficient. In the second row of Fig.~\ref{fig:comparisonforTempValuesHalf-MBB_BeamN0.5Lbisvaried}, increasing $\lb$ to $H/10 = 40$ mm amplifies the micropolar effects relative to $\lb = H/20 = 20$ mm. The enhanced micropolarity leads to the elimination of thinner internal members and promotes material redistribution toward the sides, resulting in fewer holes. Moreover, the temperature in the optimized domain decreases for designs with a higher $\lb$ value. Structure becomes stiffer by reducing the elastic strain energy of the system, which can be seen from the plot of Fig.~\ref{fig:HalfMBBeam} (a). In the third row of Fig.~\ref{fig:comparisonforTempValuesHalf-MBB_BeamN0.5Lbisvaried}, for $\lb= H/8 = 50$ mm, optimized design is shown. It is evident that as the micropolarity in the domain increases, it reduces the size and number of holes. Material shifts toward the loading point, and the length of the strut increases. The temperature in the optimized domain decreases with a higher thermal gradient. From Fig.~\ref{fig:HalfMBBeam}, it can also be observed that as the $\Tb$ increases, the elastic strain energy (\textit{i.e.}, the objective)  increases monotonically. Structure becomes more stiffer and thermally efficient, which can be seen from the plot shown in Fig.~\ref{fig:HalfMBBeam} (a) by reducing the elastic strain energy of the optimized design further. Across all $\lb$ values, increasing $\Tb$ drives material upward, leaving larger voids in the lower region of the beam. The upward material shift is most pronounced at small $\lb$, where the design is more sensitive to local stiffness variations, and becomes less significant for larger $\lb$, where microstructural bending resistance dominates. In all cases, the thinner vertical leg observed near the roller support of the beam reflects the optimization process allocating less material to regions carrying lower mechanical load under thermal softening, thereby minimizing weight while maintaining performance.
The quantitative effect of $\lb$ on structural performance is shown in Fig.~\ref{fig:HalfMBBeam} (a), where the elastic strain energy (\textit{i.e.}, the objective) is plotted for different values of $\Tb$. A clear trend emerges: increasing $\lb$ reduces the elastic strain energy, indicating enhanced stiffness. The reduction in strain energy reflects the increased micropolar effects for larger bending length scales, which improve the beam's resistance to thermo-mechanical stresses. The combined effect of the bending length scale $\lb$ and the micropolar coupling number $N$ is illustrated in Fig.~\ref{fig:HalfMBBeam} (b) for applied bottom-face temperature $\Tb = 1000\degree$C. At small values of $H/\lb$ (\textit{e.g.}, 2), the normalized objective decreases substantially, indicating a significant stiffness gain due to enhanced micropolarity (see Fig.~\ref{fig:HalfMBBeam} (b)). As $H/\lb$ increases toward 100, the normalized objective approaches that of the non-polar model, reflecting the reduced influence of micropolarity at larger specimen. Similarly, increasing the micropolar coupling number $N$ from $0.0$ to $0.5$ leads to a marked reduction in normalized strain energy. Beyond $N = 0.5$, the reduction becomes marginal, indicating that most of the stiffness benefit of micropolarity is attained at $N = 0.5$.  Overall, these results highlight the strong interplay between $\lb$ and $N$ in the optimized design and stiffness response under thermo-mechanical loading. 

\section{Conclusions}
\label{sec:conclusions}
In this article, we presented a level set-based topology optimization framework for micropolar solids subjected to thermo-mechanical loading. The formulation is based on a compliance and elastic strain energy minimization problem with a fixed volume fraction constraint and incorporates microstructural effects through the micropolar continuum theory. Numerical studies on both a fixed beam and a half-MBB beam were conducted to investigate how micropolarity and temperature influence the resulting optimal topologies.
The results show that the proposed model effectively redistributes material to enhance thermal resistance by lowering the temperature within the domain. In the fixed beam, material is removed from the heated bottom region and shifts toward the cooler upper region, enhancing thermal resistance. In the half-MBB beam, rising temperature promotes the formation of more cellular and thermally efficient structures, with material migrating toward cooler regions and voids shrinking to mitigate thermal strain. In both cases, increasing the micropolar parameter $N$ causes material to accumulate near the loading zones and reduces structural voids, with major topological changes observed between $N = 0.0$ and $N = 0.5$, and minimal variation beyond $N = 0.5$. Unlike the classical continuum model, which cannot capture intrinsic size effects and exhibits scale-independent topologies, the proposed micropolar formulation introduces a bending length scale $\lb$ and the micropolar coupling number $N$. This enables the prediction of size-dependent stiffening behavior and smoother, more connected topologies. Overall, the proposed approach successfully overcomes the limitations of the classical model by capturing size-dependent effects in thermo-mechanical topology optimization, providing a robust framework for designing thermally and mechanically efficient micropolar structures.

The present framework effectively captures the influence of micropolarity in thermo-mechanical topology optimization but is currently limited to steady-state heat transfer and linear elastic behavior. Future work will aim to extend the formulation to include plastic and ductile material responses, transient thermal effects, and stress-constrained or failure-resistant designs. Additionally, the model will be generalized to multiphysics applications such as coupled electro-mechanical and electro-thermo-mechanical systems, enabling the design of multifunctional materials with improved performance under complex loading conditions.

\section*{Data accessibility}

Source codes for the implementation of the proposed model can be downloaded as Jupyter Notebook files from the repository: \url{https://github.com/MdMasiurRahaman/MicropolarLevelSetTO}.


\section*{Acknowledgments}

The second author, Mr. Ayyappan Unnikrishna Pillai, acknowledges the financial support through the Prime Minister Research Fellowship (PMRF) from the Government of India.
\newpage
\bibliographystyle{elsarticle-num-names}
\bibliography{References.bib}
\appendix

\section{Augmented Lagrangian method for topology optimization} 
\label{sec:AugmentLag}
In topology optimization, one effective approach to handle constraints is the augmented Lagrangian method. This technique reformulates a constrained optimization problem into a sequence of unconstrained problems by simultaneously penalizing constraint violations and updating the associated Lagrange multipliers. For a shape optimization problem, the augmented Lagrangian functional can be expressed as:
\begin{equation}
    \min_{{\Omega} \subset \mathcal{D}} \, \mathcal{L}({\Omega}) = \mathcal{J}({\Omega}) - \sum_{i = 1}^{m} \left[ \lambda_i^L \mathcal{C}_i({\Omega}) + \frac{1}{2} \Lambda_i \left( \mathcal{C}_i({\Omega}) \right)^2 \right],
    \label{eq:Augmented_Lag}
\end{equation}
where $\mathcal{J}({\Omega})$ is the objective functional, $\mathcal{C}_i({\Omega})$ denote the constraints, $\lambda_i^L$ are the Lagrange multipliers, $\Lambda_i$ are penalty parameters, and $m$ is the total number of constraints. The shape derivative of the augmented Lagrangian functional is defined as:
\begin{equation}
    \mathcal{L}'({\Omega}) = \mathcal{J}'({\Omega}) - \sum_{i = 1}^{m} \left[ \lambda_i^L - \Lambda_i \mathcal{C}_i({\Omega}) \right] \mathcal{C}_i'({\Omega}).
    \label{eq:ALderv}
\end{equation}
The augmented Lagrangian method proceeds iteratively. At an iteration $q$, the domain ${\Omega}^{q}$ is updated to reduce $\mathcal{L}({\Omega}^q)$, and the multipliers $\lambda_i^L$, and the penalty parameter $\Lambda_i$ are updated as follows:
\begin{gather}
    \lambda_i^{L,q+1} = \lambda_i^{L,q} - \Lambda_i \, \mathcal{C}_i({\Omega}^{q}), \\
\Lambda_{i}^{q+1} = \min\left( \xi \Lambda_{i}^{q}, \Lambda_{\text{max}} \right),
\end{gather}
with $\xi > 1$ (assumed as $\xi = 1.1$ in this work) and $\Lambda_{\text{max}}$ set to a large value ($1 \times 10^{10}$ herein). To determine a new descent direction at iteration $q$, following \citet{wegert2025gridaptopopt}, the Hilbertian extension-regularization framework is employed. Specifically, we solve $l_{\Omega}^{q+1}$ such that:
\begin{equation}
    \langle l_{\Omega}^{q+1}, w \rangle_{\mathbb{H}} =  \mathcal{L}'(\Omega^{q})(-w \boldn) \quad \forall w \in \mathbb{H},
\label{eq:HilbertExtension}
\end{equation} 
where $\mathbb{H}$ is a Hilbert space over $\mathcal{D}$ with inner product $\langle \cdot, \cdot \rangle_{\mathbb{H}}$. The solution $l_{\Omega}^{q+1}$ extends the shape sensitivity from $\partial\Omega$ to the entire domain $\mathcal{D}$ and guarantees a descent direction for $\mathcal{J}({\Omega}^q)$ with the regularity of $\mathbb{H}$. In this work, $\mathbb{H}$ is chosen as $\mathbb{H}^{1}(\mathcal{D})$, since it provides sufficient regularity for a smooth and stable velocity field in Hamilton-Jacobi evolution. The corresponding inner product is defined as:
\begin{equation}
    \langle u, v \rangle_{\mathbb{H}} = \int_\mathcal{D} \left( \alpha_r^2 \nabla u \cdot \nabla v + u v \right) \, \mathrm{d}V,
\end{equation}
where $\alpha_r$ is the regularization length scale. In this work it is assumed as $4\, N_\mathrm{s}\, \gamma_\mathrm{h} \, h_{\mathrm{e}}$. Here, $N_{\mathrm{s}}$ is maximum number of step used for solving the Hamilton-Jacobi Eq.~\eqref{eq:HJ_eq}, $\gamma_{\mathrm{h}}$ is the time–step coefficient used in the Hamilton–Jacobi update, and $h_{\mathrm{e}}$ is the maximum element size.

\section{Algorithm for the implementation of the proposed method}
\label{sec:TopOptAlgo}
We present our optimization algorithm for solving any micropolar solids subjected to elastic and thermo-elastic loading. The augmented Lagrangian method is used to convert the constrained optimization problem into an unconstrained one. The complete set of steps is summarized in Algorithm~\ref{alg:levelset}.

\newcommand{\Step}[2]{%
  \hangindent=3em
  \hangafter=1
  \noindent\textbf{#1\ }#2\par\vspace{0.1em}
}

\begin{algorithm}[H]
\caption{Level set-based topology optimization for micropolar solids subjected to thermo-elastic loading}
\label{alg:levelset}

\textbf{Initialization:}
Given the computational domain $\mathcal{D}$, and an initial
level set function $\phi^0$ representing the starting shape $\Omega^0 \subset \mathcal{D}$.

\medskip

\textbf{for} $q = 0,1,2,\ldots,q_{\max}$ until convergence \textbf{do}

\hspace{1.7em}\Step{1.}{For the current shape $\Omega^q$, solve Eq.~\eqref{eq:cc} to obtain the state field $\mathbf{U}^q$.}

\hspace{1.7em}\Step{2.}{Evaluate the objective and constraint
shape sensitivities $\mathcal{J}'(\Omega^q)$ and $\mathcal{C}'_i(\Omega^q)$ via Eq.~\eqref{eq:dFdphiFinal}.}

\hspace{1.7em}\Step{3.}{Solve the Hilbertian extension–regularisation problem \eqref{eq:HilbertExtension} using the augmented Lagrangian shape derivative \eqref{eq:ALderv}.}

\hspace{1.7em}\Step{4.}{Solve the Hamilton-Jacobi equation \eqref{eq:HJ_eq}  followed by the reinitialization equation \eqref{eq:HJ_Signed}, giving the updated level set function $\phi^{q+1}$.}

\hspace{1.7em}\Step{5.}{Construct the new domain as
$$\Omega^{q+1} := \{\, \mathbf{x} \in \mathcal{D} : \phi^{q+1}(\mathcal{T},\mathbf{x}) < 0 \,\}.$$}

\medskip
\textbf{return} the optimized shape $\Omega^{q+1}$, and $\phi^{q+1}$.

\textbf{end for}

\end{algorithm}

\section{Validation of the proposed micropolar topology optimization with elastic topology optimization as a special case}
\label{sec:Validationmicro-polar}
\begin{figure}[H]
    \centering
    \includegraphics[width=0.5\linewidth]{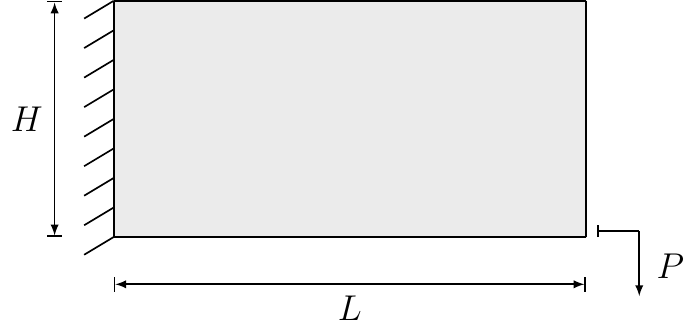}
    \caption{Problem setup from \cite{rovati2007optimal}: Cantilever beam with a downward load applied at the bottom edge. Dimensions: $L = 25$~mm, $H = 10$~mm. The downward load $P = 1$~N, micropolar coupling number $N = 0.8$, and bending length scale $\lb$ are varied.}
\label{fig:CantBeamDownLoad}
\end{figure}

\begin{figure}[H]
\centering
\begin{subfigure}[t]{0.48\linewidth}
\centering
\caption{Optimized design for $\lb = 0.005L$ (proposed method).}
\includegraphics[width=\linewidth]{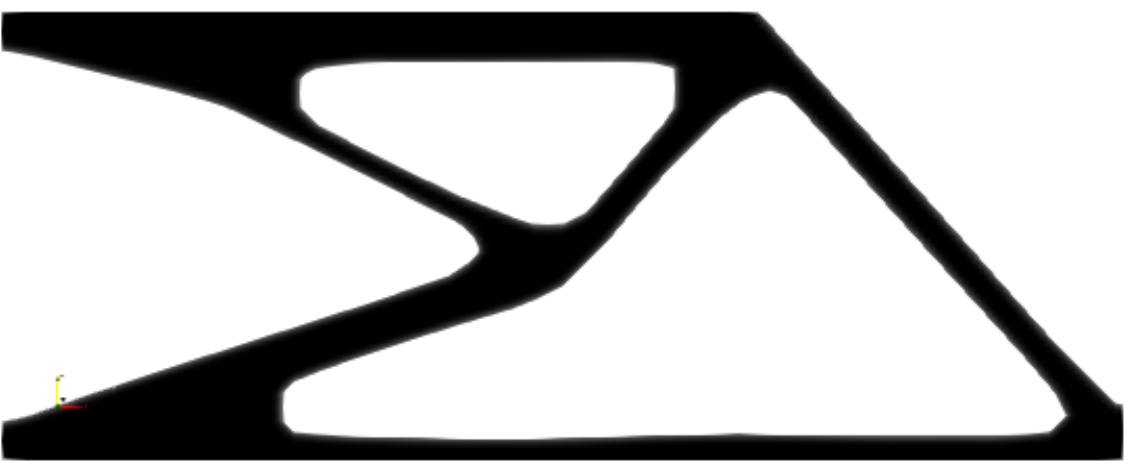}
\label{fig:CantBeamPDF0005}
\end{subfigure}
\begin{subfigure}[t]{0.48\linewidth}
\centering
\caption{Optimized design for $\lb = 0.25L$ (proposed method).}
\includegraphics[width=\linewidth]{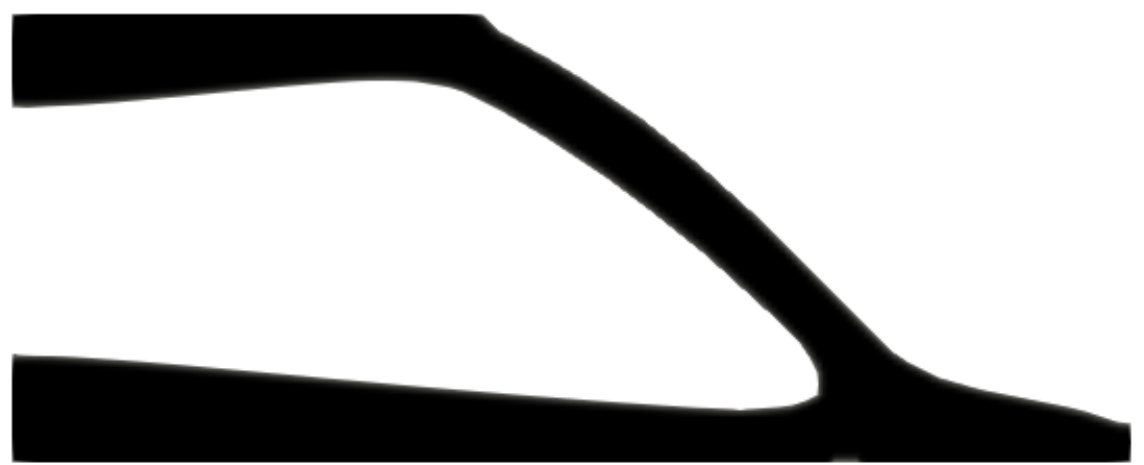}
\label{fig:CantBeamPDF025}
\end{subfigure}
\caption{The optimized designs for the cantilever beam problem. The results obtained from the proposed framework show an excellent match with those reported in Fig.~20 of \citet{rovati2007optimal}.}
\label{fig:CantBeam025}
\end{figure}

To validate the proposed micropolar topology optimization framework, we consider a benchmark problem on elastic topology optimization of a cantilever beam, where a downward point load is applied along the bottom edge as illustrated in Fig.~\ref{fig:CantBeamDownLoad}. The geometry, loading conditions, and material properties are considered from \citet{rovati2007optimal}. The beam has length $L = 25$~mm, height $H = 10$~mm, aspect ratio $(L/H)$ of the beam is $2.5$, and is subjected to a downward load $P = 1$~N under plane strain conditions. The micropolar coupling number $N = 0.8$, while the bending length scale $\lb$ is varied between $\lb = 0.005L = 0.125~\text{mm}$ and $\lb = 0.25L = 6.25~\text{mm}$. Volume fraction $V_f$ is considered as $0.3$. The numerical values considered for the material parameters of the beam are summarized in Table~\ref{tab:ElasticMatProp}.
For the topology optimization problem in the purely elastic case (\textit{i.e.}, without thermal effects), the optimized topologies are obtained from the finite element solution of Eq.~\eqref{eq:TopOptElasticProblem}. The optimized topologies obtained from the present methodology (see Fig.~\ref{fig:CantBeam025}) show excellent agreement with those reported in \citet{rovati2007optimal}. The results confirm that the proposed micropolar topology optimization accurately reproduces the established solutions for both small and large bending length scales. This demonstrates the correctness of the formulation and its ability to consistently reduce to the classical elastic topology optimization problem as a special case.

\end{document}